\documentclass[usenatbib,a4paper]{mn2e}
\usepackage{amsmath}
\usepackage{amssymb}
\usepackage{amsfonts}
\usepackage {multirow}
\usepackage {graphicx}
\usepackage {natbib}
\usepackage{txfonts}
\usepackage{graphicx}
\usepackage{soul}
\usepackage{float}
\usepackage[hyperindex,breaklinks=true, colorlinks, citecolor=blue]{hyperref}
\usepackage{tablefootnote}
\usepackage{eurosym}
\usepackage{ulem}
\usepackage[usenames,dvipsnames]{xcolor}
\usepackage{booktabs}

\def\setsymbol#1#2{\expandafter\def\csname #1\endcsname{#2}}
\def\getsymbol#1{\csname #1\endcsname}

\newbox\tablebox    \newdimen\tablewidth
\def\leaderfil{\leaders\hbox to 5pt{\hss.\hss}\hfil}
\def\tablenote#1 #2\par{\begingroup \parindent=0.8em
    \abovedisplayshortskip=0pt\belowdisplayshortskip=0pt
    \noindent
    $$\hss\vbox{\hsize\tablewidth \hangindent=\parindent \hangafter=1 \noindent
    \hbox to \parindent{$^#1$\hss}\strut#2\strut\par}\hss$$
    \endgroup}

\def\L2{\ifmmode L_2\else $L_2$\fi}

\def\DeltaT{\ifmmode \Delta T\else $\Delta T$\fi}
\def\deltat{\ifmmode \Delta t\else $\Delta t$\fi}
\def\fknee{\ifmmode f_{\rm knee}\else $f_{\rm knee}$\fi}
\def\Fmax{\ifmmode F_{\rm max}\else $F_{\rm max}$\fi}
\def\solar{\ifmmode{\rm M}_{\mathord\odot}\else${\rm M}_{\mathord\odot}$\fi}
\def\Msolar{\ifmmode{\rm M}_{\mathord\odot}\else${\rm M}_{\mathord\odot}$\fi}
\def\Lsolar{\ifmmode{\rm L}_{\mathord\odot}\else${\rm L}_{\mathord\odot}$\fi}
\def\inv{\ifmmode^{-1}\else$^{-1}$\fi}
\def\mo{\ifmmode^{-1}\else$^{-1}$\fi}
\def\sup#1{\ifmmode ^{\rm #1}\else $^{\rm #1}$\fi}
\def\expo#1{\ifmmode \times 10^{#1}\else $\times 10^{#1}$\fi}
\def\,{\thinspace}
\def\lsim{\mathrel{\raise .4ex\hbox{\rlap{$<$}\lower 1.2ex\hbox{$\sim$}}}}
\def\gsim{\mathrel{\raise .4ex\hbox{\rlap{$>$}\lower 1.2ex\hbox{$\sim$}}}}

\def\simprop{\mathrel{\raise .4ex\hbox{\rlap{$\propto$}\lower 1.2ex\hbox{$\sim$}}}}
\def\deg{\ifmmode^\circ\else$^\circ$\fi}
\def\pdeg{\ifmmode $\setbox0=\hbox{$^{\circ}$}\rlap{\hskip.11\wd0 .}$^{\circ}
          \else \setbox0=\hbox{$^{\circ}$}\rlap{\hskip.11\wd0 .}$^{\circ}$\fi}
\def\arcs{\ifmmode {^{\scriptstyle\prime\prime}}
          \else $^{\scriptstyle\prime\prime}$\fi}
\def\arcm{\ifmmode {^{\scriptstyle\prime}}
          \else $^{\scriptstyle\prime}$\fi}
\newdimen\sa  \newdimen\sb
\def\parcs{\sa=.07em \sb=.03em
     \ifmmode \hbox{\rlap{.}}^{\scriptstyle\prime\kern -\sb\prime}\hbox{\kern -\sa}
     \else \rlap{.}$^{\scriptstyle\prime\kern -\sb\prime}$\kern -\sa\fi}
\def\parcm{\sa=.08em \sb=.03em
     \ifmmode \hbox{\rlap{.}\kern\sa}^{\scriptstyle\prime}\hbox{\kern-\sb}
     \else \rlap{.}\kern\sa$^{\scriptstyle\prime}$\kern-\sb\fi}
\def\ra[#1 #2 #3.#4]{#1\sup{h}#2\sup{m}#3\sup{s}\llap.#4}
\def\dec[#1 #2 #3.#4]{#1\deg#2\arcm#3\arcs\llap.#4}
\def\deco[#1 #2 #3]{#1\deg#2\arcm#3\arcs}
\def\rra[#1 #2]{#1\sup{h}#2\sup{m}}

\def\dots{\relax\ifmmode \ldots\else $\ldots$\fi}
\def\WHzsr{\ifmmode $W\,Hz\mo\,sr\mo$\else W\,Hz\mo\,sr\mo\fi}
\def\mHz{\ifmmode $\,mHz$\else \,mHz\fi}
\def\GHz{\ifmmode $\,GHz$\else \,GHz\fi}
\def\mKs{\ifmmode $\,mK\,s$^{1/2}\else \,mK\,s$^{1/2}$\fi}
\def\muKs{\ifmmode \,\mu$K\,s$^{1/2}\else \,$\mu$K\,s$^{1/2}$\fi}
\def\muKRJs{\ifmmode \,\mu$K$_{\rm RJ}$\,s$^{1/2}\else \,$\mu$K$_{\rm RJ}$\,s$^{1/2}$\fi}
\def\muKHz{\ifmmode \,\mu$K\,Hz$^{-1/2}\else \,$\mu$K\,Hz$^{-1/2}$\fi}
\def\MJysr{\ifmmode \,$MJy\,sr\mo$\else \,MJy\,sr\mo\fi}
\def\MJysrmK{\ifmmode \,$MJy\,sr\mo$\,mK$_{\rm CMB}\mo\else \,MJy\,sr\mo\,mK$_{\rm CMB}\mo$\fi}
\def\microns{\ifmmode \,\mu$m$\else \,$\mu$m\fi}

\def\muK{\ifmmode \,\mu$K$\else \,$\mu$\hbox{K}\fi}
\def\microK{\ifmmode \,\mu$K$\else \,$\mu$\hbox{K}\fi}
\def\muW{\ifmmode \,\mu$W$\else \,$\mu$\hbox{W}\fi}
\def\kms{\ifmmode $\,km\,s$^{-1}\else \,km\,s$^{-1}$\fi}
\def\kmsMpc{\ifmmode $\,\kms\,Mpc\mo$\else \,\kms\,Mpc\mo\fi}
\providecommand{\sorthelp}[1]{}

\newcommand{\tC}{\boldsymbol{\rm C}}

\def\reff@jnl#1{{\rm#1\/}}

\def\aj{\reff@jnl{AJ}}                  
\def\araa{\reff@jnl{ARA\&A}}            
\def\apj{\reff@jnl{ApJ}}                
\def\apjl{\reff@jnl{ApJ}}               
\def\apjs{\reff@jnl{ApJS}}              
\def\ao{\reff@jnl{Appl.Optics}}         
\def\apss{\reff@jnl{Ap\&SS}}            
\def\aap{\reff@jnl{A\&A}}               
\def\aapr{\reff@jnl{A\&A~Rev.}}         
\def\aaps{\reff@jnl{A\&AS}}             
\def\azh{\reff@jnl{AZh}}                        
\def\baas{\reff@jnl{BAAS}}              
\def\jcap{\reff@jnl{JCAP}}              
\def\jrasc{\reff@jnl{JRASC}}            
\def\memras{\reff@jnl{MmRAS}}           
\def\mnras{\reff@jnl{MNRAS}}            
\def\pra{\reff@jnl{Phys.Rev.A}}         
\def\prb{\reff@jnl{Phys.Rev.B}}         
\def\prc{\reff@jnl{Phys.Rev.C}}         
\def\prd{\reff@jnl{Phys.Rev.D}}         
\def\prl{\reff@jnl{Phys.Rev.Lett}}      
\def\pasp{\reff@jnl{PASP}}              
\def\pasj{\reff@jnl{PASJ}}              
\def\qjras{\reff@jnl{QJRAS}}            
\def\skytel{\reff@jnl{S\&T}}            
\def\solphys{\reff@jnl{Solar~Phys.}}    
\def\sovast{\reff@jnl{Soviet~Ast.}}     
 \def\ssr{\reff@jnl{Space~Sci.Rev.}}    
\def\zap{\reff@jnl{ZAp}}                
\def\nat{\reff@jnl{Nature}}             
\def\procspie{\reff@jnl{Proceedings of the SPIE}}             

\usepackage{xspace}
\newcommand{\planck}{\textit{Planck}\xspace}

\newcommand{\litebird}{\textit{LiteBIRD}\xspace}

\newcommand{\pico}{\textit{PICO}\xspace}

\def\ben{\begin{enumerate}}
\def\een{\end{enumerate}}
\def\bi{\begin{itemize}}
\def\ei{\end{itemize}}
\def\be{\begin{equation}}
\def\ee{\end{equation}}
\def\bea{\begin{eqnarray}}
\def\eea{\end{eqnarray}}
\def\ba{\begin{align}}
\def\ea{\end{align}}

\def\bdw{\boldsymbol{w }}
\def\bda{\boldsymbol{a }}
\def\bdb{\boldsymbol{b }}
\def\bdd{\boldsymbol{d }}
\def\bde{\boldsymbol{e }}

\def\bdn{\boldsymbol{n }}

\def\bdw{\boldsymbol{w }}

\def\bdf{\boldsymbol{f }}

\newcommand{\tN}{\boldsymbol{{\rm N}}}
\newcommand{\tA}{\boldsymbol{{\rm A}}}

\newcommand{\tB}{\boldsymbol{{\rm B}}}

\newcommand{\healpix}{{\tt HEALPix}}
\newcommand{\psm}{{\tt PSM}}

\makeatletter
\newcommand\footnoteref[1]{\protected@xdef\@thefnmark{\ref{#1}}\@footnotemark}
\makeatother

\title[Constrained moment ILC for $B$-modes]
{Peeling off foregrounds with the constrained moment ILC method to unveil primordial CMB $B$-modes}
\author[Remazeilles, Rotti and Chluba]{Mathieu Remazeilles\thanks{E-mail:~\href{mailto:mathieu.remazeilles@manchester.ac.uk}{\textcolor{black}{mathieu.remazeilles@manchester.ac.uk}}}, Aditya Rotti\thanks{E-mail:~\href{mailto:aditya.rotti@manchester.ac.uk}{\textcolor{black}{aditya.rotti@manchester.ac.uk}}} and Jens Chluba\thanks{E-mail:~\href{mailto:jens.chluba@manchester.ac.uk}{\textcolor{black}{jens.chluba@manchester.ac.uk}}}
\\
Jodrell Bank Centre for Astrophysics, School of Physics and Astronomy, The University of Manchester, Oxford Road, Manchester, M13 9PL, U.K.
}

\begin{document}

\date{\vspace{-6mm}{Accepted  --. Received }}

\maketitle

\begin{abstract}
Galactic foregrounds are the main obstacle to observations of the cosmic microwave background (CMB) $B$-mode polarization. In addition to obscuring the inflationary $B$-mode signal by several orders of magnitude, Galactic foregrounds have non-trivial spectral signatures that are partially unknown and distorted by averaging effects along the line-of-sight, within the pixel/beam window, and by various analysis choices (e.g., spherical harmonic transforms and filters). Statistical moment expansion methods provide a powerful tool for modeling the effective Galactic foreground emission resulting from these averaging effects in CMB observations, while blind component separation treatments can handle unknown foregrounds.
In this work, we combine these two approaches to develop a new {\it semi-blind} component separation method at the intersection of parametric and blind methods, called {\it constrained moment ILC} (\texttt{cMILC}). This method adds several constraints to the standard ILC method to de-project the main statistical moments of the Galactic foreground emission. 
Applications to maps are performed in needlet space and when compared to the \texttt{NILC} method, this helps significantly reducing residual foreground contamination (bias, variance, and skewness) in the reconstructed CMB $B$-mode map, power spectrum, and tensor-to-scalar ratio. We consider sky-simulations for experimental settings similar to those of \litebird and \pico, illustrating which trade-offs between residual foreground biases and degradation of the constraint on $r$ can be expected within the new \texttt{cMILC} framework. We also outline several directions that require more work in preparation for the coming analysis challenges.
\end{abstract}


\section{Introduction}
\label{sec:intro}

The search for the primordial $B$-mode polarization (curl-like pattern) of the cosmic microwave background (CMB) radiation is recognised as one of the ultimate challenges in CMB cosmology. The primordial CMB $B$-mode signal would indeed be a clear signature of the primordial gravitational waves of quantum origin predicted by inflation \citep[e.g.][]{Starobinsky1983}, when the Universe underwent ultra-rapid accelerated expansion just about ${10^{-35}}$ seconds after the Big Bang. The amplitude of the primordial CMB $B$-mode power spectrum, parametrised using the tensor-to-scalar ratio $r$, would allow determining the energy scale of inflation: ${V^{1/4}\simeq (r/0.01)^{1/4}\times 10^{16}}$\,GeV \citep{knox2002}, and there by distinguish main classes of early Universe models \citep[e.g.,][]{Baumann2009, Martin2014P}. 

One of the main goals for next-generation CMB experiments is a statistically significant detection of primordial CMB $B$-modes down to $r \simeq 10^{-3}$. From space, this effort is led by \litebird \citep{Litebird2019}, whose launch is planned by JAXA for early 2028. Another space concept is \pico~\citep{PICO2019}, which is currently in NASA's concept study phase. In addition, ambitious future missions with a broad range of capabilities including CMB spectroscopy are being discussed for the ESA Vogage 2050 program  \citep{Jacques2019Voyage, Basu2019Voyage, Chluba2019Voyage}. 
From the ground, the Simons Observatory \citep[SO;][]{SO2019}, whose survey should start in 2021, and the CMB-S4 \citep{CMB-S4_2016} project are relentlessly moving forward.
For all these efforts, CMB foregrounds pose a real challenge because the $B$-mode signal is extremely faint ($\leq 50$\, nK r.m.s fluctuations), with intense polarized Galactic emission (mostly synchrotron and thermal dust emission) obscuring it by several orders of magnitude. In addition, gravitational lensing effects on the CMB by the large-scale structure and several inst rumental systematics create spurious contamination to CMB $B$-modes. 
Component separation methods are thus critical, since the residual foreground contamination in the recovered CMB $B$-mode map will set the ultimate uncertainty limit with which $r$ can be determined  \citep{Remazeilles2018JCAP}. 

Several component separation methods have been developed in the literature to disentangle the primary CMB signal from the foregrounds: 
\texttt{Commander} \citep{Eriksen2008}, \texttt{SMICA} \citep{Delabrouille2003,Cardoso2008}, \mbox{\texttt{NILC} \citep{Delabrouille2009}}, \texttt{SEVEM} \citep{Fernandez-Cobos2012}, and \texttt{GNILC} \citep{Remazeilles2011b}, to name only those that have been used for the data product release of the ESA's \planck\ satellite \citep{Planck2018_IV}. 
These methods (which have been successfully applied to \planck\ temperature and $E$-mode data for which the signal-to-foregrounds ratio was relatively large) are now being recycled, along with most recent algorithms like \texttt{xForecast} \citep{Stompor2016,Errard2016},  \texttt{BFoRe} \citep{Alonso2017} and tailored template fitting \citep[\texttt{Delta-map};][]{Ichiki2019}, for new faint signal-to-foregrounds regimes in $B$-mode forecasts \cite[e.g.,][]{Remazeilles2018JCAP}. 

Some of these methods, like the Bayesian fitting method \texttt{Commander }and maximum-likelihood fitting method \texttt{xForecast}, are \textit{parametric} since they rely on astrophysical models of the foreground emission, while other methods, like the variance-minimization method \texttt{NILC}, are \textit{blind} in the sense that no explicit assumption is made on foregrounds. Each of them have their advantages and weaknesses, but consistency between blind and parametric methods on CMB $B$-mode reconstruction is desired for claiming a robust detection.
In this respect, many lessons have been learned \citep{Flauger2014,BicepKeckPlanck2015}, and we now understand the importance of further developing component separation methods for $B$-modes that allow us to eliminate biases arising from residual foregrounds.
In addition, the optimal method depends on the observable that is targeted, and there may be no {\it one-rules-it-all} approach.

Component separation methods rely on the distinct spectral signatures of the various components of emission (CMB, foregrounds) to disentangle them in multi-frequency sky observations. However, Galactic foregrounds have non-trivial spectral energy distributions (SEDs), whose exact properties are still unknown at the sensitivity levels of $r\simeq 10^{-3}$.  In parametric approaches, a slight mismodelling of the foregrounds may lead to large biases on $r$ due to huge amplitude disparity between foregrounds and primordial $B$-modes \citep{Remazeilles2016, Hensley2018}. 
Similarly, the thermal dust emission from the interstellar medium is expected to be partially \textit{decorrelated} across frequencies, i.e. the dust spectral parameters (spectral index and temperature) may not only vary across the sky but also across frequencies, because of the averaging of multiple cloud contributions of different spectral indices and temperatures along the line-of-sight \citep{Tassis2015}. Thus, while still sufficient at \planck sensitivity, the common modelling of the dust SED as a unique modified blackbody across frequencies may break at the targeted $B$-mode sensitivity, and lead to significant biases on $r$ after component separation.

To minimize the effects of unknown foreground complexities, blind analysis methods provide a powerful remedy, while averaging effects of known SEDs can in principle be modeled parametrically through moment expansion methods \citep{Stolyarov2005, Chluba2017}. Aside from line-of-sight averaging effects, similar spectral averaging effects arise from the limited beam/pixel resolution of the sky maps and various analysis choices, such as spherical harmonic decomposition, downgrading the resolution, or generally when applying filters \citep{Chluba2017}. The effective SED of the foreground emission in a given pixel/beam may therefore differ from the expected SED shape in each line-of-sight, since the former is the average over several line-of-sight SEDs.  

\begin{figure}
  \begin{center}
    \includegraphics[width=\columnwidth]{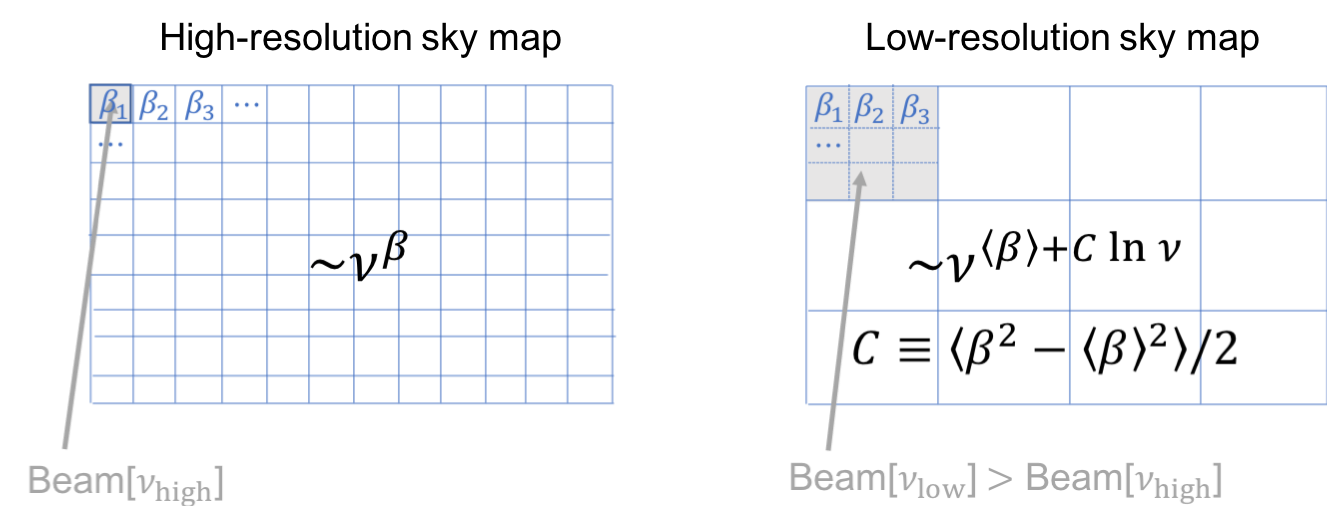}
   \end{center}
\caption{Mismodelling of the foregrounds across frequency due to beam averaging in multi-resolution sky maps: a synchrotron power-law spectrum at high angular resolution becomes a \textit{curved} power-law at lower resolution, with the curvature given by the second-order moment of the spectral index.}
\label{Fig:sketch}
\end{figure}

As an illustration (see Fig.~\ref{Fig:sketch}), let us assume that, in each single pixel of a high-resolution sky map, the synchrotron emission  can be accurately described by a single power-law, $f_{\rm sync}(\nu)\simeq \nu^{\,\beta}$, across frequencies $\nu$, with a pixel-dependent spectral index $\beta(p)$ varying across the sky. While this SED model might be approximately valid at the given map resolution and sensitivity, it can actually break for maps of lower angular resolution, for which the effective SED is now the average of multiple power-laws within a larger beam:
\begin{align}
&\langle f_{\rm sync}(\nu) \rangle 
= \langle \nu^{\,\beta(p)}\rangle
\cr
&\simeq \left< {\rm e}^{\langle\beta\rangle \ln \nu} +  \left(\beta -\langle \beta \rangle\right){{\rm d}  {\rm e}^{\beta \ln \nu} \over {\rm d}\beta}\vert_{\beta=\langle \beta \rangle} + {1\over 2}\left(\beta -\langle \beta \rangle\right)^2{{\rm d}^2  {\rm e}^{\beta \ln \nu} \over {\rm d}\beta^2}\vert_{\beta=\langle \beta \rangle} \right>
\cr
&\simeq  {\rm e}^{\langle\beta\rangle \ln \nu}
\left< 
1 +  \left(\beta -\langle\beta\rangle\right)\ln \nu + {1\over 2}\left(\beta -\langle\beta\rangle\right)^2 \left(\ln \nu\right)^2  \right>
\cr
&\simeq {\rm e}^{\langle\beta\rangle \ln \nu}\left( 1 + {1\over 2}\langle \beta^2 -\langle\beta\rangle^2\rangle \left(\ln \nu\right)^2  \right)
\simeq \nu^{\,\langle\beta\rangle + C\ln\nu}.
\label{eq:decorr}
\end{align}
Here, we defined the beam-average $\langle Y(p,\nu) \rangle\equiv \langle Y(p,\nu) \rangle_{p\,\in\, {\rm beam}[\nu]}$.
As Eq.~\eqref{eq:decorr} shows, the average SED is no longer a power-law but a \textit{curved} power-law at lower angular resolution, with an effective curvature ${C={1\over 2}\langle \beta(p)^2 -\langle\beta\rangle^2\rangle}$ given by the variance (second-order moment) of the spectral index within the beam \citep{Chluba2017}. In the last step, we resummed the series, which assumes that all higher order moment are Gaussian, which in general may not be valid.
Therefore, while the spectral intensity of the synchrotron might be accurately modelled by a power-law at one resolution, the power-law model is no longer valid at lower angular resolution. This results in an effective decorrelation of the foreground emission across frequencies.
For similar reasons, omitting effective curvature due to beam averaging in the SED modelling of thermal dust can induce non-negligible biases on $r\lesssim 10^{-3}$ \citep{Remazeilles2018JCAP}. Modelling as many moments of the foreground emission as possible, given the sensitivity limits of an experiment, thus provides a natural way for capturing some of the additional complexity of the CMB foreground contamination.

With this in mind, in this work we develop a {\it semi-blind} component separation method, called {\it constrained moment ILC} (\texttt{cMILC}), which operates at the intersection of blind ILC methods and parametric foreground modelling methods. We extend the \texttt{NILC} method by adding several nulling constraints on the effective SEDs of the main moments of the dust and synchrotron emissions in order to remove the bulk of the residual foreground contamination in the recovered CMB $B$-mode map. This is achieved by combining the moment expansion technique \citep{Chluba2017} with the Constrained ILC method \citep{Remazeilles2011}. While here the focus is on $B$-mode signal, this type of semi-blind approach has broader applicability and was recently introduced for the extraction of the relativistic Sunyaev-Zeldovich (SZ) effect \citep{Remazeilles2019b} and CMB spectral distortions \citep{Rotti2021}. In harmonic space, the power of moment methods was furthermore recently explored on \planck data \citep{Mangilli2019}. Moment expansion techniques have also been used in the modeling of SZ signals \citep{Chluba2012moments}.
First-order moments have also been recently used to augment internal template fitting methods for $B$-mode foreground cleaning \citep{Ichiki2019}.

The paper is organised as follows. In Sect.~\ref{sec:basics}, we review the basics of the ILC and Constrained ILC methods. In Sect.~\ref{sec:method}, we present our new semi-blind component separation method \texttt{cMILC}, which plugs the foreground moment expansion technique described in Sect.~\ref{subsec:moments} into the Constrained ILC method, as outlined in Sect.~\ref{subsec:semiblind}. We discuss the statistical properties of residual foregrounds and noise in Sect.~\ref{subsec:stat}. We then present the results of our $B$-mode analysis on sky simulations for experimental settings similar to \litebird and \pico in Sect.~\ref{sec:analysis}. We discuss some directions for optimisation of the method in Sect~\ref{sec:disc}, and draw our conclusions in Sect.~\ref{sec:conc}.

\section{Basics of ILC methods}
\label{sec:basics}

\subsection{The standard ILC}
\label{subsec:ilc}

The standard internal linear combination \citep[ILC;][]{WMAP2003,Tegmark2003,Eriksen2004,Delabrouille2009,Basak2012} is a blind component separation method as there is no attempt at directly modelling the foregrounds. Hence, the sky observations $d_\nu(p)$, at each frequency $\nu$ and for each pixel $p$ are written as
\begin{align}
\label{eq: sky_ilc}
d_\nu(p) = a_\nu\, s(p) + n_\nu(p),
\end{align}
where $a_\nu$ is the known spectral response (SED)\footnote{Modelled by the first temperature derivative of the blackbody spectrum.}  of the CMB anisotropies $s(p)$ at frequency $\nu$, while $n_\nu(p)$ is the un-modelled overall contamination from foregrounds and instrumental noise. 

For convenience with the algebra, we recast Eq.~\eqref{eq: sky_ilc} in a $n_f\times 1$ vector form, where $n_f$ is the number of frequency channels:
\begin{align}
\bdd(p) = \bda s(p) + \bdn(p).
\end{align}
Here, $\bdd(p)=\{d_\nu(p)\}_\nu$ collects the set of available frequency maps, $\bda=\{a_\nu\}_\nu$ is the known CMB SED vector, $s(p)$ corresponds to the unknown CMB anisotropies that we aim at extracting, and 
${\bdn(p)=\{n_\nu(p)\}_\nu}$  collects the un-modelled foregrounds and noise that we aim at mitigating in the reconstructed CMB map. 

The standard ILC estimate, $\hat{s}(p)$, of the CMB anisotropies is obtained by forming a weighted linear combination of the frequency maps
\begin{align}
\hat{s} = \sum_\nu w_\nu\, d_\nu \equiv \bdw^{\rm T}\bdd
\end{align}
that is of minimum variance, i.e. 
\begin{align}
\label{eq:varmin}
{\partial \over \partial \bdw}\langle \hat{s}^{\,2} \rangle = 0.
\end{align}
Here, the ILC weights $\bdw=\{w_\nu\}_\nu$ assigned to the frequency maps are constrained to provide unit response to the CMB SED:
\begin{align}
\label{eq:wa1}
\bdw^{\rm T}\bda = 1
\end{align}
in order to guarantee the full conservation of the CMB signal $s(p)$ in the variance minimization. This ensures that the ILC estimate
\begin{align}
\label{eq:ilc}
\hat{s} = \bdw^{\rm T}\bdd = \bdw^{\rm T}\left(\bda s + \bdn\right) = s + \bdw^{\rm T}\bdn
\end{align}
does not alter the CMB signal $s$ [i.e. no multiplicative error thanks to the constraint Eq.~\eqref{eq:wa1}], while the variance of the residual foreground and noise $\bdw^{\rm T}\bdn$ is minimized by Eq.~\eqref{eq:varmin}. 

The variance of the ILC map, $\langle \hat{s}^{\,2}\rangle$, relates to the covariance of the data $\bdd$ as
\begin{align}
\label{eq:ilcvar}
\langle\hat{s}^{\,2}\rangle = \bdw^{\rm T} \langle \bdd \bdd^{\rm T} \rangle \bdw = \bdw^{\rm T} \tC\, \bdw,
\end{align}
where $\tC =  \langle \bdd \bdd^{\rm T} \rangle$ is the  $n_f\times n_f$ covariance matrix of the data whose elements are ${\rm C}_{\nu\nu'} = \langle d_\nu d_{\nu'}\rangle$ for any pair of frequencies $\left(\nu,\nu'\right)$. An empirical estimate of the data covariance matrix in each pixel $p$ can be obtained by ergodicity:
\begin{align}
\label{eq:empirical}
\tC(p) = {1\over N_p} \sum_{p' \in\, \mathcal{D}(p)} \bdd (p') \bdd(p')^{\rm T},
\end{align}
where $\mathcal{D}(p)$ is a domain of pixels surrounding pixel $p$, and $N_p$ the number of pixels in this domain.

The ILC weights $\bdw$ are thus the solution of the constrained variance-minimization problem:
\begin{equation}
\label{eq:ilcm}
\begin{cases}
{\partial \over \partial \bdw} \left(\bdw^{\rm T} \tC\, \bdw\right) = 0,\\[1.5mm]
\bdw^{\rm T}\bda = 1,
\end{cases}
\end{equation}
which can be addressed with a Lagrange multiplier $\lambda$ by solving
\begin{align}
{\partial \over \partial \bdw} \left[\bdw^{\rm T} \tC\, \bdw  +\lambda\left(1 - \bdw^{\rm T}\bda\right)\right]=0.
\end{align}
This yields ${\bdw^T = \lambda \bda^{\rm T} \tC^{-1}}$, while $\bdw^{\rm T}\bda = 1$ leads to ${\lambda = \left(\bda^{\rm T} \tC^{-1}\,\bda\right)^{-1}}$, so that the standard ILC weights are given by
\begin{align}
\label{eq:ilcw}
\bdw^{\rm T} = \left(\bda^{\rm T} \tC^{-1}\,\bda\right)^{-1}\,\bda^{\rm T} \tC^{-1}.
\end{align}

\subsection{The Constrained ILC (\texttt{cILC})}
\label{subsec:cilc}
While the variance of the global contamination (i.e. foregrounds plus noise) is minimized by the standard ILC estimate, this is not necessarily the solution with minimal foreground variance. Instead of minimizing the variance of the global contamination, the Constrained ILC (\texttt{cILC}) method \citep{Remazeilles2011} extends the standard ILC to fully eliminate selected foreground residuals by using prior knowledge on their SED and adding \textit{orthogonality} constraints to the variance-minimization problem:
\begin{equation}
\label{eq:cilc}
\begin{cases}
{\partial \over \partial \bdw} \left(\bdw^{\rm T} \tC\, \bdw\right) = 0 
\\
\bdw^{\rm T}  \bda = 1 
\\
\bdw^{\rm T}  \bdb_1 = 0 
\\
\vdots 
\\
\bdw^{\rm T}  \bdb_m = 0, 
\end{cases}
\end{equation}
where $\bdb_1, \cdots, \bdb_m$ are the SED vectors of $m$ modelled foreground components, to which the \texttt{cILC} weights $\bdw$ are orthogonal.
If we write the sky observations as
\begin{align}
\label{eq: sky_cilc}
\bdd(p) = \bda\, s(p)\, +\, \sum_{i=1}^m \bdb_i \,g_i(p)\, + \bdn(p),
\end{align}
where $\{g_i(p)\}_{1\leq i \leq m}$ are the $m$ foreground components whose SEDs $\{\bdb_i\}_{1\leq i \leq m}$ are modelled, while $\bdn(p)$ includes un-modelled foregrounds and instrumental noise, then the orthogonality constraints $\bdw^T\bdb_i=0$ of the \texttt{cILC} method in Eq.~\eqref{eq:cilc} guarantee the full cancellation of the residual foreground contamination from the $m$ modelled components in the reconstructed CMB map:
\begin{align}
\hat{s}(p)
& =\left(\bdw^{\rm T}  \bda\right)\,s(p)\, +\, \sum_{i=1}^m \left(\bdw^{\rm T} \bdb_i\right) \,g_i(p)\, + \bdw^{\rm T}\bdn(p) \cr
& = s(p) + \bdw^{\rm T}\bdn(p).
\end{align}
This significantly differs from the ILC method (Eq.~\ref{eq:ilcm}) which suffers from additional foreground residuals as $\sum_{i=1}^m \left(\bdw^{\rm T} \bdb_i\right) \,g_i(p) \neq 0$ in this case. However, the penalty of adding nulling constraints on some foregrounds in the \texttt{cILC} is an increase of the noise residuals $\bdw^{\rm T}\bdn(p)$ compared to the standard ILC. But there is a sweet spot between noise cost and reduction of foreground biases, as we show in this work. This optimization furthermore depends on the observable that is targeted, which in our case is $r$ (see Sect.~\ref{subsec:results}).

If we introduce the $n_f \times (m+1)$ SED matrix ${\tA =[\bda\, \bdb_1\, \cdots \bdb_m]}$, whose columns collect the SED vectors of the modelled components, and the $1 \times (m+1)$ transposed vector  ${\bde^{\rm T}=[1\, 0\, \cdots\, 0]}$, then the system of equations~\eqref{eq:cilc} can be rewritten in the compact form:
\begin{equation}
\label{eq:cilc2}
\begin{cases}
{\partial \over \partial \bdw} \left(\bdw^{\rm T} \tC\, \bdw\right) = 0 \\[1.5mm]
\bdw^{\rm T}  \tA = \bde^{\rm T}.
\end{cases}
\end{equation}
Hence, the problem is equivalent to solving
\begin{align}
{\partial \over \partial \bdw} \left[\bdw^{\rm T} \tC\, \bdw  +\Lambda^{\rm T}\left(\bde - \tA^{\rm T}\bdw\right)\right]=0,
\end{align}
where $\Lambda^{\rm T} = [\lambda\, \mu_1\, \cdots\, \mu_m]$ is a vector of $(m+1)$ Lagrange multipliers. This yields to ${\bdw^{\rm T}=\Lambda^{\rm T}\tA^{\rm T}\tC^{-1}}$, while the constraints ${\bdw^{\rm T}\tA=\bde^{\rm T}}$ give ${\Lambda^{\rm T}=\bde^{\rm T}\left(\tA^{\rm T}\tC^{-1}\tA\right)^{-1}}$. The \texttt{cILC} weights are then given by
\begin{align}
\label{eq:cilcw}
\bdw^{\rm T}=\bde^{\rm T}\left(\tA^{\rm T}\tC^{-1}\tA\right)^{-1}\tA^{\rm T}\tC^{-1},
\end{align}
which is a generalization of Eq.~\eqref{eq:ilcw}.
Equation ~\eqref{eq:cilcw} has been previously derived in the literature for the extraction of the relativistic SZ effect \citep{Remazeilles2019b} and CMB spectral distortions \citep{Rotti2021}. 
Similar multidimensional ILC filters have been derived for other component separation methods like \texttt{GNILC} \citep{Remazeilles2011b} and \texttt{MILCA} \citep{Hurier2013}, the main difference between them lying in the form of the mixing matrix $\tA$.

It should be noted that for a single orthogonality constraint, i.e. ${\tA =[\bda\, \bdb]}$ and ${\bde^{\rm T}=[1\, 0]}$, the general expression Eq.~\eqref{eq:cilcw} reduces to 
\begin{align}
\label{eq:cilcw2d}
\bdw^{\rm T}={\left(\bdb^{\rm T}\tC^{-1}\bdb\right)\bda^{\rm T}\tC^{-1} - \left(\bda^{\rm T}\tC^{-1}\bdb\right)\bdb^{\rm T}\tC^{-1}\over \left(\bda^{\rm T}\tC^{-1}\bda\right)\left(\bdb^{\rm T}\tC^{-1}\bdb\right)-\left(\bda^{\rm T}\tC^{-1}\bdb\right)^2},
\end{align}
which was introduced by \cite{Remazeilles2011} to null SZ cluster residuals in primary CMB maps, when $\bda$ and $\bdb$ are the SEDs of CMB and thermal SZ, respectively. Finally, in the absence of any orthogonality constraint, i.e. $\tA\equiv \bda$ and $\bde\equiv 1$, the expression Eq.~\eqref{eq:cilcw} reduces to the standard ILC given in Eq.~\eqref{eq:ilcw}.

\section{Semi-blind component separation method}
\label{sec:method}

We now work out our new \textit{semi-blind} component separation method for $B$-modes, the constrained moment ILC (\texttt{cMILC}), which combines foreground moment expansion \citep{Chluba2017} and \texttt{cILC} method \citep{Remazeilles2011}.

\subsection{Moment expansion of the foreground emission}
\label{subsec:moments}

The spectral intensity\footnote{By "intensity", we mean either $I$, $Q$, $U$, $E$, or $B$, without loss of generality.} of the Galactic foreground emission depends on several spectral parameters $\boldsymbol{\beta}(p)\equiv\{\beta_i(p)\}_{1\leq i \leq n}$ (e.g. spectral indices, temperatures) that vary across the sky (hence the $p$ dependence for 'pixel') and along the line-of-sight:
\begin{align}
I(\nu,p)=A_{\nu_0}(p)\,f\left(\nu,\boldsymbol{\beta}(p)\right).
\end{align}
$A_{\nu_0}(p)$ is the amplitude of the foreground emission at some pivot frequency $\nu_0$, and $f\left(\nu,\boldsymbol{\beta}(p)\right)$ is the spectral energy distribution (SED) which varies both across frequencies $\nu$ and across the sky, depending on the local spectral parameters $\boldsymbol{\beta}(p)\equiv\{\beta_i(p)\}_{1\leq i \leq n}$.

As stressed by \cite{Chluba2017}, we can expand the foreground SED, $f\left(\nu,\boldsymbol{\beta}(p)\right)$, in terms of \textit{statistical moments} around some fixed pivot parameters $\overline{\boldsymbol{\beta}}\equiv\{\overline{\beta}_i\}_i$:
\begin{flalign}
&f \left(\nu,\boldsymbol{\beta}(p)\right)& \cr
=&\sum_k \sum_{\alpha_1+\cdots+\alpha_n=k}{\left(\beta_1(p)-\overline{\beta}_1\right)^{\alpha_1}\cdots\left(\beta_n(p)-\overline{\beta}_n\right)^{\alpha_n}\over \alpha_1!\cdots\alpha_n!}{\partial^{\,k} f\left(\nu,\overline{\boldsymbol{\beta}}\right)\over\partial\overline{\beta}_1^{\,\alpha_1}\cdots \partial\overline{\beta}_n^{\,\alpha_n}}\cdot&
\end{flalign}
This is nothing but a multi-dimensional Taylor series in all the spectral parameters, however, it gives the flexibility to describe various physical and observational averaging processes.
The moment expansion highlights that the foreground emission can be captured by a set of \textit{moment components} in the sky:
\begin{align}
m_{\alpha_1,\cdots,\alpha_n}(p)=A_{\nu_0}(p)\, {\left(\beta_1(p)-\overline{\beta}_1\right)^{\alpha_1}\cdots\left(\beta_n(p)-\overline{\beta}_n\right)^{\alpha_n}\over \alpha_1!\cdots\alpha_n!},
\end{align}
each of them having a \textit{uniform} SED across the sky given by
\begin{align}
{\partial^{\,\alpha_1+\cdots+\alpha_n} f\over\partial\beta_1^{\,\alpha_1}\cdots \partial\beta_n^{\,\alpha_n}}\left(\nu,\boldsymbol{\beta}(p)=\overline{\boldsymbol{\beta}}\right).
\end{align}
Below, we will consider thermal dust and synchrotron as the main polarized Galactic foregrounds for $B$-modes. The SED of the synchrotron emission is known to be close to a power-law 
\begin{align}
f_{\rm sync}\left(\nu, \beta_s(p)\right) = \left({\nu \over \nu_s}\right)^{\beta_s(p)}, 
\end{align}
for data in Rayleigh-Jeans brightness temperature units, with an average spectral index of $\langle \beta_s(p)\rangle_p \simeq -3$ over the sky  \citep{Kogut2007,Miville-Deschenes2008,Krachmalnicoff2018,Planck2018_IV}. Conversely, the SED of the thermal dust emission is known to be close to a modified blackbody
\begin{align}
f_{\rm dust}\left(\nu, \beta_d(p), T_d(p)\right) = \left({\nu \over \nu_d}\right)^{\beta_d(p)+1} {1\over \exp\left({h\nu\over kT_d(p)}\right)-1},
\end{align}
 for data in Rayleigh-Jeans brightness temperature units, with an average spectral index of $\langle \beta_d(p)\rangle_p \simeq 1.5$ and an average temperature of $\langle T_d(p)\rangle_p \simeq 20$\,K \citep{planck2013_XI,Planck2015_X,Planck2016GNILC}.

Hence, the moment expansion of the synchrotron emission up to second order yields\footnote{We will be using the notation $\partial_{\bar{\beta}} f(\bar{\beta}) \equiv \frac{\partial f(\beta)}{\partial \beta} \Big{\vert}_{\beta=\bar{\beta}}$ throughout the paper.}
\begin{align}
I_{\rm sync}(\nu,p)
&=A_{\nu_s}(p) f_{\rm sync} \left(\nu,\overline{\beta}_s\right)\cr
		&\qquad+ A_{\nu_s}(p)\,\Delta \beta_s(p)\,\partial_{\overline{\beta}_s} f_{\rm sync} \left(\nu,\overline{\beta}_s\right)
		\\ \nonumber
		&\qquad\qquad+{1\over 2} \,A_{\nu_s}(p)\,\Delta \beta^2_s(p)\,\partial^2_{\overline{\beta}_s} f_{\rm sync} \left(\nu,\overline{\beta}_s\right)+\mathcal{O}\left(\beta_s^3\right),
 \end{align}
where $\Delta \beta_s(p)=\beta_s(p) - \overline{\beta}_s$ and
\begin{subequations}
\begin{flalign}
\label{eq:sync_0}
f_{\rm sync} \left(\nu,\overline{\beta}_s\right) &= \left({\nu\over\nu_s}\right)^{\overline{\beta}_s},\\
\label{eq:sync_1}
\partial_{\overline{\beta}_s} f_{\rm sync} \left(\nu,\overline{\beta}_s\right) &= \ln\left({\nu\over\nu_s}\right)f_{\rm sync} \left(\nu,\overline{\beta}_s\right),\\ 
\label{eq:sync_2}
\partial^2_{\overline{\beta}_s} f_{\rm sync} \left(\nu,\overline{\beta}_s\right) &= \left[\ln\left({\nu\over\nu_s}\right)\right]^2f_{\rm sync} \left(\nu,\overline{\beta}_s\right) 
\end{flalign}
\end{subequations}
are the SEDs of zeroth-order (Eq.~\ref{eq:sync_0}), first-order (Eq.~\ref{eq:sync_1}), and second-order (Eq.~\ref{eq:sync_2}) moment components of the synchrotron. 

\begin{figure}
  \begin{center}
    \includegraphics[width=1\columnwidth]{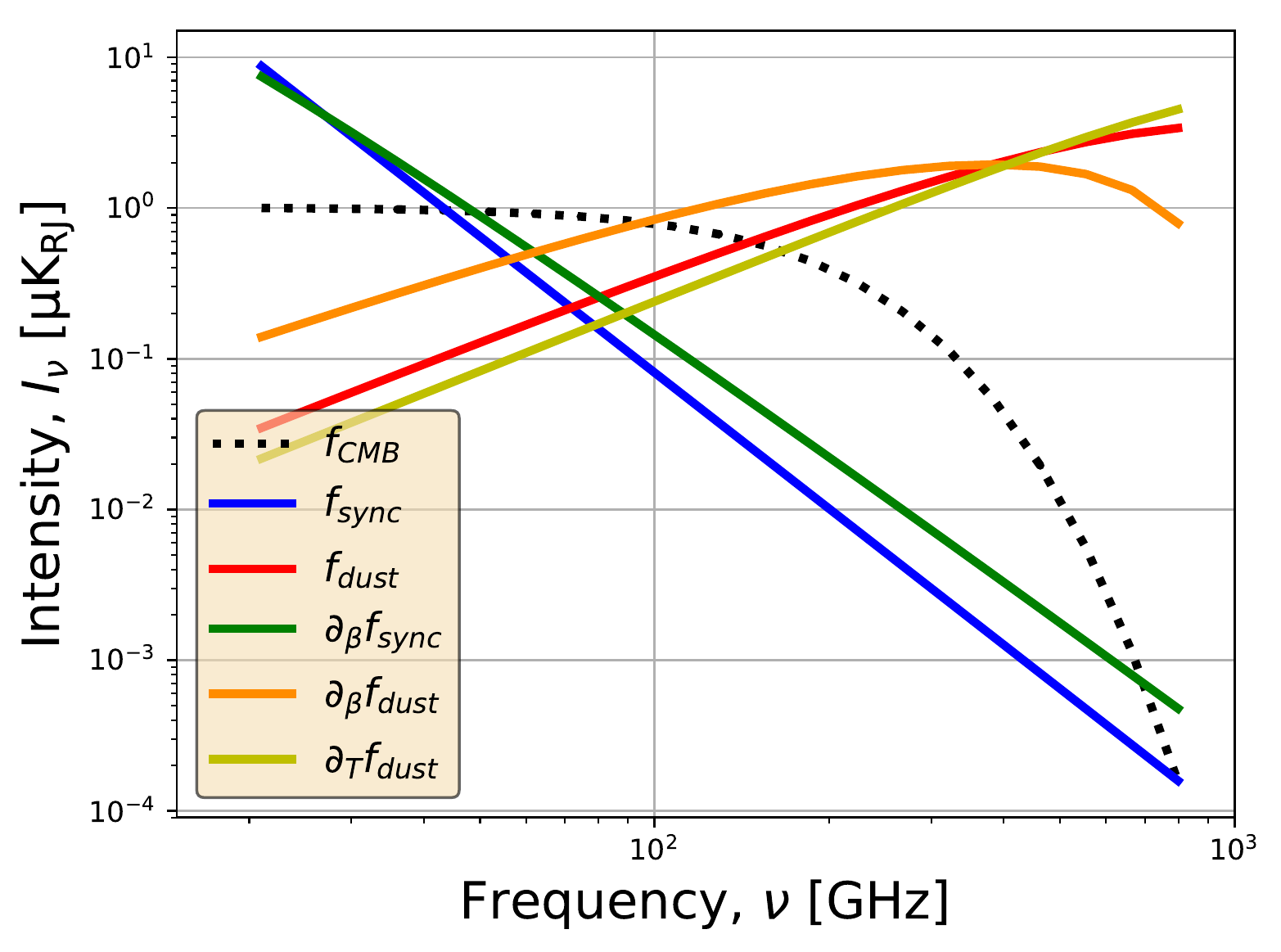}
   \end{center}
\caption{Spectral energy distribution (SED) of the zeroth- and first-order moments of the dust and synchrotron (\textit{coloured solid lines}), along with the SED of the CMB (\textit{black dotted line}), across frequencies. The overall normalisation of the SEDs is arbitrary.}
\label{Fig:moments_sed}
\end{figure}

Similarly, the moment expansion of the thermal dust emission up to second order yields 
\begin{align}
I_{\rm dust}(\nu,p)=\,&\,A_{\nu_d}(p)  f_{\rm dust} \left(\nu,\overline{\beta}_d\right)\cr
		&+A_{\nu_d}(p)\,\Delta \beta_d(p)\;\partial_{\overline{\beta}_d} f_{\rm dust} \left(\nu,\overline{\beta}_d, \overline{T}_{\!d}\right)\cr
		&+A_{\nu_d}(p)\,\Delta T_d(p)\;\partial_{\overline{T}_d} f_{\rm dust} \left(\nu,\overline{\beta}_d, \overline{T}_{\!d}\right)\cr
		&+{1\over 2}\,A_{\nu_d}(p)\,\Delta \beta^2_d(p)\;
		\partial^2_{\overline{\beta}_d} f_{\rm dust} \left(\nu,\overline{\beta}_d, \overline{T}_{\!d}\right)\cr
		&+A_{\nu_d}(p)\,
		\Delta \beta_d(p)
		\Delta T_d(p)\;
		\partial_{\overline{\beta}_d}\partial_{\overline{T}_d} f_{\rm dust} \left(\nu,\overline{\beta}_d, \overline{T}_{\!d}\right)\cr
		&+{1\over 2}\,A_{\nu_d}(p)\,
		\Delta T^2_d(p)\;
		\partial^2_{\overline{T}_d} f_{\rm dust} \left(\nu,\overline{\beta}_d, \overline{T}_{\!d}\right)\cr
		&+\mathcal{O}\left(\Delta\beta_d^3, \Delta T_d^3\right),
\end{align}
where $\Delta \beta_d(p)=\beta_d(p) - \overline{\beta}_d$, $\Delta T_d(p)=T_d(p) - \overline{T}_d$ and
\begin{subequations}
\begin{flalign}
\label{eq:dust_0}
&f_{\rm dust} \left(\nu,\overline{\beta}_d, \overline{T}_{\!d}\right) = \left({\nu \over \nu_d}\right)^{\overline{\beta}_d+1} {1\over \exp\left({\bar{x}}\right)-1},\\
\label{eq:dust_1b}
&\partial_{\overline{\beta}_d} f_{\rm dust} \left(\nu,\overline{\beta}_d, \overline{T}_{\!d}\right) = \ln\left({\nu\over\nu_d}\right)f_{\rm dust} \left(\nu,\overline{\beta}_d, \overline{T}_{\!d}\right),\\
\label{eq:dust_1t}
&\partial_{\overline{T}_d} f_{\rm dust} \left(\nu,\overline{\beta}_d, \overline{T}_{\!d}\right) = {\bar{x}\over\overline{T}_{\!d}}{ \exp \left( {\bar{x}} \right) \over \exp \left( {\bar{x}} \right) - 1}f_{\rm dust} \left(\nu,\overline{\beta}_d, \overline{T}_{\!d}\right),\\
\label{eq:dust_2b}
&\partial^2_{\overline{\beta}_d} f_{\rm dust} \left(\nu,\overline{\beta}_d, \overline{T}_{\!d}\right) = \left[\ln\left({\nu\over\nu_d}\right)\right]^2f_{\rm dust} \left(\nu,\overline{\beta}_d, \overline{T}_{\!d}\right),\\
\label{eq:dust_2t}
&\partial^2_{\overline{T}_d} f_{\rm dust} \left(\nu,\overline{\beta}_d, \overline{T}_{\!d}\right) = \left[\bar{x}\coth\left({\bar{x}\over 2}\right) - 2\right] {1\over \overline{T}_{\!d}}\partial_{\overline{T}_d} f_{\rm dust} \left(\nu,\overline{\beta}_d, \overline{T}_{\!d}\right),\\
\label{eq:dust_2bt}
&\partial_{\overline{\beta}_d}\partial_{\overline{T}_d} f_{\rm dust} \left(\nu,\overline{\beta}_d, \overline{T}_{\!d}\right) = \ln\left({\nu\over\nu_d}\right)\partial_{\overline{T}_d} f_{\rm dust} \left(\nu,\overline{\beta}_d, \overline{T}_{\!d}\right) 
\end{flalign}
\end{subequations}
are the SEDs of zeroth-order (Eq.~\ref{eq:dust_0}), first-order (Eqs.~\ref{eq:dust_1b}-\ref{eq:dust_1t}), and second-order (Eqs.~\ref{eq:dust_2b}-\ref{eq:dust_2t}-\ref{eq:dust_2bt}) moment components of the dust, and $\bar{x}\equiv h\nu / k \overline{T}_{\!d}$. The SED shapes of zeroth- and first-order moments of the dust and synchrotron are plotted in Fig.~\ref{Fig:moments_sed}.

By choosing pivot parameters that are close to the expected mean of the spectral parameters over the sky, ${\overline{\beta}_s = \langle \beta_s(p)\rangle_p \simeq -3}$, ${\overline{\beta}_d = \langle \beta_d(p)\rangle_p \simeq 1.5}$, and ${\overline{T}_{\!d} = \langle T_d(p)\rangle_p \simeq 20}$\,K, we ensure that the bulk of the Galactic foreground emission projects onto the zeroth-, first-, and second-order moments. 
However, a detailed optimization is more complicated as we briefly discuss in Sect.~\ref{subsec:optimal_pivot}.

\subsection{The constrained moment ILC (\texttt{cMILC})}
\label{subsec:semiblind}

Our \textit{semi-blind} component separation method \texttt{cMILC}  aims at recovering the CMB $B$-mode polarization signal from sky observations by adding nulling constraints on the main foreground moments in the \texttt{cILC} framework in order to improve the removal of the Galactic foreground contamination compared to standard ILC methods, which rely only on simple variance minimization. We thus build the \texttt{cMILC} method by enforcing the following constraints:
\begin{equation}
\label{eq:cmilc}
\begin{cases}
{\partial \over \partial \bdw} \left(\bdw^{\rm T} \tC\, \bdw\right) = 0 
\\[1.5mm]
\bdw^{\rm T} \cdot  \bda = 1 \\[1.5mm]
\bdw^{\rm T} \cdot  \bdf_{sync}\left(\overline{\beta}_s\right) = 0 \\[1.5mm]
\bdw^{\rm T} \cdot  \partial_{\overline{\beta}_s}\bdf_{sync}\left(\overline{\beta}_s\right) = 0\\[1.5mm]
\bdw^{\rm T} \cdot  \partial^2_{\overline{\beta}_s}\bdf_{sync}\left(\overline{\beta}_s\right) = 0\\[1.5mm]
\bdw^{\rm T} \cdot  \bdf_{dust}\left(\overline{\beta}_d, \overline{T}_{\!d}\right) = 0 \\[1.5mm]
\bdw^{\rm T} \cdot  \partial_{\overline{\beta}_d}\bdf_{dust}\left(\overline{\beta}_d, \overline{T}_{\!d}\right) = 0\\[1.5mm]
\bdw^{\rm T} \cdot  \partial_{\overline{T}_d}\bdf_{dust}\left(\overline{\beta}_d, \overline{T}_{\!d}\right) = 0\\[1.5mm]
\bdw^{\rm T} \cdot  \partial^2_{\overline{\beta}_d}\bdf_{dust}\left(\overline{\beta}_d, \overline{T}_{\!d}\right) = 0\\[1.5mm]
\bdw^{\rm T} \cdot  \partial^2_{\overline{T}_d}\bdf_{dust}\left(\overline{\beta}_d, \overline{T}_{\!d}\right) = 0\\[1.5mm]
\bdw^{\rm T} \cdot  \partial_{\overline{\beta}_d}\partial_{\overline{T}_d}\bdf_{dust}\left(\overline{\beta}_d, \overline{T}_{\!d}\right) = 0.
\end{cases}
\end{equation}
It is important to stress that \texttt{cMILC} does not attempt to fit the foregrounds, which are spatially correlated and uncertain components, but instead aims at \textit{deprojecting} the dominant foreground moments in sky maps without altering the CMB component. 
Thus, bulk of the Galactic foreground contamination that is spatially correlated with the deprojected moments is eliminated from the recovered CMB $B$-mode power spectrum. Any remaining unconstrained foreground contribution is just variance-minimised, like in blind methods.\footnote{Hence, the exact pivots parameters do not need to be absolutely known (Fig.~\ref{Fig:varpivot}), although pivots can help optimizing the method (see Sect.~\ref{subsec:optimal_pivot}).}

The \texttt{cMILC} weights are thus again given by Eq.~\eqref{eq:cilcw}:
\begin{align}
\label{eq:semiblindw}
\bdw^{\rm T}_{\rm cMILC}=\bde^{\rm T}\left(\tA^{\rm T}\tC^{-1}\tA\right)^{-1}\tA^{\rm T}\tC^{-1},
\end{align}
where, when limiting constraints to zeroth- and first-order moments as an example, the SED matrix $\tA$ is a $n_f \times 6$ matrix given by
\begin{align}
\label{eq:sedmatrix}
\tA =
\begin{pmatrix}
\bda & \bdf_{sync} & \bdf_{dust} &  \partial_{\overline{\beta}_s}\bdf_{sync}  & \partial_{\overline{\beta}_d}\bdf_{dust}& \partial_{\overline{T}_d}\bdf_{dust} 
\end{pmatrix},
\end{align}
and the vector $\bde$ is given by
\begin{align}
\bde^{\rm T} =
\begin{pmatrix}
1 & 0 & 0 & 0 & 0 & 0
\end{pmatrix}.
\end{align}
The SED matrix $\tA$ and vector $\bde$ can be augmented by including higher-order moments, depending on the sensitivity floor of the experiment. 
We stress that by applying the Woodbury formula to Eq.~\eqref{eq:covar}, the \texttt{cILC}/\texttt{cMILC} filter Eq.~\eqref{eq:semiblindw} reduces to $\bde^{\rm T}(\tA^{\rm T}\tN^{-1}\tA)^{-1}\tA^{\rm T}\tN^{-1}$, where, unlike maximum-likelihood parametric fitting methods \citep[e.g.][]{Stompor2016}, the covariance matrix $\tN$ here does not only account for the noise but also for the \textit{unconstrained} foregrounds (e.g. unmodelled foregrounds or higher-order moments) that may be omitted in the mixing matrix $\tA$. The data covariance matrix $\tC$ in Eq.~\eqref{eq:semiblindw} thus allows for blind variance minimization of any remaining unparameterized foregrounds but also of possible residuals that would arise from suboptimal pivot choices (see Fig.~\ref{Fig:varpivot} in Appendix~\ref{sec:flexi}).

For our analysis, we will be applying several \texttt{cMILC} filters (see  Table~\ref{tab:nom}) to the sky simulations by considering more and more columns in the SED matrix $\tA$, starting from the standard ILC method without nulling constraints (i.e. $\tA\equiv \bda$), and ending with the fully-constrained moment ILC with all the nulling constraints on zeroth-, first-, and second-order foreground moments (i.e. with a $n_f \times 10$ SED matrix $\tA$). This progressive approach will help to highlight how increasing the number of moment constraints affects the trade-offs between foreground and noise residuals (bias, variance, and higher-order statistics) in the recovered CMB $B$-mode map, power spectrum, and tensor-to-scalar ratio. 
Below we will see that the optimal number of moments is a function of the experimental sensitivity and spectral coverage.

\subsection{Needlet implementation}
\label{subsec:needlets}

ILC filtering can be implemented directly in pixel space on the sky maps $d_\nu(p)$ \citep[e.g.][]{WMAP2003,Eriksen2004} or in harmonic space on spherical harmonic coefficients $a_\nu(\ell,m)$ of the sky maps \citep[e.g.][]{Tegmark2003}. Since pixel-space ILC methods are fully local in pixel space, they are nonlocal in harmonic space because of uncertainty principle, hence all multipoles are given the same weights. Similarly, harmonic-space ILC methods being fully local in multipole space are nonlocal in pixel space and thus provide same weights to all pixels of a sky map. Nonlocal weights in either pixel or multipole space are not optimal for component separation, since the local conditions of contamination by foregrounds and noise vary both across the sky and across multipoles. 

Because of their localisation properties both in pixel space and harmonic space, spherical wavelets such as \textit{needlets}  \citep{Narcowich2006,Guilloux2009} provide a powerful alternative to allow the ILC weights to adjust themselves depending on local conditions of contamination both across the sky and across angular scales \citep{Delabrouille2009}.  Therefore, in this work we implement both ILC and \texttt{cMILC} methods on the same needlet frame, so that hereafter the standard ILC method will be referred to as \texttt{NILC} for needlet ILC \citep{Delabrouille2009}, while our needlet-based constrained moment ILC method is still referred to as \texttt{cMILC}.

\begin{figure}
  \begin{center}
    \includegraphics[width=\columnwidth]{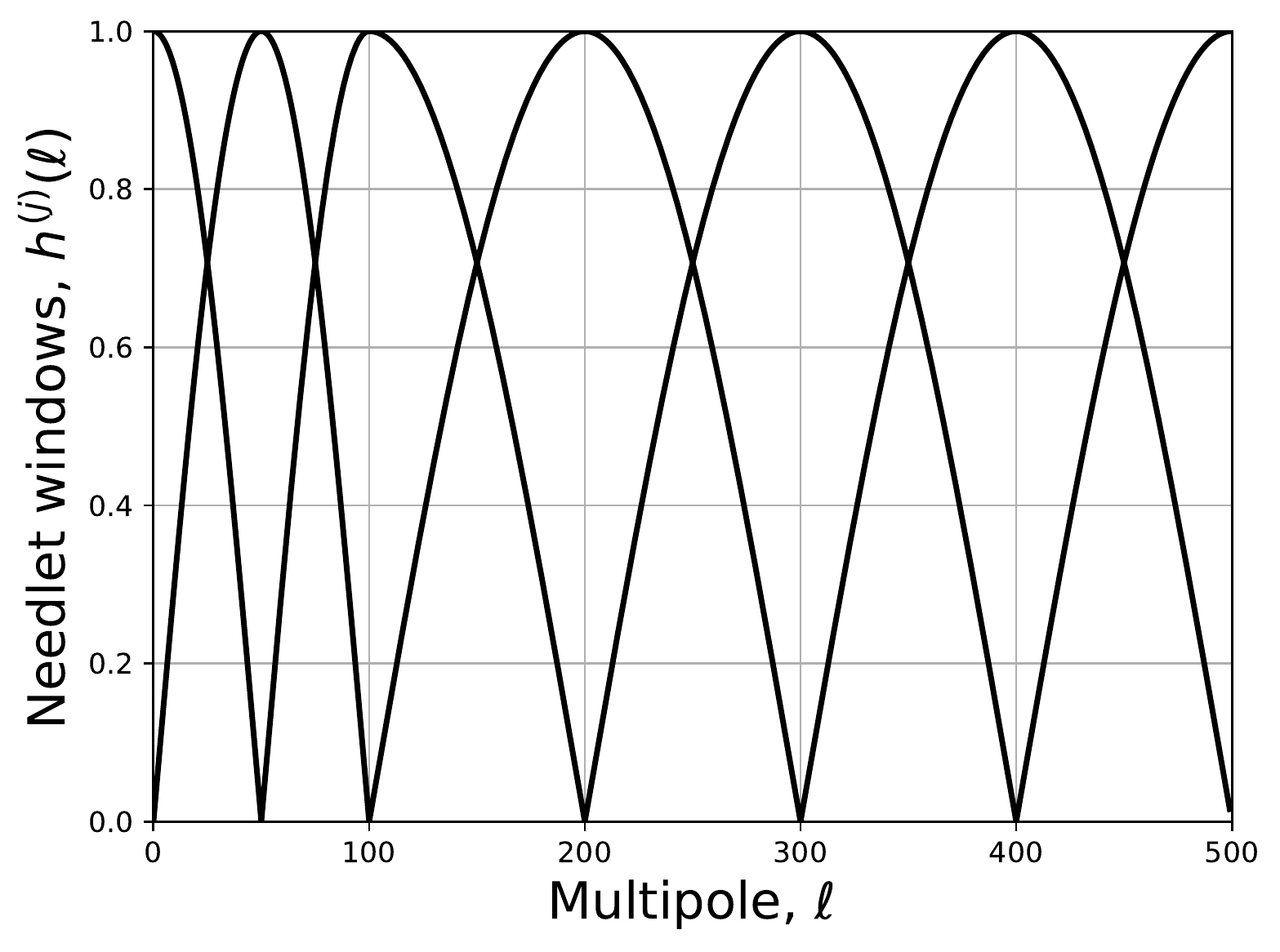}
   \end{center}
\caption{Needlet windows in harmonic space that are used in the analysis.}
\label{Fig:needlets}
\end{figure}

Needlet decomposition is performed on $B$-mode sky maps at each frequency as follows. We first define a set of seven bandpass windows $\{h^{(j)}(\ell)\}_{1\leq j \leq 7}$ of cosine shape in harmonic space (Fig.~\ref{Fig:needlets}), following the prescription of \cite{Basak2013}. Each needlet window isolates a subrange of multipoles or angular scales, while the whole set satisfies the property ${\sum_{j=1}^7 (h^{(j)}(\ell))^2 = 1}$ in order to not lose any power in the data processing. 
The number and width of needlet windows were chosen to allow decent localization in harmonic space while conserving sufficient localization in pixel space for efficient foreground cleaning, by following a dyadic scheme with wider windows at high multipoles to ensure more localization in pixel space for the cleaning of small-scale fluctuations in sky maps. The number, width and shape of the needlet windows are free parameters and different prescriptions have been used in the literature depending on the observable of interest \citep[e.g.][]{planck2015_XII}, thus leaving room for further optimisation (see Sect.~\ref{sec:disc}).
The spherical harmonic coefficients $a_\nu(\ell,m)$ of the frequency maps are then bandpass filtered by the needlet windows as ${\tilde{a}^{(j)}_\nu(\ell,m)=h^{(j)}(\ell)\,a_\nu(\ell,m)}$. The inverse spherical harmonic transform of the bandpass-filtered harmonic coefficients $\tilde{a}^{(j)}_\nu(\ell,m)$ then yields seven needlet maps $\tilde{d}^{\,(j)}_\nu(p)$ for each frequency. Each needlet map thus contains typical anisotropies of specific angular scales selected by the needlet window. The ILC weights at all frequencies are then computed for each needlet scale $(j)$ independently, so that we obtain seven ILC CMB maps ${\widehat{s}^{\,(j)}(p)=\sum_\nu w_\nu^{(j)}(p)\tilde{d}^{\,(j)}_\nu(p)}$ for each needlet frame. The spherical harmonic coefficients, $\widehat{s}^{\,(j)}(\ell,m)$, of the seven needlet ILC CMB maps are then bandpass-filtered again as ${\widehat{z}^{\,(j)}(\ell,m)=h^{(j)}(\ell)\,\widehat{s}^{\,(j)}(\ell,m)}$, and inverse spherical-harmonic transformed into maps $\widehat{z}^{\,(j)}(p)$, which are finally coadded to form the complete \texttt{NILC} or \texttt{cMILC} CMB map ${\widehat{s}(p) = \sum_{j=1}^7\widehat{z}^{\,(j)}(p)}$.

\vspace{-4mm}
\subsection{Statistics}
\label{subsec:stat}

In this section, we first compare the statistics of the constrained ILC (\texttt{cILC}, \texttt{cMILC}) and standard ILC (\texttt{NILC}) approaches, by generalizing the formulas of the ILC variance \citep[e.g.][]{Tegmark2003, Delabrouille2009} to our \texttt{cMILC} filter. We then illustrate our findings on simulations in Sect.~\ref{subsubsec:stats}. We stress that, while the standard ILC is the solution of minimum \textit{overall} variance (foregrounds plus noise), \texttt{cMILC} leads to the solution of minimum \textit{foreground} variance. In particular, \texttt{cMILC} beats the standard ILC in terms of residual foreground contamination (bias, variance, and skewness) thanks to nulling constraints, but at the cost of larger residual noise contamination due to the increased volume of the parameter space. However, there is an optimal trade-off point for \texttt{cMILC}, where the noise penalty is largely compensated by the significant reduction of the residual foreground bias. 
Aside from experimental parameters, this sweet spot depends on the observable and here we consider $r$ as figure of merit (see Figs.~\ref{Fig:r_stat} and \ref{Fig:r_stat_mhd} in Sect.~\ref{subsec:results}), illustrating the gains for both the variance and the bias on $r$ through the cumulative systematic and statistical error $\delta r={\sqrt{(r-r^{\,\rm true})^2+\sigma^2(r)}}$.

\vspace{-3mm}
\subsubsection{Overall variance}

Using Eq.~\eqref{eq:ilcvar} and Eq.~\eqref{eq:ilcw}, we recover the standard expression \citep[e.g.][]{Tegmark2003} for the variance of the ILC map\footnote{Note that in the derivation of the standard formulas Eqs.~\eqref{eq:varilc}-\eqref{eq:varcilc0} for the variance we make the assumption that empirical correlations between the weights $\bdw(p)$ and the data $d(p)$ can be neglected to first order, and thus the weights can be pulled out of $\langle \cdots \rangle$ in Eq.~\eqref{eq:ilcvar}.}:
\begin{align}
\label{eq:varilc}
\sigma_{\rm ILC}^2  &\equiv\langle \hat{s}^2_{\rm ILC}\rangle = \bdw_{\rm ILC} ^{\rm T}\tC\, \bdw_{\rm ILC} = {1\over \bda^{\rm T} \tC^{-1} \bda},
\end{align}
where $\tC=\langle \bdd \bdd^T \rangle$ is the $n_f\times n_f$ covariance matrix of the data, and $\bda$ is the SED for CMB.
Similarly, by using Eq.~\eqref{eq:semiblindw} we can derive the expression for the variance of the \texttt{cMILC} map, $\sigma^2_{\rm cMILC}\equiv\langle \hat{s}^2_{\rm cMILC}\rangle$:
\begin{align}
\label{eq:varcilc0}
\sigma_{\rm cMILC}^2 &= \bdw_{\rm cMILC} ^{\rm T} \tC\, \bdw_{\rm cMILC} = \bde^T( \tA^{\rm T} \tC^{-1}  \tA)^{-1}\bde,
\end{align}
where ${\tA =[\bda\, \bdb_1\, \cdots \bdb_m]}$ is the $n_f \times (m+1)$ SED matrix for the CMB and the constrained foreground moments (e.g. Eq.~\ref{eq:sedmatrix}), and ${\bde^{\rm T}=[1\, 0\, \cdots\, 0]}$.

By defining the submatrix ${\tB =[\bdb_1\, \cdots \bdb_m]}$ as the $n_f \times m$ SED matrix for the modelled foreground moments only, the full SED matrix $\tA$ can be written in the block matrix form:
\begin{align}
\tA=
\begin{pmatrix}
\bda & \tB
\end{pmatrix},
\end{align}
so that the variance of the \texttt{cMILC} map Eq.~\eqref{eq:varcilc0} reads as
\begin{align}
\sigma_{\rm cMILC}^2=\bde^{\rm T}
\begin{pmatrix}
\bda^{\rm T} \tC^{-1} \bda & \bda^{\rm T} \tC^{-1}\tB\\
\tB^{\rm T}\tC^{-1}\bda & \tB^{\rm T}\tC^{-1}\tB
\end{pmatrix}^{-1}
\bde.
\end{align}
Using block matrix inversion, the variance of the \texttt{cMILC} map reduces to
\begin{align}
\label{eq:varcilc}
\sigma_{\rm cMILC}^2 &={\sigma_{\rm ILC}^2 \over 1 \, - \, \sigma^2_{\rm ILC}\,\bda^{\rm T} \tC^{-1}\tB\,\left(\tB^{\rm T}\tC^{-1}\tB\right)^{-1}\tB^{\rm T}\tC^{-1}\bda}.
\end{align}
Therefore, the overall variance of the \texttt{cMILC} map is 
 larger than the overall variance of the standard ILC map, with an amplification factor of
\begin{align}
\label{eq:overall}
{\sigma_{\rm cMILC}^2 \over \sigma_{\rm ILC}^2} = {1\over 1 \, - \, \sigma^2_{\rm ILC}\,\bda^{\rm T} \tC^{-1}\tB\,\left(\tB^{\rm T}\tC^{-1}\tB\right)^{-1}\tB^{\rm T}\tC^{-1}\bda} \geq 1.
\end{align}
This increase of the overall variance is expected, since the volume of the parameter space (number of modelled components) is larger for the \texttt{cMILC} than for the standard ILC. 
However, while the standard ILC is the solution of minimum overall variance, it is not the solution of minimum foreground variance. As we show below, \texttt{cMILC} yields lower foreground contamination than standard ILC.

\subsubsection{Foregrounds residual variance}

Let us model the data with independent contributions from the various components:
\begin{align}
\bdd(p) = \bda\, s(p)\, +\, \sum_{i=1}^m \bdb_i \,g_i(p)\, + \bdn(p),
\end{align}
where $s(p)$ is the CMB signal and $\bda$ its SED, $\{g_i(p)\}_{1\leq i\leq m}$ the $m$ foreground components whose SEDs $\{\bdb_i\}_{1\leq i\leq m}$ are constrained in \texttt{cMILC}, while not in the ILC, and $\bdn(p)$ the contamination from both instrumental noise and un-modelled foregrounds. The covariance matrix of the data thus reads as 
\begin{align}
\label{eq:covar}
\tC = \bda\, \sigma_s^2\, \bda^{\rm T}\, +\, \sum_{i=1}^m \bdb_i\, \sigma_{g_i}^2\, \bdb_i^{\rm T}\,+\,\tN,
\end{align}
where ${\sigma_s^2=\langle s(p)^2\rangle}$ is the variance of intrinsic CMB anisotropies, ${\sigma_{g_i}^2=\langle g_i(p)^2\rangle}$ the variance of the modelled foregrounds, and ${\tN=\langle \bdn(p)\bdn(p)^T\rangle}$ the covariance matrix of the noise and un-modelled foregrounds. 

By inserting Eq.~\eqref{eq:covar} into Eq. \eqref{eq:varilc}, the contribution from the different components to the overall variance of the standard ILC map becomes explicit:
\begin{align}
\label{eq:fgilc}
\sigma^2_{\rm ILC}
                              &= \bdw_{\rm ILC}^{\rm T}\left( \bda\, \sigma_s^2\, \bda^{\rm T}\, +\, \sum_{i=1}^m \bdb_i\, \sigma_{g_i}^2\, \bdb_i^{\rm T}\,+\,\tN\right)\, \bdw_{\rm ILC}\cr
                              &=\sigma_s^2 + \sum_{i=1}^m \left(\bdw_{\rm ILC}^{\rm T}\bdb_i\right)^2 \sigma_{g_i}^2 + \bdw_{\rm ILC}^{\rm T} \tN\,\bdw_{\rm ILC},
\end{align}
since  $\bdw_{\rm ILC}^{\rm T}\bda =1$, but $\bdw_{\rm ILC}^{\rm T}\bdb_i \neq 0$. 
In contrast to the ILC solution, the variance of the \texttt{cMILC} map Eq. \eqref{eq:varcilc0} reduces to:
\begin{align}
\label{eq:fgcilc}
\sigma^2_{\rm cMILC}
&= \bdw_{\rm cMILC}^{\rm T}\left( \bda\, \sigma_s^2\, \bda^{\rm T}\, +\, \sum_{i=1}^m \bdb_i\, \sigma_{g_i}^2\, \bdb_i^{\rm T}\,+\,\tN\right)\, \bdw_{\rm cMILC}\cr
&=\sigma_s^2 + \bdw_{\rm cMILC}^{\rm T} \tN\,\bdw_{\rm cMILC},
\end{align}
since $\bdw_{\rm cMILC}^{\rm T}\bda =1$ and $\bdw_{\rm cMILC}^{\rm T}\bdb_i =0$. 
Therefore, the variance of the \texttt{cMILC} map is cleared of any contribution from the constrained foregrounds (Eq.~\ref{eq:fgcilc}), while some residual variance from those foregrounds still contributes to the variance of the standard ILC map (Eq.~\ref{eq:fgilc}). If we now consider that the unconstrained foreground moments are those which are below the noise of the instrument, then $\bdw_{\rm ILC}^{\rm T} \tN\,\bdw_{\rm ILC}$ and $\bdw_{\rm cMILC}^{\rm T} \tN\,\bdw_{\rm cMILC}$ are just the noise variance contributions to the standard ILC and \texttt{cMILC} maps, and hence \texttt{cMILC}  moves towards the solution with minimum residual foreground contamination, unlike the standard ILC.
Consequently, the inequality Eq.~\eqref{eq:overall} on overall variance implies that the residual noise variance must be larger in the \texttt{cMILC} map than in the standard ILC map, i.e. ${\bdw_{\rm cMILC}^{\rm T} \tN\,\bdw_{\rm cMILC} \geq \bdw_{\rm ILC}^{\rm T} \tN\,\bdw_{\rm ILC}}$.

To conclude, \texttt{cMILC} reduces the variance and bias from residual foreground contamination, while it has larger noise variance than the standard ILC. By including more and more foreground moments in \texttt{cMILC}, the residual noise variance increases in the CMB map, but in contrast the residual foreground variance decreases towards the minimum foreground variance solution, thus largely beating the standard ILC.
These theoretical expectations are confirmed by our analysis on sky simulations (see Fig.~\ref{Fig:sky_stat}).

While foreground residuals and noise residuals both cause biases on the CMB $B$-mode power spectrum, and hence on the inferred tensor-to-scalar ratio $r$, the noise bias can be corrected for (i.e., by Jackknife), such that any bias on a detection arises from foreground residuals, not from instrumental noise residuals. In this respect, the \texttt{cMILC} method focuses on nulling foreground biases at the expense of increased noise variance.

\vspace{-1mm}
\subsubsection{Higher-order statistics}
By realizing that CMB and instrumental noise are mostly Gaussian fields compared to Galactic foregrounds which are highly non-Gaussian, it becomes clear that \texttt{cMILC} will get rid of most non-Gaussian residuals in the CMB map through nulling constraints, while the standard ILC will not. 
As an illustration, the skewness of the reconstructed CMB map reads:
\begin{align}
\label{eq:skew}
\langle \hat{s}^3 \rangle 
&=\langle s^3 \rangle\,+\, \sum_{i,j,k=1}^m \left(\bdw^{\rm T}\bdb_i\right)\left(\bdw^{\rm T}\bdb_j\right)\left(\bdw^{\rm T}\bdb_k\right) \langle g_i\,g_j\,g_k\rangle\,+\, \langle \left(\bdw^{\rm T}\bdn\right)^3\rangle
\nonumber \\
&\simeq \sum_{i,j,k=1}^m \left(\bdw^{\rm T}\bdb_i\right)\left(\bdw^{\rm T}\bdb_j\right)\left(\bdw^{\rm T}\bdb_k\right) \langle g_i\,g_j\,g_k\,\rangle,
\end{align}
where in the last line we assumed that CMB and noise are mostly Gaussian and have negligible skewness, i.e.  ${\langle s^3\rangle \simeq 0}$ and ${\langle \bdn^3\rangle \simeq 0}$. By construction, the \texttt{cMILC} weights give $\bdw_{\rm cMILC}^{\rm T}\bdb_i=0$, such that the skewness due to foreground residuals in Eq.~\eqref{eq:skew} should reduce to zero for the \texttt{cMILC} map, while this is not the case for the standard ILC map. These expectations are also confirmed by our analysis on sky simulations (see Fig.~\ref{Fig:sky_stat}). 
While this work focuses on $B$-modes, we refer to Sect.~\ref{sec:disc} for a brief discussion of the potential benefits of reducing non-Gaussian residuals for primordial non-Gaussianity and CMB lensing analyses.

\vspace{-3mm}
\section{Analysis}
\label{sec:analysis}

\begin{table*}
\caption{Nomenclature of the \texttt{NILC} and \texttt{cMILC} methods for several combinations of constraints on various moments of the foreground emission.}
\label{tab:nom}
\begin{tabular}{llc}
\toprule
    Case &  Moments (SED) &  Parameters \\
\midrule
 NILC &  $a_{\rm CMB}$ &  1 \\
 cMILC01 &  $a_{\rm CMB}$ ; $f_{\rm sync}$ &  2 \\
 cMILC02 &  $a_{\rm CMB}$ ; $f_{\rm dust}$ &  2 \\
 cMILC03 &  $a_{\rm CMB}$ ; $f_{\rm sync}$ ; $f_{\rm dust}$ &  3 \\
 cMILC04 &  $a_{\rm CMB}$ ; $f_{\rm dust}$ ; $\partial_\beta\,f_{\rm dust}$ &  3 \\
 cMILC05 &  $a_{\rm CMB}$ ; $f_{\rm sync}$ ; $f_{\rm dust}$ ; $\partial_\beta\,f_{\rm sync}$ &  4 \\
 cMILC06 &  $a_{\rm CMB}$ ; $f_{\rm sync}$ ; $f_{\rm dust}$ ; $\partial_\beta\,f_{\rm dust}$ &  4 \\
 cMILC07 &  $a_{\rm CMB}$ ; $f_{\rm sync}$ ; $f_{\rm dust}$ ; $\partial_\beta\,f_{\rm sync}$ ; $\partial_\beta\,f_{\rm dust}$ &  5 \\
 cMILC08 &  $a_{\rm CMB}$ ; $f_{\rm sync}$ ; $f_{\rm dust}$ ; $\partial_\beta\,f_{\rm sync}$ ; $\partial_\beta\,f_{\rm dust}$ ; $\partial_T\,f_{\rm dust}$ &  6 \\
 cMILC09 &  $a_{\rm CMB}$ ; $f_{\rm sync}$ ; $f_{\rm dust}$ ; $\partial_\beta\,f_{\rm sync}$ ; $\partial_\beta\,f_{\rm dust}$ ; $\partial_T\,f_{\rm dust}$ ; $\partial^2_T\,f_{\rm dust}$ &  7 \\
 cMILC10 &  $a_{\rm CMB}$ ; $f_{\rm sync}$ ; $f_{\rm dust}$ ; $\partial_\beta\,f_{\rm sync}$ ; $\partial_\beta\,f_{\rm dust}$ ; $\partial_T\,f_{\rm dust}$ ; $\partial^2_\beta\,f_{\rm sync}$ ; $\partial^2_T\,f_{\rm dust}$ &  8 \\
 cMILC11 &  $a_{\rm CMB}$ ; $f_{\rm sync}$ ; $f_{\rm dust}$ ; $\partial_\beta\,f_{\rm sync}$ ; $\partial_\beta\,f_{\rm dust}$ ; $\partial_T\,f_{\rm dust}$ ; $\partial^2_\beta\,f_{\rm sync}$ ; $\partial^2_T\,f_{\rm dust}$ ; $\partial_\beta\partial_T\,f_{\rm dust}$ &  9 \\
 \midrule
 \multirow{2}{*}{cMILC12} &  $a_{\rm CMB}$ ; $f_{\rm sync}$ ; $f_{\rm dust}$ ; $\partial_\beta\,f_{\rm sync}$ ; $\partial_T\,f_{\rm dust}$\, at low $\ell$\,\qquad \multirow{2}{*}{(hybrid case; see Sect.~\ref{sec:disc})} &  \multirow{2}{*}{5} \\
                &  $a_{\rm CMB}$ ; $f_{\rm sync}$ ; $f_{\rm dust}$ ; $\partial_\beta\,f_{\rm dust}$  ; $\partial_T\,f_{\rm dust}$\, at high $\ell$ &   \\
\bottomrule
\end{tabular}
\end{table*}

\subsection{Sky simulations}
\label{subsec:sims}
In our sky simulations, we consider different models of the foreground emission, capitalizing on existing tools. %
For \pico, we use the public \pico sky simulations\footnote{\url{https://zzz.physics.umn.edu/ipsig/20180424_dc_maps} (models 91 and 96).}, which have been delivered on NERSC as an open data analysis challenge. These include both the \texttt{PySM} {\tt d1s1} model \citep{PySM2017}, which is consistent with \planck observations, and a more complicated dust model derived from MHD simulations \citep{Kritsuk2018} that has non-trivial spectral dependence and accounts for line-of-sight effects, as we discuss in Sect.~\ref{subsubsec:mhd}.
For \litebird, since there is no available public simulation, we replicated the \texttt{PySM} {\tt d1s1} model of the public \pico simulation with specifications given in Table~\ref{tab:ltb}.

We first evaluate the performance of our semi-blind component separation method \texttt{cMILC} on the \texttt{PySM} sky simulations {\tt d1s1} for both  \litebird and \pico experiments, and compare the results of \texttt{cMILC} with those obtained with the standard \texttt{NILC} method.
The simulated sky maps in polarization for the model {\tt d1s1} include several components of emission: CMB anisotropies with a tensor-to-scalar ratio of $r=0$, an optical depth to reionization of $\tau =0.054$ and full lensing contamination ($A_{\rm L}=1$); Galactic thermal dust emission as a modified blackbody component with varying spectral index and temperature across the sky, based on the \textit{Planck} \texttt{Commander} maps \citep{Planck2015_X}; and Galactic synchrotron emission as a power-law component with varying spectral index across the sky, based on the template maps from \cite{Miville-Deschenes2008}. For more details regarding PySM and its implementation we refer to \cite{PySM2017}.

The sky emission is integrated over either \litebird (Table~\ref{tab:ltb}) or \pico frequency bands (Table~\ref{tab:pico}), for which current simulations assume $\delta$-function bandpasses.
Each frequency map has a \healpix\ $N_{\rm side}=512$ pixelisation scheme \citep{Gorski2005}, and is smoothed with a Gaussian beam of FWHM values listed in Tables~\ref{tab:ltb}-\ref{tab:pico}. Gaussian white noise map realisations of typical r.m.s. values listed in the aforementioned tables were then co-added to the sky maps. Therefore, we have 15 sky maps ranging from $40$\,GHz to $402$\,GHz for the \litebird-like simulation, and 21 sky maps ranging from $20$\,GHz to $800$\,GHz for the \pico simulation. 

We perform foreground removal by applying the \texttt{NILC} and \texttt{cMILC} methods to these two sets of sky maps, and study the impact on CMB $B$-mode reconstruction (map, power spectrum, and tensor-to-scalar ratio) of deprojecting more and more foreground moments with \texttt{cMILC}, with combinations as listed in Table~\ref{tab:nom}. For our analysis, we first transform the Stokes $Q,U$ full-sky maps into full-sky $B$-mode maps at each frequency, then apply the \texttt{NILC} and \texttt{cMILC} methods on the set of full-sky $B$-mode observations, thus avoiding $E$-to-$B$ leakage inherent to masking procedures. To obtain numerically stable results, it was important to carefully consider the precision settings of the linear algebra routines. The CMB $B$-mode map reconstruction is performed at $40'$ angular resolution for both \litebird and \pico. As a main figure of merit we use $r$, but we expect other observables such as $\sum m_\nu$ and $N_{\rm eff}$ to also benefit from \texttt{cMILC} (see discussion in Sect.~\ref{sec:disc}).

\subsection{Results}
\label{subsec:results}

\subsubsection{Visual inspection of recovered maps in the BICEP2 region}
\label{subsubsec:maps}

\begin{figure*}
  \begin{center}
    \includegraphics[width=0.7\columnwidth]{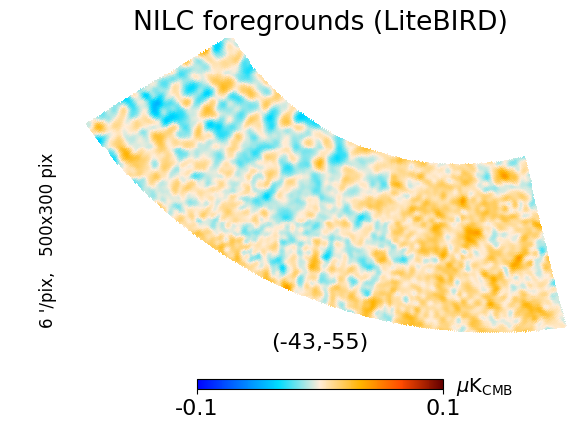}~
     \includegraphics[width=0.7\columnwidth]{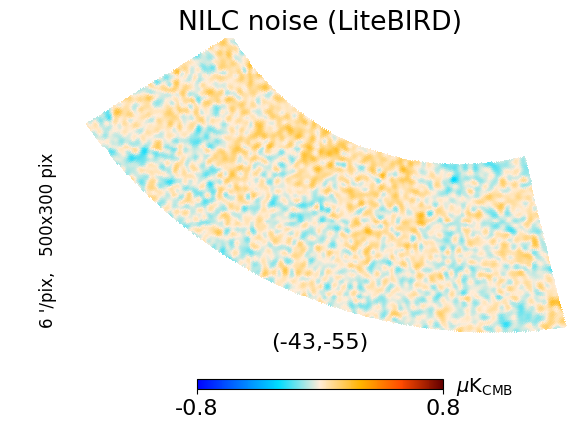}~
      \includegraphics[width=0.7\columnwidth]{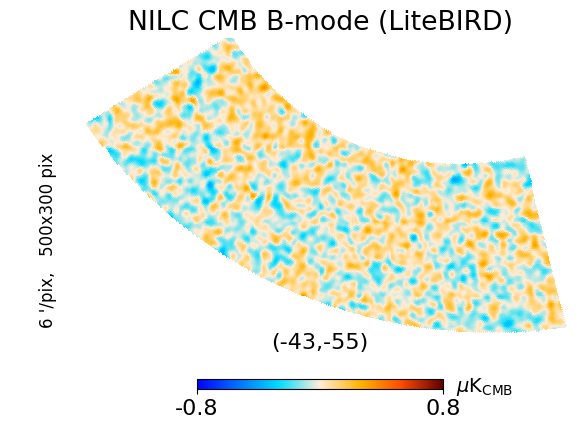}
    \\[1.5mm]
     \includegraphics[width=0.7\columnwidth]{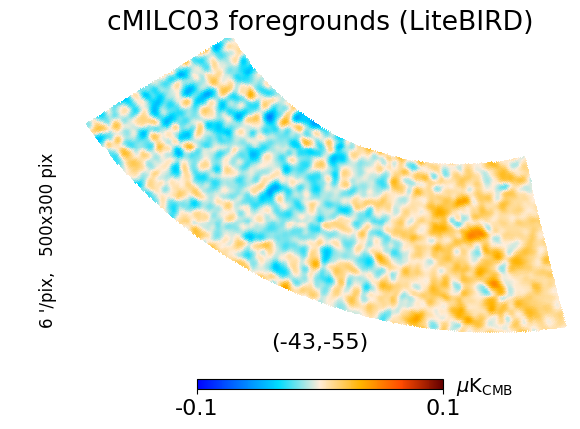}~
     \includegraphics[width=0.7\columnwidth]{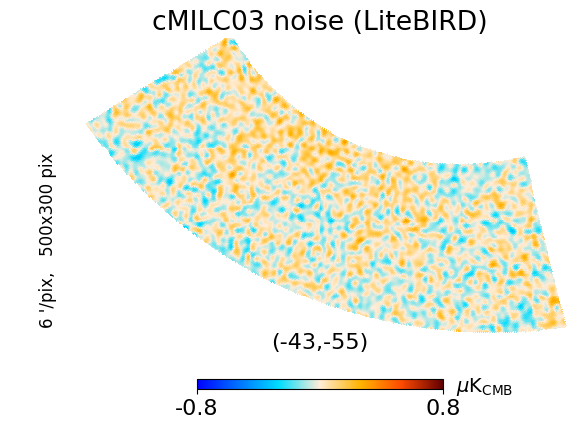}~
      \includegraphics[width=0.7\columnwidth]{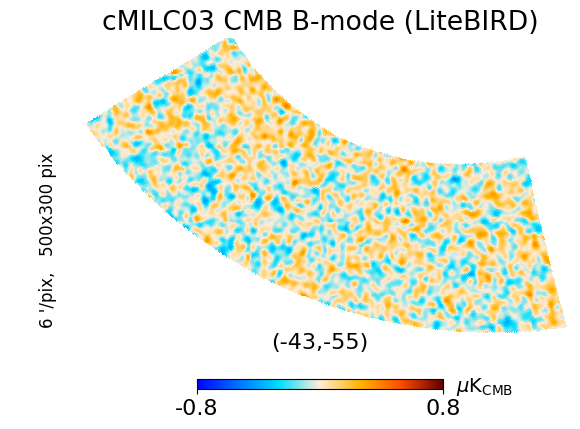}
    \\[1.5mm]
     \includegraphics[width=0.7\columnwidth]{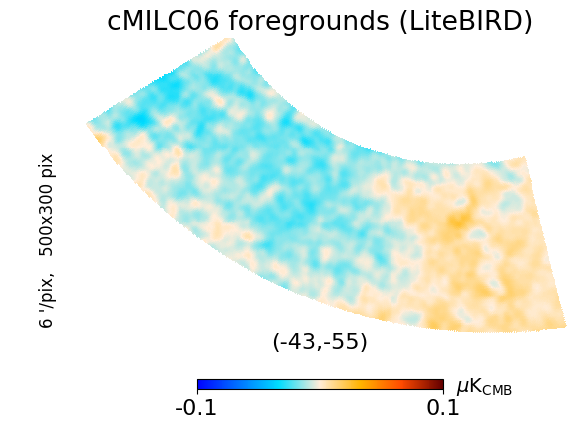}~
     \includegraphics[width=0.7\columnwidth]{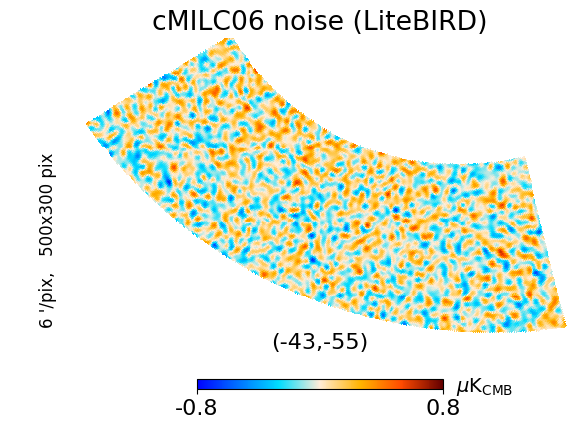}~
      \includegraphics[width=0.7\columnwidth]{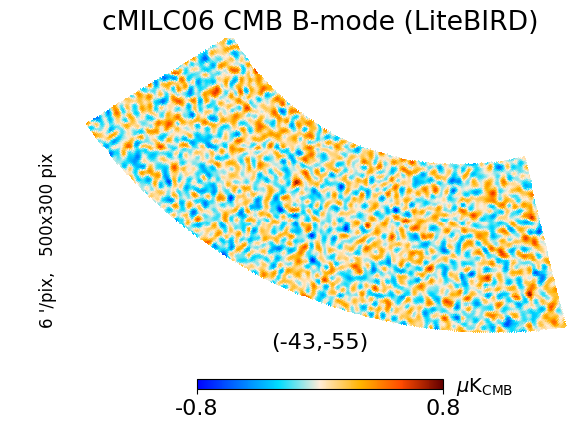}
   \\[1.5mm]
     \includegraphics[width=0.7\columnwidth]{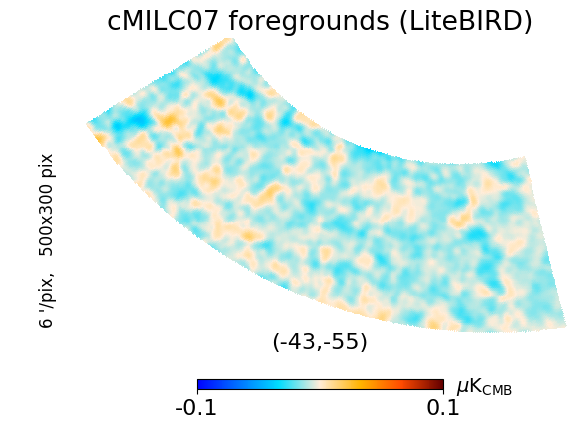}~
     \includegraphics[width=0.7\columnwidth]{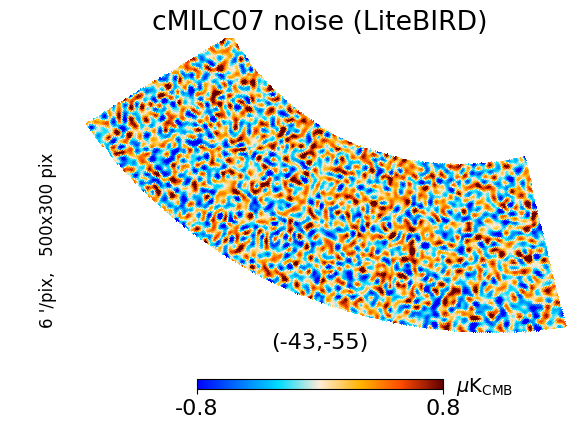}~
      \includegraphics[width=0.7\columnwidth]{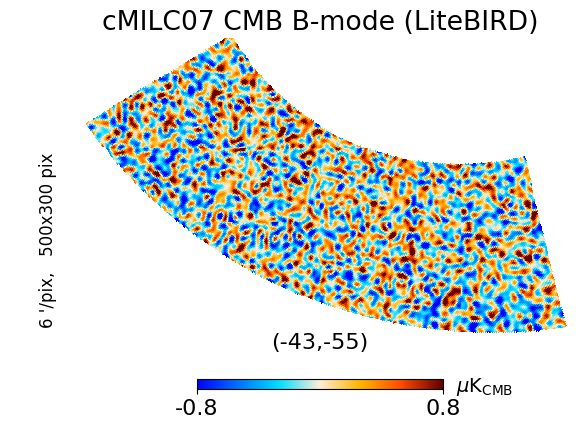}
      \\[1.5mm]
     \includegraphics[width=0.7\columnwidth]{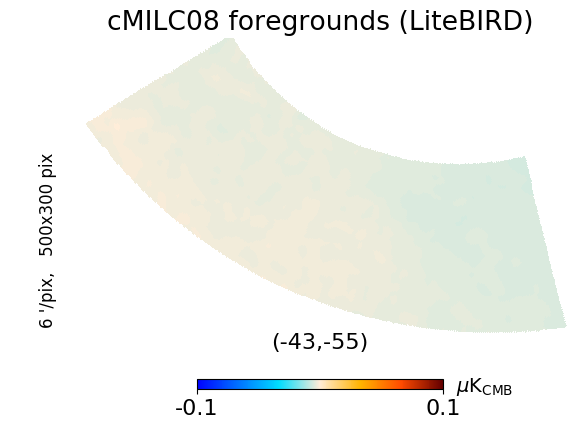}~
     \includegraphics[width=0.7\columnwidth]{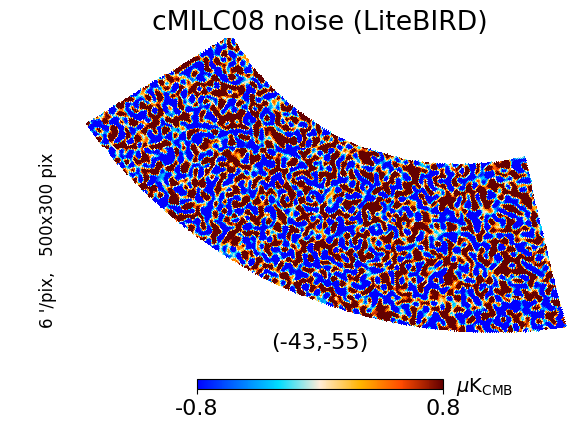}~
      \includegraphics[width=0.7\columnwidth]{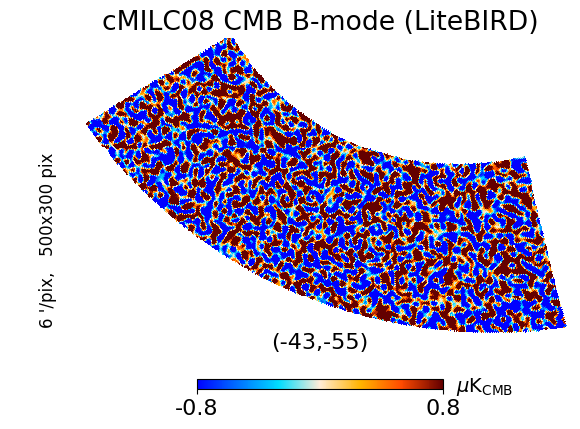}  
 \end{center}
\caption{\textit{Top to bottom rows}: \texttt{cMILC} results for \litebird in the BICEP2 region when when deprojecting more and more foreground moments. \textit{Left column}: residual foregrounds. \textit{Middle column}: residual noise. \textit{Right column}: reconstructed CMB $B$-mode map.  Deprojecting moments with \texttt{cMILC} significantly reduces the residual foreground contamination in the recovered CMB $B$-mode map, although this comes along with a noise penalty. Among these maps, CMILC06 is the optimal choice in terms of trade-off between residual foreground bias and noise penalty on the tensor-to-scalar ratio $r$ (see Sect.~\ref{subsubsec:r}).}
\label{Fig:bicep_field_ltb}
\end{figure*}

\begin{figure*}
  \begin{center}
    \includegraphics[width=0.7\columnwidth]{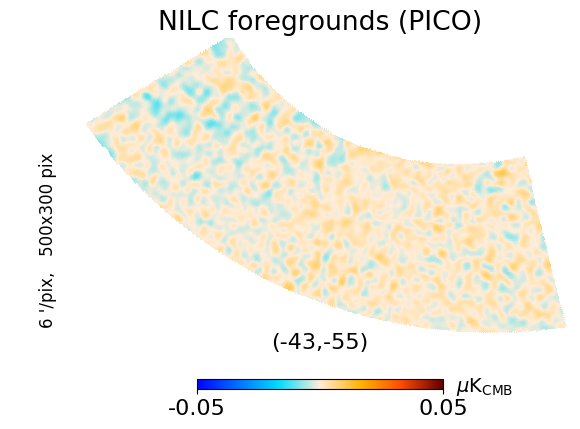}~
     \includegraphics[width=0.7\columnwidth]{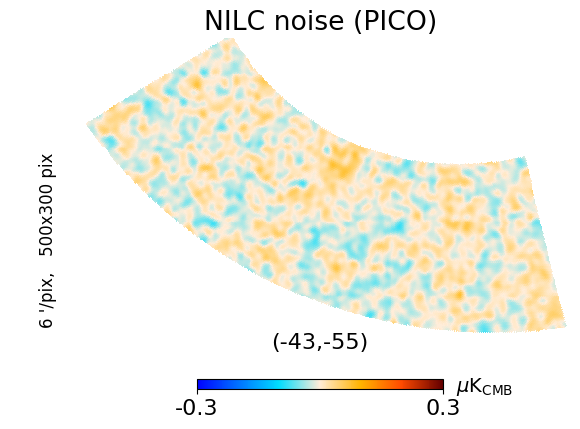}~
      \includegraphics[width=0.7\columnwidth]{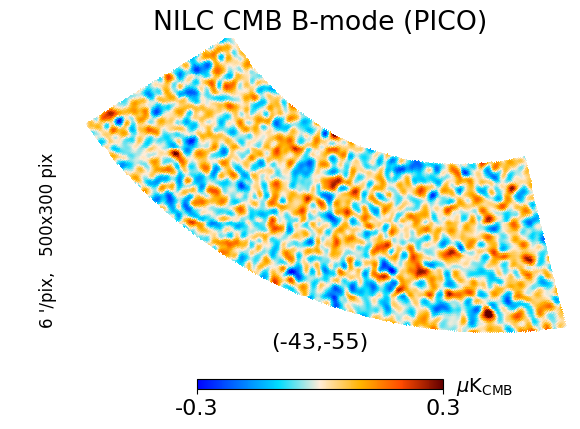}
    \\[1.5mm]
     \includegraphics[width=0.7\columnwidth]{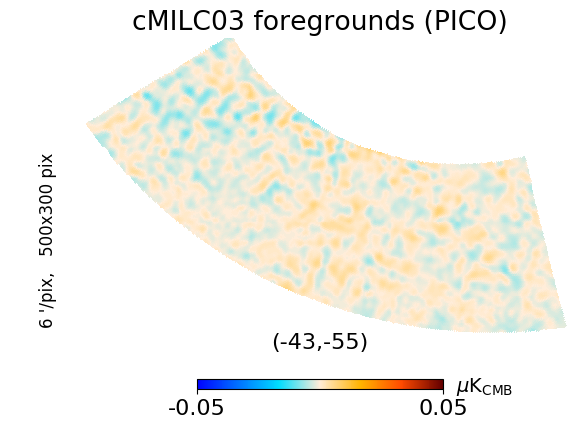}~
     \includegraphics[width=0.7\columnwidth]{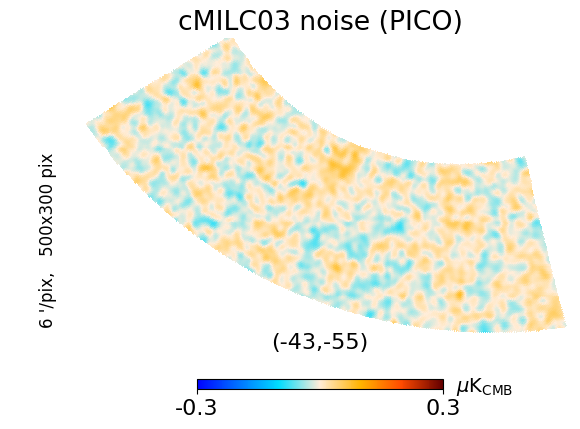}~
      \includegraphics[width=0.7\columnwidth]{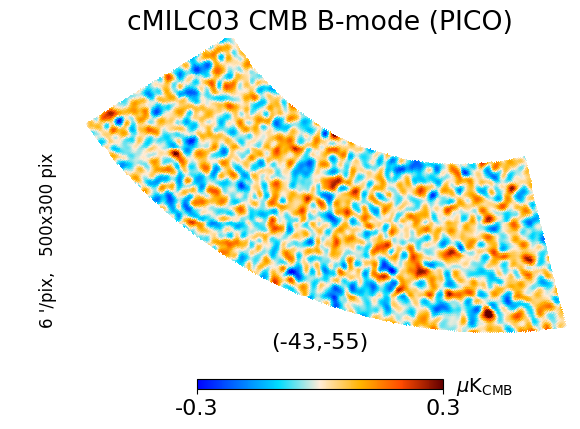}
    \\[1.5mm]
     \includegraphics[width=0.7\columnwidth]{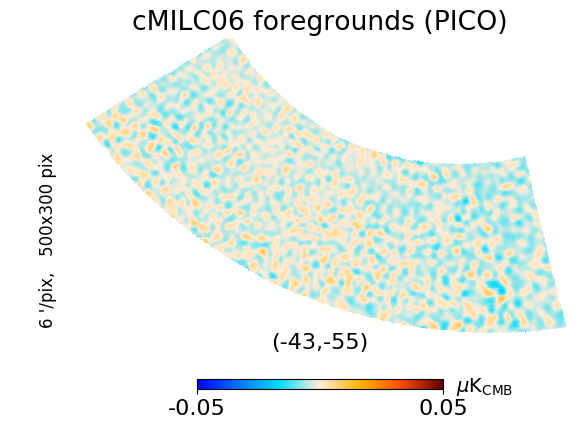}~
     \includegraphics[width=0.7\columnwidth]{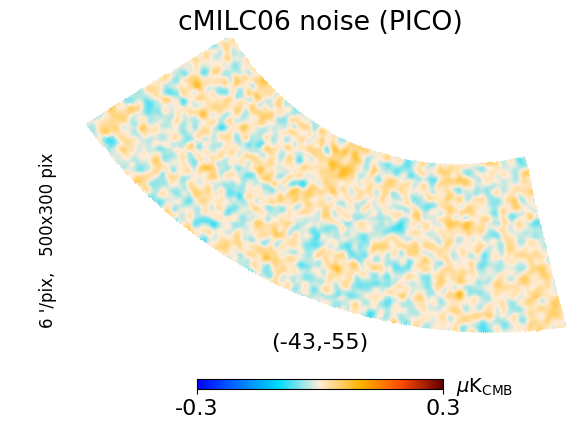}~
      \includegraphics[width=0.7\columnwidth]{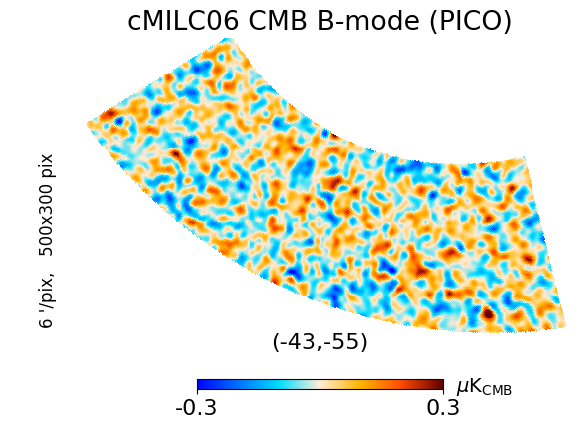}
       \\[1.5mm]
     \includegraphics[width=0.7\columnwidth]{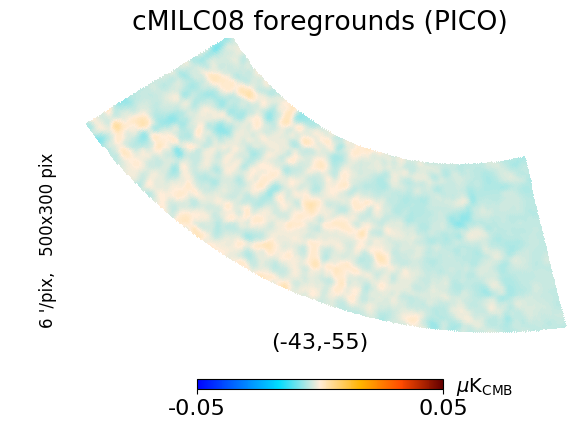}~
     \includegraphics[width=0.7\columnwidth]{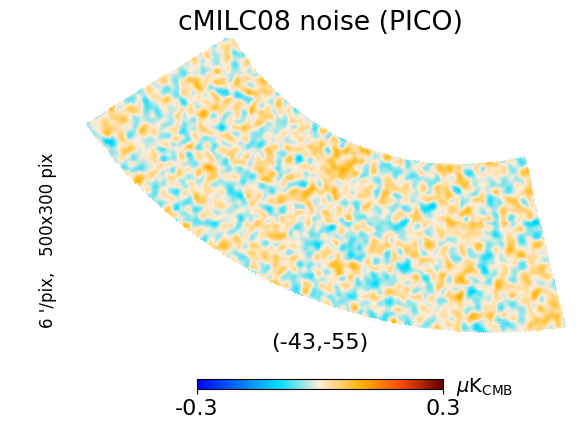}~
      \includegraphics[width=0.7\columnwidth]{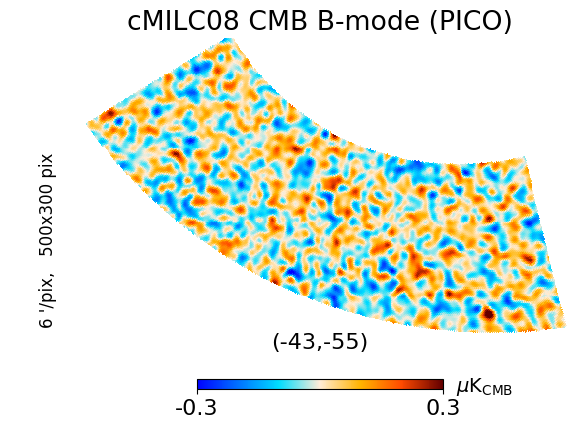}
        \\[1.5mm]
     \includegraphics[width=0.7\columnwidth]{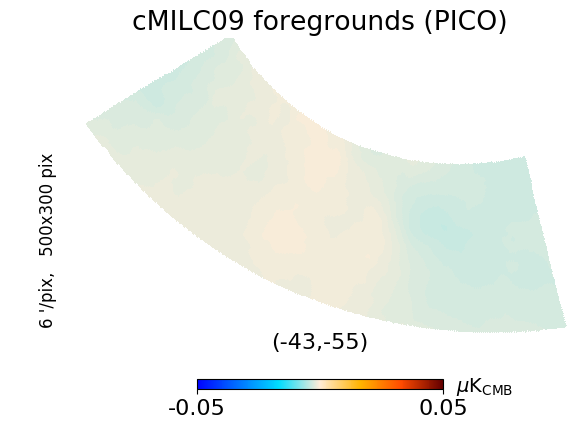}~
     \includegraphics[width=0.7\columnwidth]{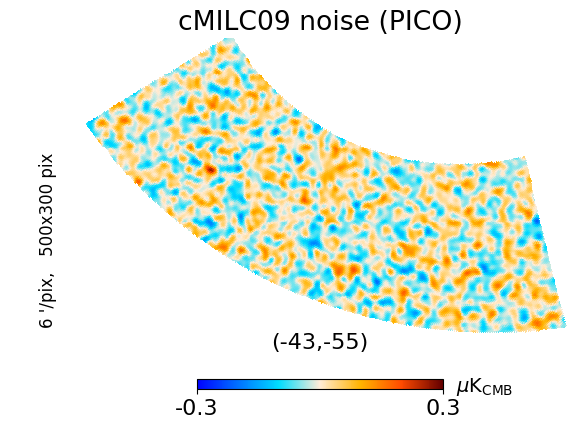}~
      \includegraphics[width=0.7\columnwidth]{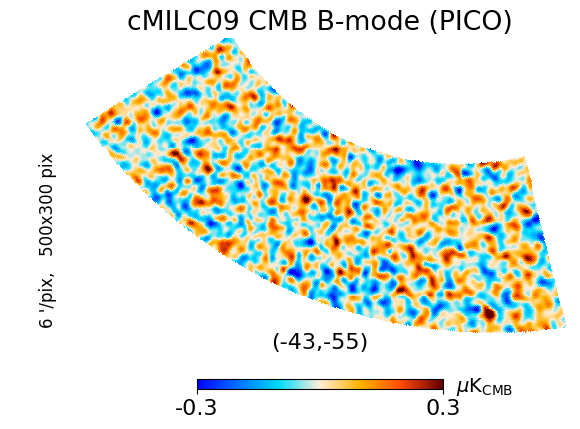}
 \end{center}
\caption{\textit{Top to bottom rows}: \texttt{cMILC} results for \pico in the BICEP2 region when deprojecting more and more foreground moments. \textit{Left column}: residual foregrounds. \textit{Middle column}: residual noise. \textit{Right column}: reconstructed CMB $B$-mode map.  Deprojecting moments with \texttt{cMILC} significantly reduces the residual foreground contamination in the recovered CMB $B$-mode map, while the noise penalty is still reasonably low for \pico. CMILC08 is the optimal choice in terms of trade-off between residual foreground bias and noise penalty on the tensor-to-scalar ratio $r$ (see Sect.~\ref{subsubsec:r}).}
\label{Fig:bicep_field_pico}
\end{figure*}

\begin{figure*}
  \begin{center}
       \includegraphics[width=0.495\textwidth]{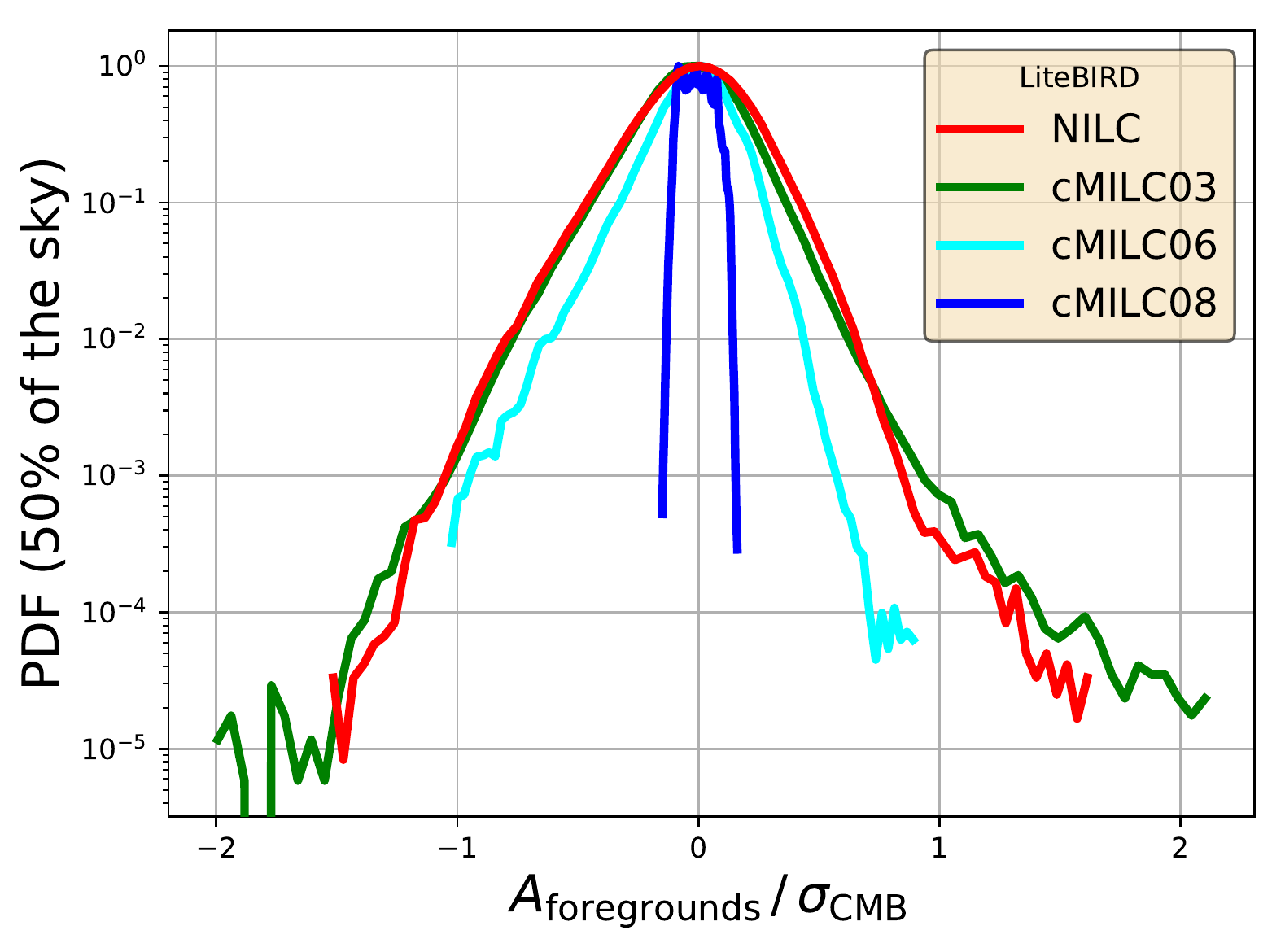}
        \includegraphics[width=0.495\textwidth]{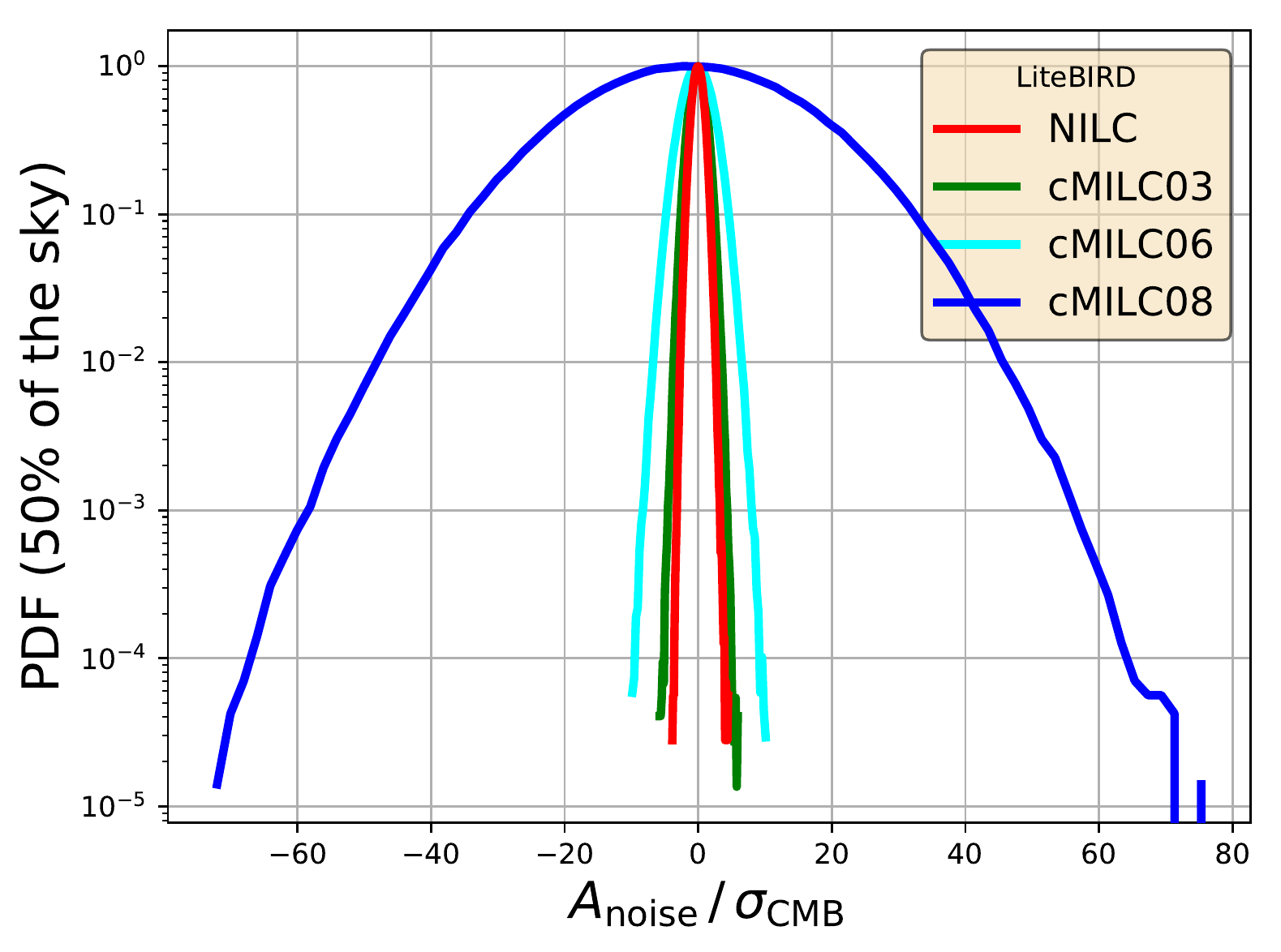}~
         \\[2mm]
      \includegraphics[width=0.495\textwidth]{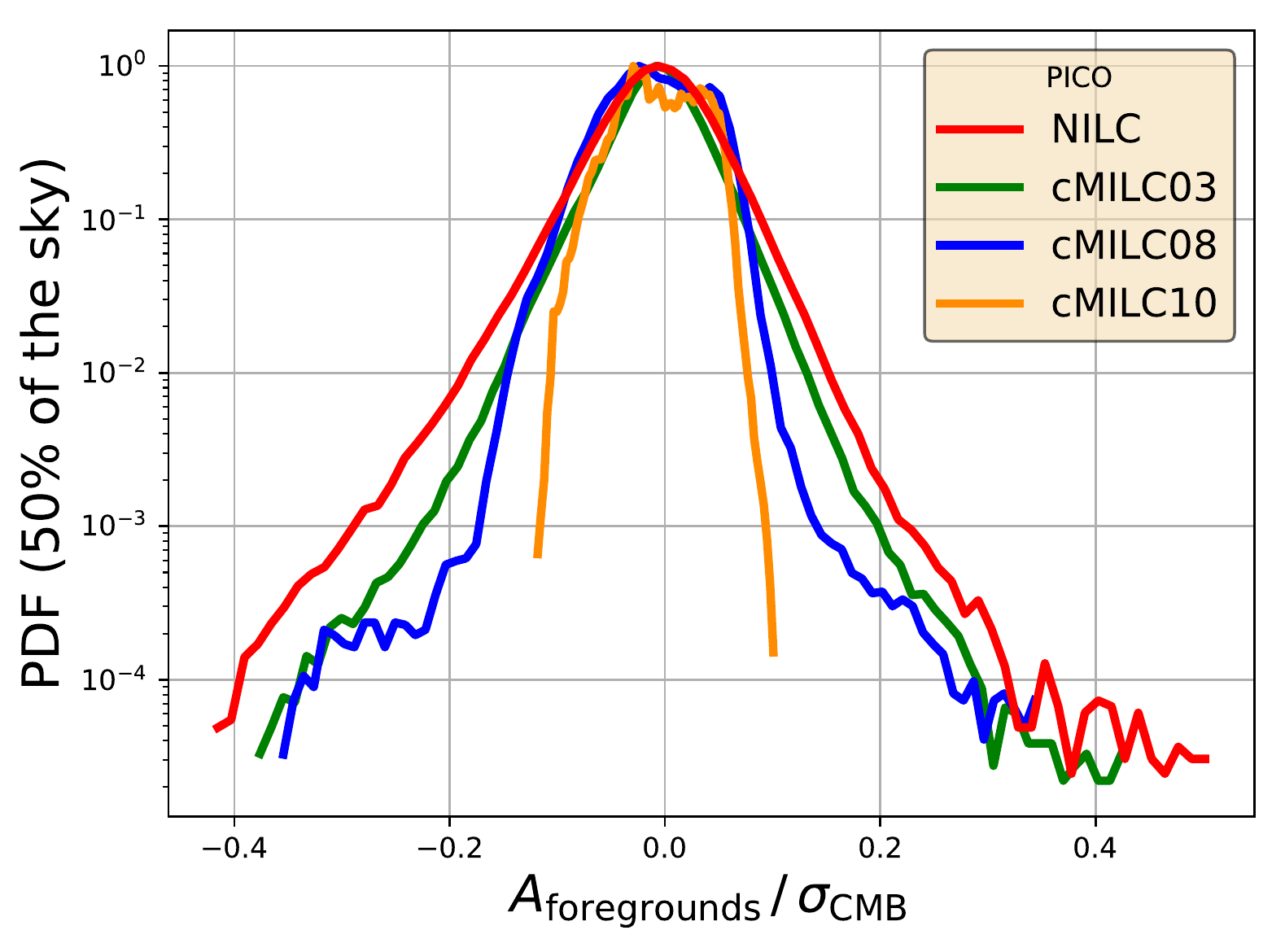}
      \includegraphics[width=0.495\textwidth]{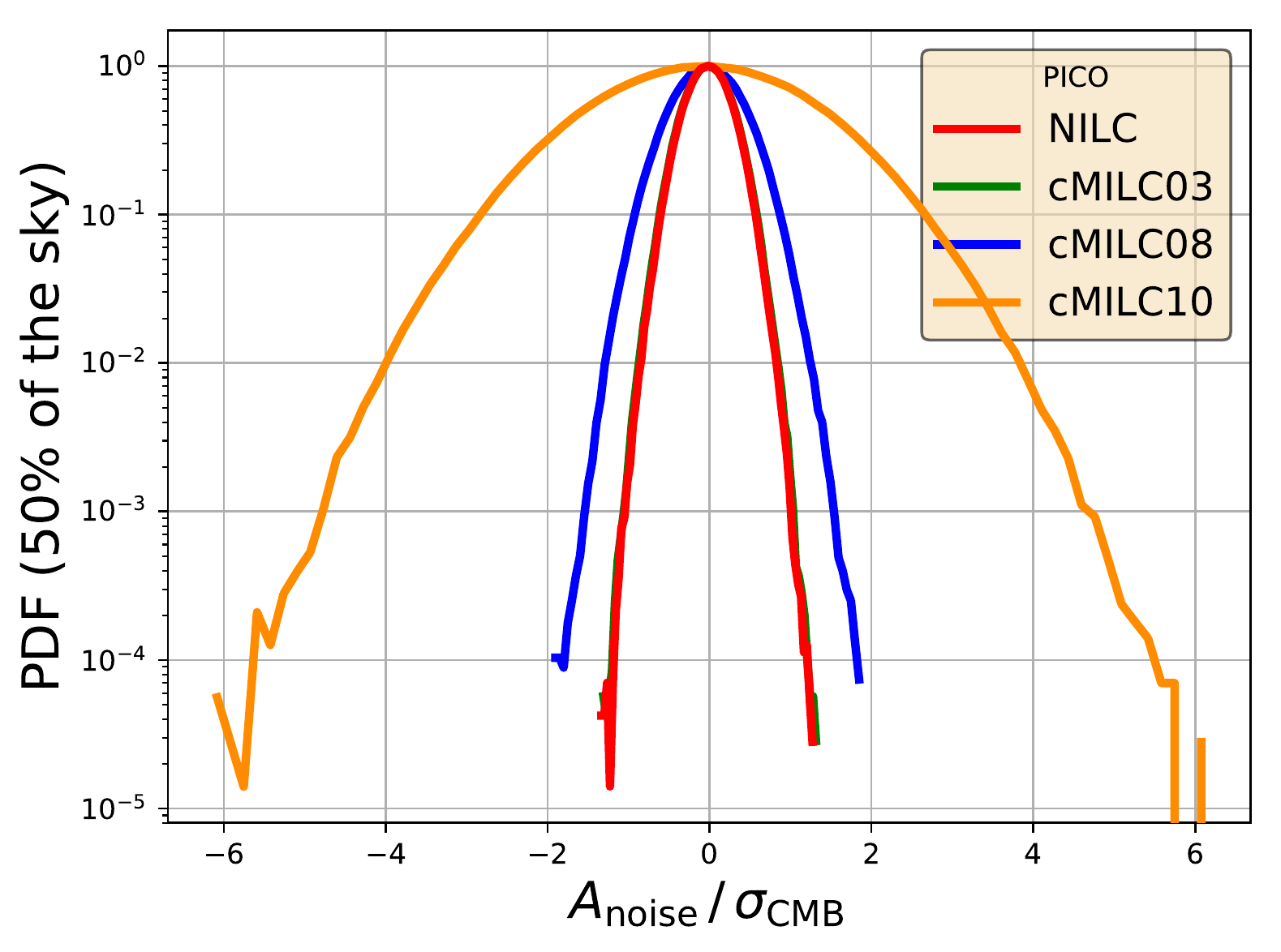}~
       \end{center}
\caption{One-point statistics (probability distribution function -- PDF) of foreground residuals (\textit{left panels}) and noise residuals (\textit{right panels}) on $f_{\rm sky}=50$\% of the sky for \litebird (\textit{upper panels}) and \pico (\textit{lower panels}). Deprojecting more and more moments with \texttt{cMILC} significantly reduces the variance and skewness of non-Gaussian foreground residuals in the recovered CMB $B$-mode map. Besides $B$-modes, this would be of great benefit also for the search for primordial non-Gaussianity and for CMB lensing reconstruction (see Sect.~\ref{sec:disc} for discussion).}
\label{Fig:sky_stat}
\end{figure*}

While the reconstruction of the CMB $B$-mode map from the \litebird and \pico sky maps by the \texttt{NILC} and \texttt{cMILC} methods is performed on the entire sky, for illustration we first inspect the quality of the reconstructed maps and their residual foreground and noise contamination in the BICEP2 region \citep{BICEP2014}. This allows us to better appreciate the improvement in foreground removal with \texttt{cMILC} in comparison to \texttt{NILC}.
To evaluate residual foreground and noise contamination in the reconstructed CMB $B$-mode map after component separation, we applied the same \texttt{NILC}/\texttt{cMILC} weights that were assigned in the reconstruction of the CMB $B$-mode map to the input foreground and noise maps of the simulations.

Figure~\ref{Fig:bicep_field_ltb} summarizes the results for \litebird in the BICEP2 region. The first row shows the \texttt{NILC} results, while the following rows show the \texttt{cMILC} results when including progressively more constraints on foreground moments, starting in the second row with cMILC03 which deprojects zeroth-order moments ${(f_{\rm sync}, f_{\rm dust})}$, and ending in the bottom row with cMILC08 which deprojects both zeroth- and first-order moments ${(f_{\rm sync}, f_{\rm dust}, \partial_{\beta_s} f_{\rm sync}, \partial_{\beta_d} f_{\rm dust}, \partial_{T_d} f_{\rm dust})}$. No further constraints on second-order moments were imposed in \texttt{cMILC} because the noise degradation prevented us from probing higher-order moments of the foreground emission. 

The residual foreground contamination left by each method is shown in the first column: clearly, the level of residual foregrounds is significantly reduced in the \texttt{cMILC} CMB $B$-mode maps when including more and more moments in the component separation. The second and third columns show respectively the residual noise fluctuations and the recovered CMB $B$-mode fluctuations for each method. We see that the improvement on foreground removal due to moment constraints is at the expense of increasing noise in the reconstructed CMB $B$-mode map, as expected. Hence, there is a clear trade-off between mitigation of residual foreground contamination and noise degradation, but also a sweet spot where the noise penalty is more than compensated by the reduction of residual foreground contamination, depending on the observable of interest. As we will see in Sect.~\ref{subsubsec:r}, for $r$ the optimal solution is given by cMILC06 for \litebird.

Figure~\ref{Fig:bicep_field_pico} summarizes the results for \pico in the BICEP2 region. As expected, the level of residual foregrounds and noise in the CMB $B$-mode map reconstruction  for \pico is further reduced compared to \litebird due to higher sensitivity and broader spectral coverage ({$20$-$800$\,GHz}). \pico also has the capability to probe second-order moments of the foreground emission. In particular, adding constraints on first- and second-order dust temperature moments, $\bdw^{\rm T} \cdot \partial_{T_d} f_{\rm dust}=0$ (cMILC08; fourth row) and $\bdw^{\rm T} \cdot \partial^2_{T_d} f_{\rm dust}=0$ (cMILC09; bottom row), significantly reduces the residual foreground contamination at small scales. Again, the left column highlights the improvement in terms of residual foreground contamination when including more and more moments in \texttt{cMILC}, with best results obtained in the bottom row (cMILC09), where several moments of dust and synchrotron up to second order ${(f_{\rm sync}, f_{\rm dust}, \partial_{\beta_s} f_{\rm sync}, \partial_{\beta_d} f_{\rm dust}, \partial_{T_d} f_{\rm dust}, \partial^2_{T_d} f_{\rm dust})}$ were deprojected.
As expected, additional moment constraints are at the expense of an increase of residual noise contamination (middle column), but remain at a reasonable level compared to the r.m.s of CMB $B$-mode fluctuations (right column) thanks to the high sensitivity of \pico. The best trade-off for $r$ among those maps is actually given by cMILC08 for \pico, as we show in Sect.~\ref{subsubsec:r}.

\subsubsection{Statistical properties of residual foregrounds and noise}
\label{subsubsec:stats}

Besides map visualisation, it is instructive to look at the statistical properties of the residual foreground and noise contaminations in the recovered CMB $B$-mode maps, which we present in Fig.~\ref{Fig:sky_stat} for \litebird (upper panels) and \pico (lower panels).

In the left panels of Fig.~\ref{Fig:sky_stat}, we computed the one-point statistics (probability distribution function -- PDF) of the residual foreground contamination on $f_{\rm sky}=50$\% of the sky for the \texttt{NILC} CMB $B$-mode map (red line) and for several \texttt{cMILC} CMB $B$-mode maps, with more and more moments being deprojected. Similarly, in the right panels of Fig.~\ref{Fig:sky_stat}, we show the PDF of the residual noise contamination for the same \texttt{NILC} and \texttt{cMILC} maps. Overall, the residual foreground and noise contamination is lower for \pico than \litebird due to higher sensitivity and broader spectral coverage. As we already stressed in Sect.~\ref{subsec:stat}, the variance of the residual noise contamination (right panels) increases when adding constraints on moments in \texttt{cMILC}  because of the increasing volume of the parameter space with respect to \texttt{NILC}. In stark contrast, the variance of the residual foreground contamination (left panels) significantly decreases by deprojecting moments with \texttt{cMILC}. 

Clearly, the PDF of foreground residuals (left panels) shows larger variance and skewness for \texttt{NILC} (red line), while adding nulling constraints on zeroth- and first-order moments of dust and synchrotron ${(f_{\rm sync}, f_{\rm dust}, \partial_{\beta_s} f_{\rm sync}, \partial_{\beta_d} f_{\rm dust}, \partial_{T_d} f_{\rm dust})}$ with \texttt{cMILC} significantly reduces the variance and skewness of the residual foreground contamination (cMILC08; blue line), therefore leaving only negligible Gaussian residuals in the recovered CMB $B$-mode map. 
More quantitatively, for \litebird the variance of residual foregrounds is $\sigma_{\rm FG}^2  = 0.05\,\sigma_{\rm CMB}^2$ for \texttt{NILC} and $\sigma_{\rm FG}^2  = 0.003\,\sigma_{\rm CMB}^2$ for cMILC08, hence a reduction of the foreground variance by $94$\% with \texttt{cMILC}. For \pico, the variance of residual foregrounds is $\sigma_{\rm FG}^2  = 0.0025\,\sigma_{\rm CMB}^2$ for \texttt{NILC} and $\sigma_{\rm FG}^2  = 0.0012\,\sigma_{\rm CMB}^2$ for cMILC10, hence a reduction of the foreground variance by $52$\% with \texttt{cMILC}.
Similarly, the skewness of residual foregrounds on $f_{\rm sky}=50$\% of the sky is $\langle s_{\rm FG}^3 \rangle / \sigma_{\rm FG}^3 = -0.3$ for \texttt{NILC} and $\langle s_{\rm FG}^3 \rangle / \sigma_{\rm FG}^3 = 0.18$ for cMILC08, hence a reduction of the residual skewness by $40$\% with \texttt{cMILC} (with respect to the CMB r.m.s., the residual skewness is $\langle s_{\rm FG}^3 \rangle / \sigma_{\rm CMB}^3 = -3\times 10^{-3}$ for \texttt{NILC} and $\langle s_{\rm FG}^3 \rangle / \sigma_{\rm CMB}^3 = 4\times 10^{-5}$ for cMILC08). For \pico, the skewness of residual foregrounds is $\langle s_{\rm FG}^3 \rangle / \sigma_{\rm FG}^3 = -0.3$ for \texttt{NILC} and $\langle s_{\rm FG}^3 \rangle / \sigma_{\rm FG}^3 = -0.11$ for cMILC10, hence a reduction of the residual skewness by $60$\% with \texttt{cMILC} (with respect to the CMB r.m.s., the residual skewness is $\langle s_{\rm FG}^3 \rangle / \sigma_{\rm CMB}^3 = -4\times 10^{-5}$ for \texttt{NILC} and $\langle s_{\rm FG}^3 \rangle / \sigma_{\rm CMB}^3 = -5\times 10^{-6}$ for cMILC10).
Therefore, \texttt{cMILC} leads to a solution that minimizes foreground variance and skewness, largely beating \texttt{NILC} on that front.  The trends observed in our numerical results are consistent with those predicted by the  analytical expressions in Sect.~\ref{subsec:stat}.

\subsubsection{$B$-mode power spectrum}
\label{subsubsec:ps}

\begin{figure*}
  \begin{center}
	\includegraphics[width=0.99\textwidth]{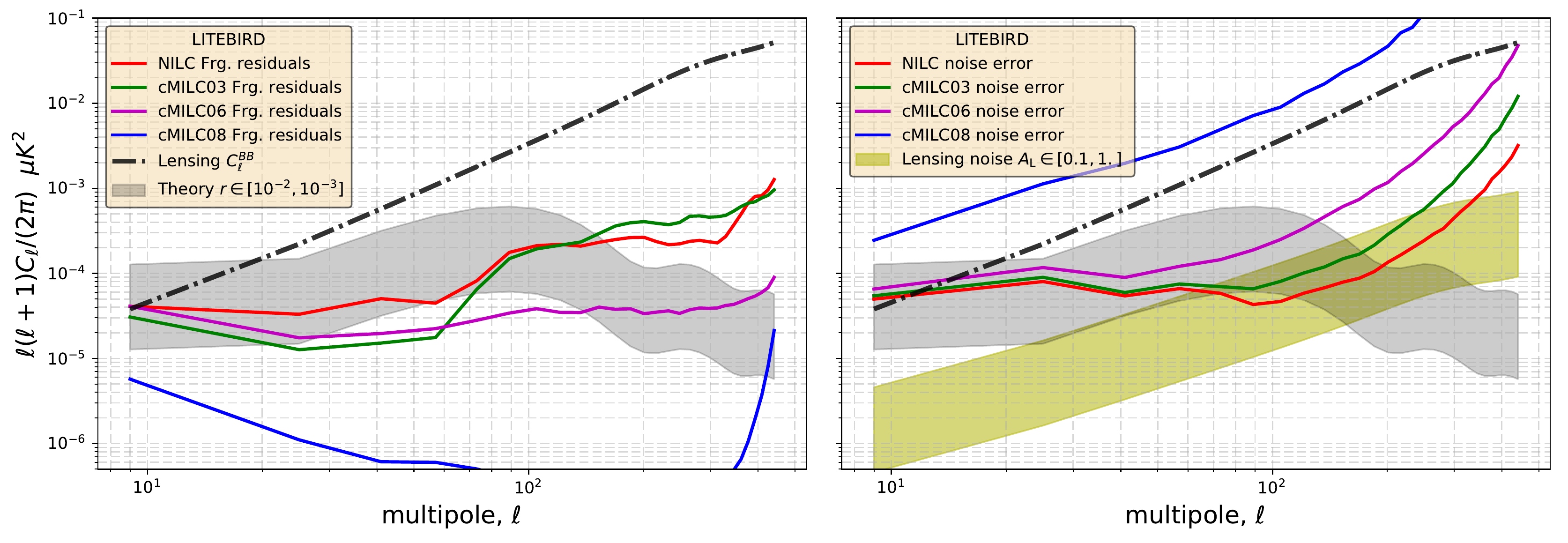}~\\
 	\includegraphics[width=0.99\textwidth]{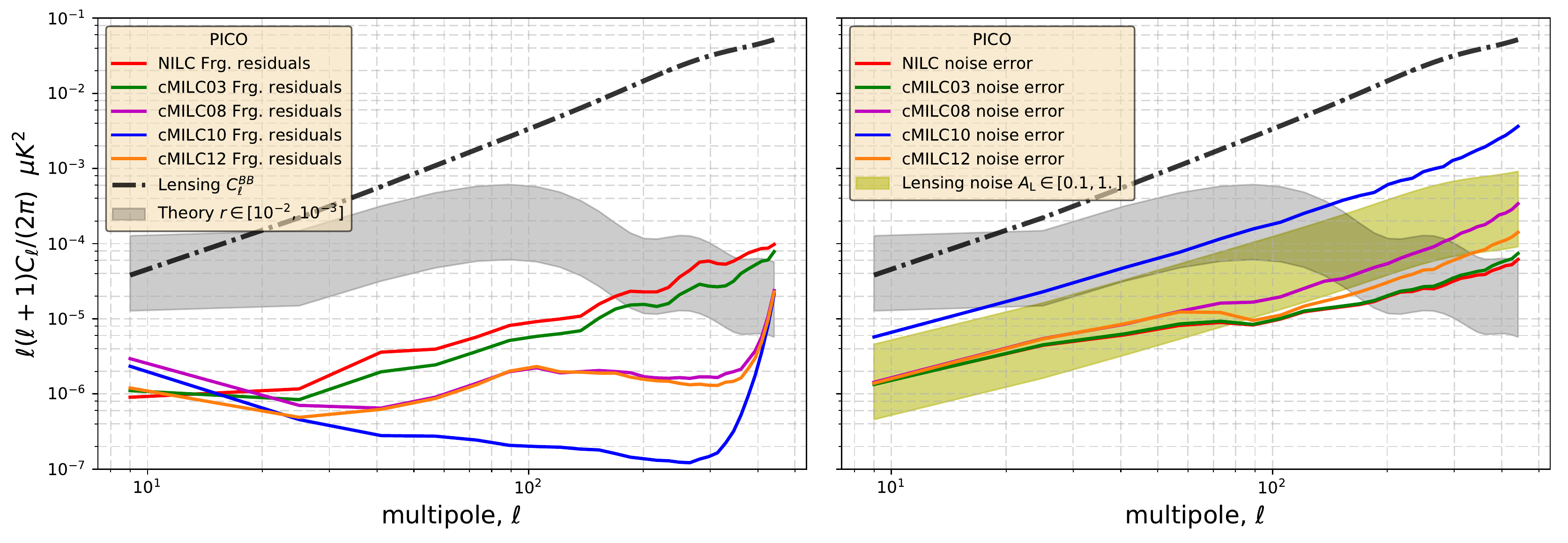}\\
     \end{center}
\caption{ \textit{Left panels}: Power spectrum of the residual foreground contamination, $C_\ell^{\rm fgds}$, when deprojecting more and more foreground moments with \texttt{cMILC}, for \litebird (upper panel) and \pico (lower panel) for the \texttt{PySM} {\tt d1s1} simulation.  \textit{Right panels}: Standard deviation of the residual noise and lensing power spectra, ${\sqrt{2 / (2\ell +1)f_{\rm sky}}C_\ell^{\rm noise/lensing}}$. The residual foreground contamination across multipoles is significantly reduced by deprojecting more and more moments with \texttt{cMILC}, as compared to \texttt{NILC}.}
\label{Fig:power_spectra}
\end{figure*}

In Fig.~\ref{Fig:power_spectra}  we compute the angular power spectrum of the residual foreground contamination ($C_\ell^{\rm fgds}$; left panels) in the recovered CMB $B$-mode maps on $f_{\rm sky}=50\%$ of the sky, along with the standard deviation of the residual noise and lensing power spectra (${\sqrt{2 / (2\ell +1)f_{\rm sky}}C_\ell^{\rm noise/lensing}}$; right panels). The results from \texttt{NILC} and \texttt{cMILC} are shown for \litebird in the upper panels and for \pico in the lower panels. The power spectra are deconvolved from the Galactic mask and the beam window function using \texttt{MASTER} \citep{MASTER2002}, and binned across multipoles with $\Delta\ell =16$. The same Galactic mask was used for the \texttt{NILC} and \texttt{cMILC} maps for fair comparison, but in principle the shape of the Galactic mask  can be optimised for each variant of \texttt{cMILC}, since the morphology of the residual foreground contamination obviously differs, as is also suggested by the studies presented in Sec~\ref{subsubsec:maps} and Sec.~\ref{subsubsec:stats}.

It is clear from the left panels of Fig.~\ref{Fig:power_spectra} that adding constraints on dust and synchrotron moments in \texttt{cMILC} significantly reduces the residual foreground contamination across a broad range of multipoles compared to \texttt{NILC}, and thus potentially prevents biases due to residual foregrounds in the recovered primordial CMB $B$-mode power spectrum. In contrast, while the residual noise bias on the power spectrum can be corrected for using Jackknife, the noise error (standard deviation of residual noise power spectrum) increases when increasing the volume of the constrained parameter space in \texttt{cMILC}, as it is shown in the right panels of Fig.~\ref{Fig:power_spectra}.

For \litebird (upper left panel), while the power spectrum of residual foreground contamination for \texttt{NILC} (red line) is reduced well below the lensing $B$-mode signal, it is still at a level equivalent to $r \gtrsim 3\times 10^{-3}$ across the full range of multipoles. Therefore, after delensing, \texttt{NILC} would not guarantee proper recovery of the primordial CMB $B$-mode power spectrum at $r=10^{-3}$. In contrast, by adding constraints on more and more dust and synchrotron moments, \texttt{cMILC} helps reducing the level of residual foreground contamination in the recovered CMB $B$-mode power spectrum down to $r < 10^{-3}$ at recombination scales. In particular, deprojecting all zeroth- and first-order moments of the dust and synchrotron (cMILC08; blue line) enables significant reduction of the residual foreground contamination to CMB $B$-modes for \litebird, down to levels well below $r = 10^{-3}$ at all multipoles.

Similarly, the lower left panel of Fig.~\ref{Fig:power_spectra} shows the power spectra of the residual foreground contamination for \pico after component separation with the \texttt{NILC} and \texttt{cMILC} methods. Due to its higher sensitivity and broader spectral coverage, \pico allows reducing foreground residuals well below the primordial CMB $B$-mode power spectrum at $r=10^{-3}$ across the full range of multipoles covering both reionization and recombination scales, even with a blind \texttt{NILC} method. Still, \texttt{cMILC} enables further reduction of the residual foreground contamination to the CMB $B$-mode power spectrum by deprojecting more and more foreground moments. We find that first- and second-order dust temperature moments (cMILC08, cMILC10, and cMILC12) lead to the lowest biases due to residual foreground contamination, reaching below the detection limit $r=5\times 10^{-4}$ of \pico \citep{PICO2019}.

While deprojecting successively more moments with \texttt{cMILC} reduces the level of residual foreground contamination across a wide range of multipoles, for \pico the overall decrease of residuals actually breaks at the largest scales $\ell \lesssim 15$, suggesting that our first-guess pivots, based on zeroth-order fits in the literature \citep[e.g.][]{Planck2015_X}, might not be appropriate for the largest angular scales, where averaging processes, and thus high-order moments, are the most significant. For sensitive experiments like \pico, high-order moments should thus be included in parametric SED fitting to revise zeroth-order pivots.  In Sect.~\ref{subsec:optimal_pivot} and Fig.~\ref{Fig:optimal_pivot}, we update the dust pivot temperature $\overline{T}_d$ to show that, with revised pivots, \texttt{cMILC} is able to reduce the residual foreground contamination consistently at all mutipoles. However, a clear computational procedure for optimizing the pivots still requires more work, as we also explain in Sect.~\ref{subsec:optimal_pivot}.

To conclude, \texttt{cMILC} provides quite spectacular results on foregrounds removal across all multipoles, thus significantly reducing biases on the CMB $B$-mode power spectrum. However, there is a clear trade-off between residual foregrounds mitigation (left panels of Fig.~\ref{Fig:power_spectra}) and noise degradation (right panels of Fig.~\ref{Fig:power_spectra}) that needs to be appreciated to find the best combination of moments in  \texttt{cMILC} that will lead to less biased but still sensitive constraints on the observable of interest, which here is $r$ (see Sect.~\ref{subsubsec:r}).

\begin{figure*}
  \begin{center}
    \includegraphics[width=0.97\textwidth]{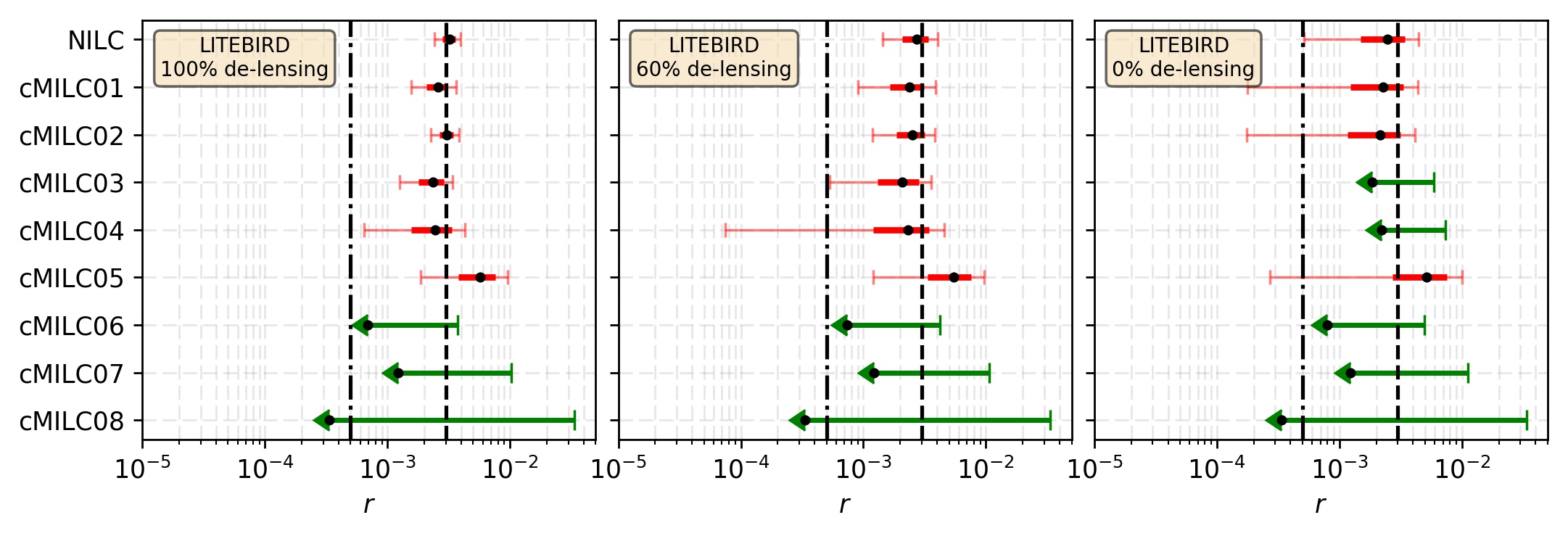}
    \\[0mm]
      \includegraphics[width=0.97\textwidth]{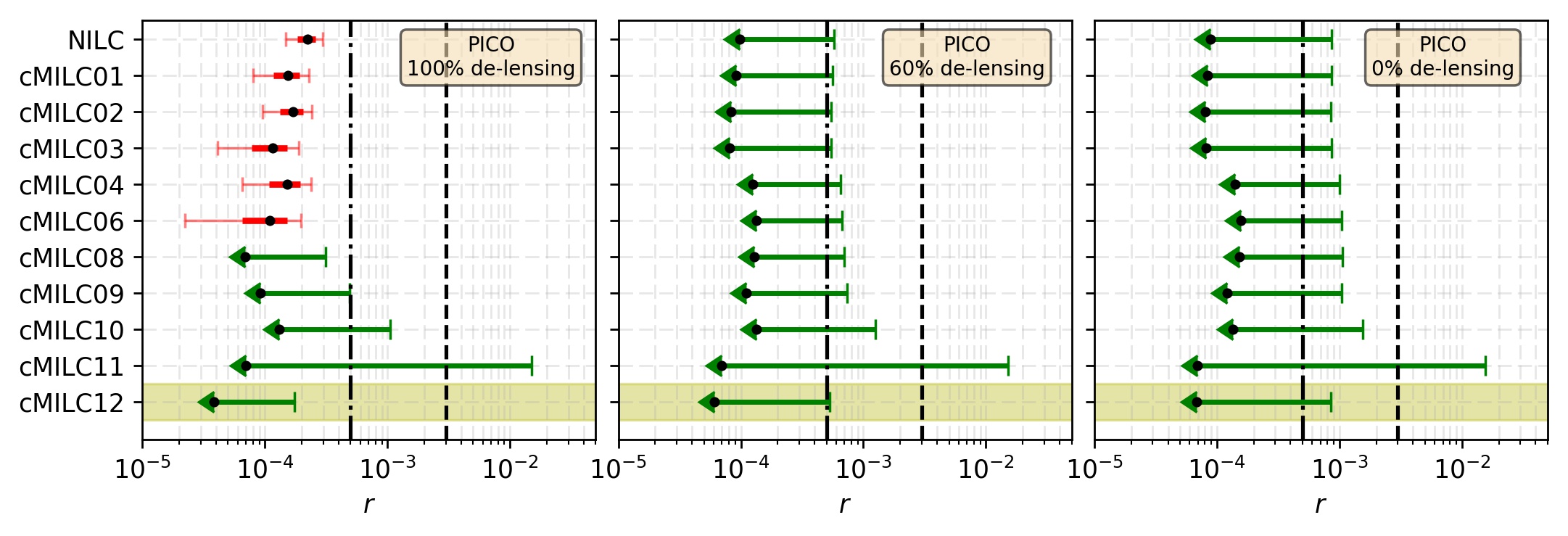}
     \end{center}
\caption{ \textit{Left panels}: Bias and uncertainty on the recovered tensor-to-scalar ratio $r$ for \litebird (\textit{upper panels}) and \pico  (\textit{lower panels}) after component separation on the \texttt{PySM} {\tt d1s1} simulation with the \texttt{NILC} and \texttt{cMILC} methods, for different levels of residual lensing contamination. Biased detections at more than $2\sigma$ due to residual foregrounds are shown in red with $1\sigma$ (bold) and $2\sigma$ (thin) error bars, while unbiased detections consistent with $r=0$ are shown in green with $2\sigma$ upper limits. The dashed vertical line marks $r=3\times 10^{-3}$ and the dash-dotted line $r=5\times 10^{-4}$. CMILC06 provides the best result on $r$ for \litebird (upper panel), by showing minimal bias due to foreground residuals, while not paying much noise penalty with respect to \texttt{NILC} in terms of $2\sigma$ upper limit.  For \pico (lower panel), cMILC08 and cMILC12 provide the lowest biases on $r$, without paying much noise penalty with respect to \texttt{NILC}.} 
\label{Fig:r_stat}
\end{figure*}

\subsubsection{Likelihood estimation of the tensor-to-scalar ratio}
\label{subsubsec:r}
In this section, we perform the likelihood estimation of the recovered probability distribution of the tensor-to-scalar ratio $r$ after foreground removal with \texttt{NILC} and \texttt{cMILC}. We emphasize two important aspects: the irreducible bias on $r$, or \textit{systematic error} arising from residual foreground contamination (noting that noise and lensing biases on the power spectrum can usually be corrected for), and the \textit{statistical uncertainty} $\sigma(r)$. The latter has contributions from cosmic variance of the primordial signal, residual foregrounds, residual lensing signals and noise.

The power spectrum of the reconstructed CMB $B$-mode map from either \texttt{NILC} or \texttt{cMILC} is
\begin{align}
\hat{C}^{\,\rm BB}_{\ell} = C^{\,\rm lens}_{\ell} + \hat{C}^{\,\rm fgds}_{\ell} + \hat{C}^{\,\rm noise}_{\ell},
\end{align}
where $C^{\,\rm lens}_{\ell_b}$ is the power spectrum of the CMB lensing $B$-mode signal ($r=0$, $A_L=1$), $\hat{C}^{\,\rm fgds}_{\ell}$ the power spectrum of residual foregrounds (left panels of Fig.~\ref{Fig:power_spectra}), and $\hat{C}^{\,\rm noise}_{\ell}$ the noise power spectrum. The reconstructed CMB $B$-mode map can be corrected for part of the cosmic variance of the lensing signal, either through internal delensing  \citep{Larsen2016,Carron2017,Millea2019} or through external lensing tracers \citep{Sherwin2015,Planck2018_VIII} such as cosmic infrared background maps \citep{Planck2016GNILC}.

For our likelihood analysis, we will assume three levels of delensing: no delensing ($A_L=1$), $60\%$ delensing ($A_L=0.4$), and full delensing ($A_L=0$), such that after component separation and delensing the measured CMB $B$-mode power spectrum is
\begin{align}
\label{eq:delens}
\tilde{C}^{\,\rm BB}_{\ell} \equiv \hat{C}^{\,\rm BB}_{\ell} - (1-A_L)\,C^{\,\rm lens}_{\ell} =A_L\,C^{\,\rm lens}_{\ell} + \hat{C}^{\,\rm fgds}_{\ell} + \hat{C}^{\,\rm noise}_{\ell}.
\end{align}
The measured CMB $B$-mode power spectrum, $\tilde{C}^{\,\rm BB}_{\ell}$, can in principle be corrected for the noise bias $\hat{C}^{\,\rm noise}_{\ell}$ and for the residual lensing bias $A_L\,C^{\,\rm lens}_{\ell}$, so that after these corrections, any irreducible bias on the measured CMB $B$-mode power spectrum arises from the residual foreground contamination:
\begin{align}
\hat{C}^{\,\rm fgds}_{\ell} = \tilde{C}^{\,\rm BB}_{\ell} - A_L\,C^{\,\rm lens}_{\ell} - \hat{C}^{\,\rm noise}_{\ell}.
\end{align}
The binned likelihood \citep{Hamimeche2008} on the tensor-to-scalar ratio $r$ is thus built as
\begin{align}
\label{eq:lkl}
-2\ln \mathcal{L}\left(r\right) =\sum_{\ell_b,\,\ell'_b} \left(\hat{C}^{\,\rm fgds}_{\ell_b} \,-\, r\,C^{\,\rm prim}_{\ell_b} \right) {\rm M}^{-1}_{\ell_b\ell'_b} \left(\hat{C}^{\,\rm fgds}_{\ell'_b} \,-\, r\,C^{\,\rm prim}_{\ell'_b}\right),
\end{align}
where $\hat{C}^{\,\rm fgds}_{\ell}$ is the power spectrum of residual foregrounds, $C^{\,\rm prim}_{\ell_b}$ is the primordial CMB $B$-mode power spectrum model for a tensor-to-scalar ratio $r=1$, and ${\rm M}_{\ell\ell^{'}}$ is the covariance matrix for a fiducial cosmological model with $r=0$. Given that the input sky simulations do not contain any primordial signal, i.e. $r=0$, the likelihood Eq.~\eqref{eq:lkl} thus computes the equivalent $r$ bias due to residual foreground contamination, and its significance with respect to the overall uncertainty due to cosmic variance of residual lensing, noise, and residual foregrounds, which are all included in the covariance matrix.

The binning, $\ell_b\pm \Delta\ell/2$ with $\Delta\ell=16$, of the power spectra mitigates correlations between different $\ell$ modes, so that off-diagonal terms of the covariance matrix can be neglected, thus leaving only the diagonal elements:
\begin{align}
\label{eq:m}
&{\rm M}_{\ell_b\ell_b} = {2\over \left(2\ell_b+1\right)f_{\rm sky}\Delta\ell} \left(\hat{C}^{\,\rm BB}_{\ell_b}-\left(1-A_L\right)C^{\,\rm lens}_{\ell_b}\right)^2,&
\end{align}
which, according to  Eq.~\eqref{eq:delens}, accounts for the cosmic variance of the residual lensing signal ${2(A_LC^{\,\rm lens}_{\ell})^2/(2\ell+1)}$, the sample variance of the residual foreground power spectrum ${2 (\hat{C}^{\,\rm fgds}_{\ell})^2/(2\ell+1)}$, the sample variance of the noise power spectrum ${2 (\hat{C}^{\,\rm noise}_{\ell})^2/(2\ell+1)}$, and their cross-terms. The likelihood results on $r$ are obtained by summing modes between ${\ell_{\rm min}=2}$ and ${\ell_{\rm max}=450}$, thus accounting for both the reionization and recombination bumps.

Figure~\ref{Fig:r_stat} summarizes our results on the recovered tensor-to-scalar ratio $r$ after component separation with \texttt{NILC} or \texttt{cMILC} for both \litebird and \pico, and for different levels of delensing. Biased detections (SNR$\geq 2$) due to significant foreground residuals are shown in red with both $1\sigma$ (thick red) and $2\sigma$ (thin red) error bars, while for unbiased (SNR$ < 2$) measurements, the 95\% upper bounds on $r$ are depicted by the thick green lines. 
As already observed on map reconstruction (Sect.~\ref{subsubsec:maps}-\ref{subsubsec:stats}) and power spectrum analysis (Sect.~\ref{subsubsec:ps}), adding constraints on an increasing number of moments with \texttt{cMILC} generally increases the uncertainty on $r$ due to noise degradation, but allows for eliminating biases on $r$ due to residual foreground contamination.
This thus can help to robustly turn high significance but {\it false} detections with \texttt{NILC} into unbiased upper limits with the optimal choice of moment number in \texttt{cMILC}.
However, the details depend on the experimental configuration and the overall level of delensing as we explain now.

The goal for full success of the \litebird mission \citep{Litebird2019} is to achieve $\delta r \lesssim 10^{-3}$, including both statistical and systematic uncertainties. For additional comparison, we thus define the simple metric
\begin{align}
\delta r \equiv \sqrt{r^2+\sigma^2(r=0)},
\end{align}
which accounts for both the \textit{statistical error} $\sigma(r=0)$ and the \textit{systematic error}, i.e. the residual bias $r$ with respect to $r=0$. For \litebird (upper panels of Fig.~\ref{Fig:r_stat}), \texttt{NILC} clearly leads to a significant bias on $r$ in our simulations, with $r = (2.5\pm 1.0)\times 10^{-3}$ in the absence of any delensing and $r = (3.2\pm 0.4)\times 10^{-3}$ in case of full delensing, which corresponds to a $3\sigma$ to $8\sigma$ bias on $r=0$ and thus a cumulative systematic and statistical error of $\delta r \gtrsim 3\times 10^{-3}$. 
Therefore, the systematic foreground residuals of \texttt{NILC} would not allow \litebird to robustly reach its goals, even if the statistical uncertainties are sufficient.
In contrast, for \texttt{cMILC} the bias on $r$ progressively decreases by deprojecting more and more foreground moments, with cMILC06, cMILC07, and cMILC08 all leading to unbiased measurements of $r=0$, i.e. $2\sigma$ upper limits. 
Among these three unbiased versions, we see that for \litebird cMILC06 (${f_{\rm sync}, f_{\rm dust}, \partial_\beta\,f_{\rm dust}}$) provides the best trade-off between bias mitigation and noise degradation on $r$, with $r = (0.7\pm 1.1)\times 10^{-3}$ consistent with unbiased measurement of $r=0$, while accumulating a systematic and statistical errors as low as ${\delta r \simeq \sigma(r=0) \simeq 10^{-3}}$. This would allow a constraint on $r$ that is in line with the goals for \litebird. 
Deprojecting dust temperature moments $\partial_T f_{\rm dust}$ (cMILC08) would further reduce residual foreground biases down to $r\lesssim 0.3\times 10^{-3}$, but the limited spectral coverage of \litebird at high frequency does not seem to provide enough sensitivity to constrain dust temperature moments and the noise degradation for cMILC08 starts blowing up. Finally, increasing the level of delensing helps reducing statistical uncertainties on $r$ for \texttt{NILC} and low-order \texttt{cMILC}. In contrast, the gain is less significant for the most-constrained versions of \texttt{cMILC} like cMILC06, because noise dominates over the lensing error for \litebird (see upper right panel of Fig.~\ref{Fig:power_spectra}).

\begin{figure*}
  \begin{center}
     \includegraphics[width=0.98\textwidth]{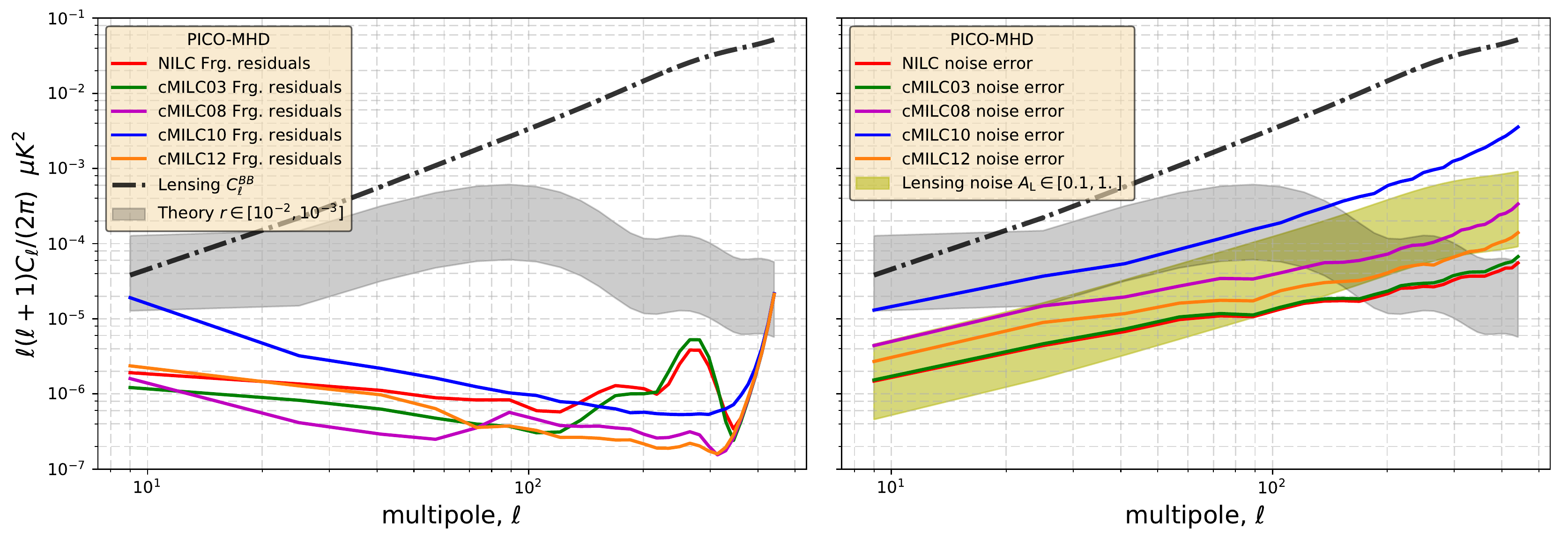}
    \\
     \includegraphics[width=0.98\textwidth]{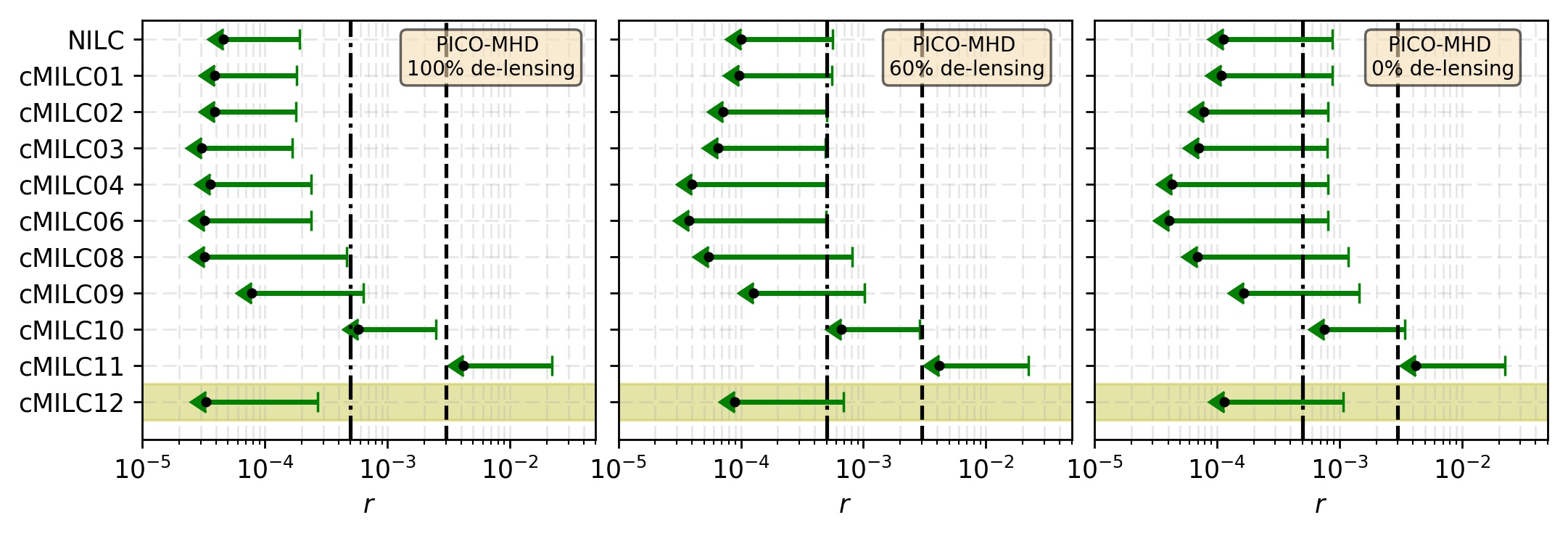}
    \end{center}
\caption{Idem Fig.~\ref{Fig:power_spectra} and \ref{Fig:r_stat}, but for the MHD foreground simulation with line-of-sight effects. By deprojecting moments arising from line-of-sight contributions, \texttt{cMILC} helps reducing biases on $r$ without paying much noise penalty with respect to \texttt{NILC} in terms of $2\sigma$ upper limit. cMILC03, cMILC04 and cMILC06 provide the best results on $r$ in terms of trade-off between residual bias and noise penalty.}
\label{Fig:r_stat_mhd}
\end{figure*}

For \pico (lower panels of Fig.~\ref{Fig:r_stat}), we can generally say that the overall degradation of the statistical error is not as large when including progressively more moments until about cMILC10. This is due to the broad frequency coverage and high channel sensitivity of \pico, which allows us to extract a lot of extra information even beyond $r$.
However, the overall performance of the method for recovery of $r$ depends significantly on the level of delensing.
Assuming full delensing (left panel), \texttt{NILC} would yield $r = (2.2\pm 0.4)\times 10^{-4}$, which corresponds to a $5\sigma$ bias on $r=0$ due to residual foreground contamination, while \texttt{cMILC} allows to reduce the bias on $r$ by progressively deprojecting more moments of the foregrounds.
For now omitting the optimised hybrid case cMILC12 (which we will return to in Sect.~\ref{subsec:hybrid}), for \pico, cMILC08 (${f_{\rm sync}, f_{\rm dust}, \partial_\beta\,f_{\rm sync}, \partial_\beta\,f_{\rm dust}, \partial_T\,f_{\rm dust}}$) provides the best results, giving $r = (0.7\pm 0.9)\times 10^{-4}$ consistent with $r=0$, and hence a cumulative systematic and statistical error as low as $\delta r\simeq 10^{-4}$. 
Due to high-frequency coverage above $500$\,GHz, \pico allows to constrain dust temperature moments $\partial_T f_{\rm dust}$ (cMILC08) with enough sensitivity to control residual foreground biases, while still mitigating the noise degradation. Nevertheless, for $A_{\rm lens}=0$, we can see that {\it all} considered methods, including \texttt{NILC}, provide tight constraints on $r$, with biases typically smaller than $r=5\times 10^{-4}$, which is the target detection of \pico \citep{PICO2019}.
For partial delensing, all the considered methods provide quite consistent results on $r$, compatible with $r=0$, since the residual lensing error dominates over the noise for \pico (see lower right panel of Fig.~\ref{Fig:power_spectra}). However, as we will show below, pivot optimization can affect this behaviour.

Overall our results indicate, \texttt{NILC} alone does not lead to unbiased estimates of $r$ in the case of \litebird, but has a good performance for \pico. By applying \texttt{cMILC}, these limitations can be overcome and even \pico's constraints on $r$ could be further improved. It also seems that with the \texttt{d1s1} simulations inclusion of first order moments suffices for constraints on $r$, even if \pico does show sensitivity to the second order moment of $T_d$.

\subsubsection{More complex foreground models: MHD simulation with line-of-sight contributions}
\label{subsubsec:mhd}

As stressed by \cite{Tassis2015}, the effective SED of the Galactic thermal dust foreground emission must be less trivial than a simple modified blackbody because of the average of multiple cloud contributions of various emissivities, temperatures, and magnetic fields orientations along the line-of-sight, thus leading to some decorrelation of the polarized dust emission across frequencies. 
The moments of the effective foreground emission resulting from line-of-sight averaging effects can in principle be deprojected with \texttt{cMILC}. In this section, we thus investigate the performance of \texttt{cMILC} on sky simulations having non-trivial foreground complexity, by using publicly available MHD-model simulations\footnote{\url{https://zzz.physics.umn.edu/ipsig/20180424_dc_maps} (model 96).} that were produced for the \pico data challenge.

In this sky simulation, the dust and synchrotron emissions are consistently derived from MHD simulations of the magnetic field turbulence in the interstellar medium \citep{Kritsuk2018,Kim2019}, with the dust model described in \cite{Hensley2015}. The MHD-model simulation is particularly interesting because of integrating multiple modified blackbodies of varying temperatures and spectral indices along the line-of-sight, so that the resulting dust SED in each pixel will not be a perfect modified blackbody, which makes this foreground model both more realistic and more challenging for parametric fitting methods aiming at $B$-mode component separation. The CMB and noise realisations in the MHD simulation are the same as those in the {\tt d1s1} simulation.

Figure~\ref{Fig:r_stat_mhd} summarises the \texttt{cMILC} results on the residual foregrounds and noise contamination to the CMB $B$-mode power spectrum for the MHD simulation (upper panels), together with the results on the recovered tensor-to-scalar ratio (lower panels). By deprojecting foregrounds moments with the semi-blind method \texttt{cMILC}, we are able to tackle the foreground complexity arising from line-of-sight effects in the MHD simulation, and again reduce the residual foreground foreground contamination further than \texttt{NILC}. The best results on $r$ for the MHD simulation are obtained with cMILC03 and cMILC06. For $60\%$ delensing, cMILC06 provides $r=\left(0.3\pm 0.8\right)\times 10^{-4}$ consistent with $r=0$, and a cumulation of systematic and statistical errors of ${\delta r = \sqrt{r^2+\sigma^2(r)} \lesssim 2.1\times 10^{-4}}$, which reduces down to ${\delta r \lesssim 8.5\times 10^{-5}}$ in the case of full delensing.

Interestingly, deprojecting the first- and second-order moments of synchrotron in cMILC08, cMILC09, cMILC10, and cMILC11 tends to degrade the results for the MHD simulation. 
We interpret this behaviour as being caused by the way the MHD simulation has been built: while the dust emission has spectral variations along the line-of-sight and across the sky, the synchrotron index is uniform in the MHD simulation (Brandon S. Hensley, private communication). This implies that higher-order moments of the synchrotron spectral index do not contribute in this simulation, and any constraints that nulls synchrotron components in \texttt{cMILC} will only increase the variance of the unconstrained foregrounds and noise. This also highlights that \texttt{cMILC} can provide a solid diagnosis of the level of spectral complexity of each foreground.

Overall our analysis shows that \texttt{cMILC} is a quite robust component separation method, which without any extra modifications also allows to handle foregrounds complexity arising from line-of-sight and other averaging effects. In addition, the effect of unmodeled spectral complexity is successfully reduced by means of the blind variance minimization within \texttt{cMILC}.

\section{Discussion}
\label{sec:disc}
While the previous sections demonstrated some general aspects of {\tt cMILC}, reaching the full potential of the method and comparison to other methods requires more work. 
Here, we first discuss some possible optimisations of \texttt{cMILC} (Sects.~\ref{subsec:optimal_pivot} and \ref{subsec:hybrid}), and then briefly mention the relevance to ground-based CMB experiments (Sect.~\ref{subsec:ground}), and other observables and figures of merit (Sect.~\ref{subsec:other}).
\newline 

\begin{figure*}
  \begin{center}
     \includegraphics[width=0.97\textwidth]{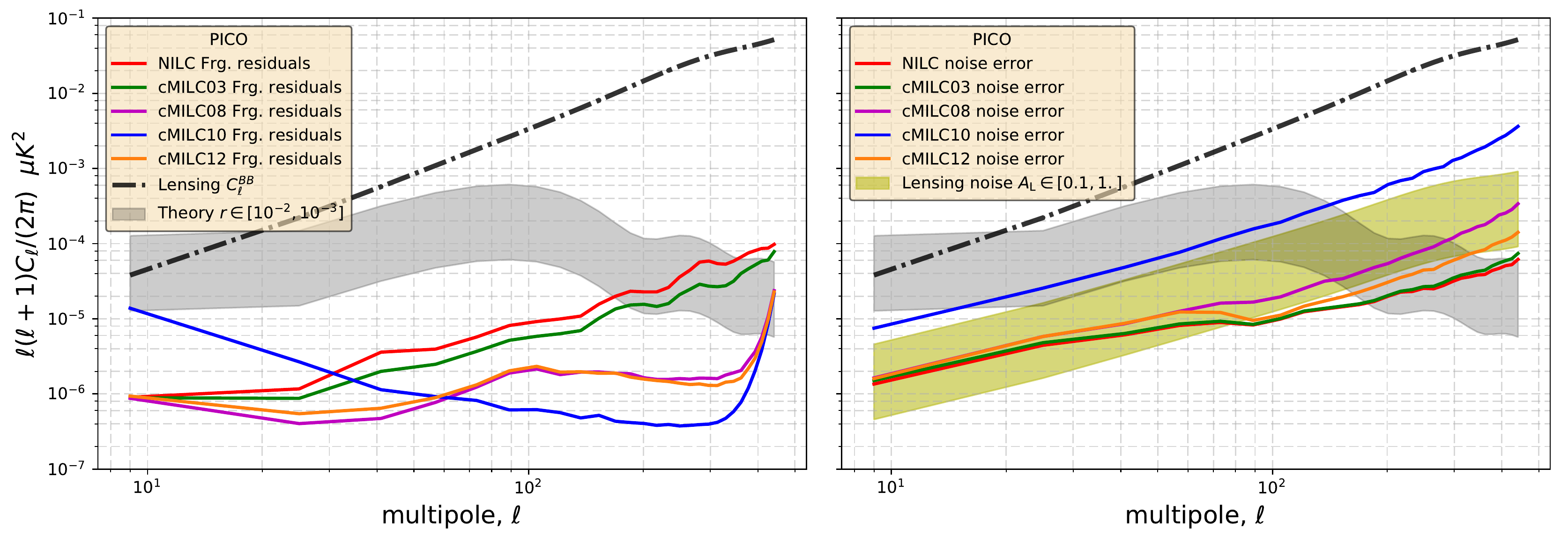}~\\
      \includegraphics[width=0.97\textwidth]{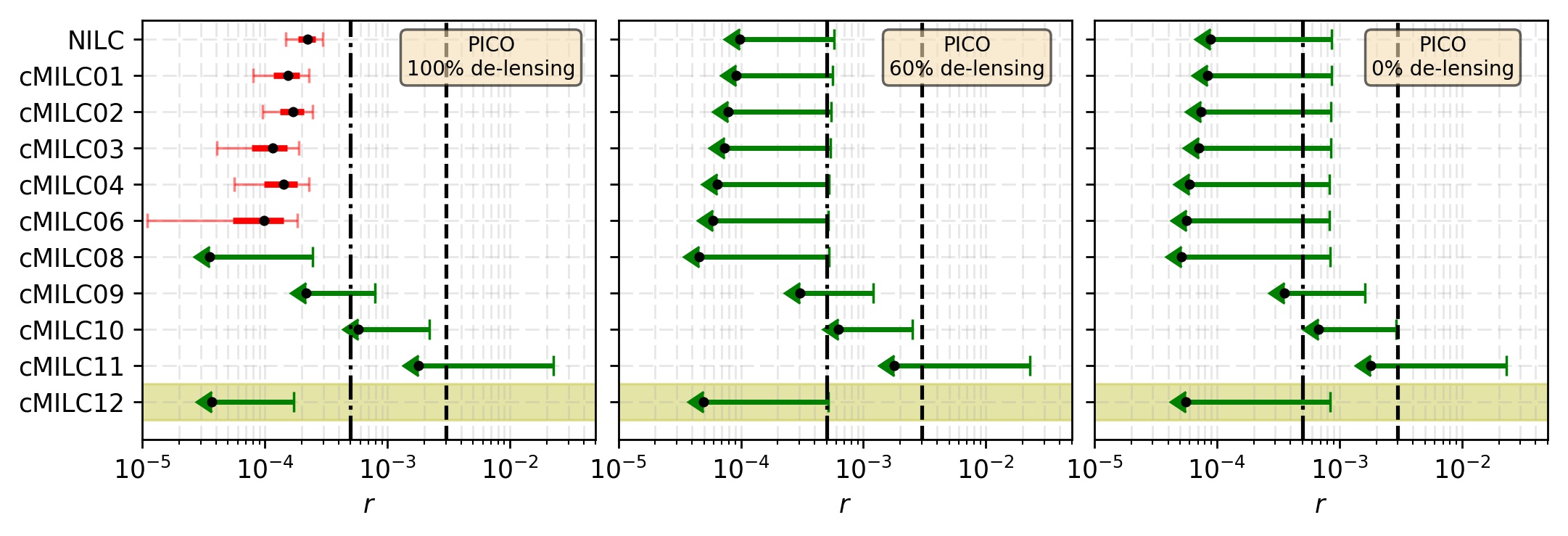}~
     \end{center}
\caption{Update of Fig.~\ref{Fig:power_spectra} and Fig.~\ref{Fig:r_stat} for \pico\ {\tt d1s1} by adopting a different pivot temperature, $\overline{T}_d=16$\,K, for \texttt{cMILC} at low $\ell$. By updating the pivot at low $\ell$, \texttt{cMILC} allows to reduce the residual foreground contamination across the whole range of multipoles, and thus progressively reduces the bias on $r$ by deprojecting more and more moments, without paying much noise penalty.}
\label{Fig:optimal_pivot}
\end{figure*}

\subsection{On the importance of optimal pivot parameters}
\label{subsec:optimal_pivot}

Throughout our analysis, we have been using first-guess pivots ${\{\overline{\beta}_s,\overline{\beta}_d,\overline{T}_d\}}$ in \texttt{cMILC} (Eq.~\ref{eq:cmilc}), based on the mean spectral index and temperature of dust and synchrotron issued from the current knowledge on intensity foregrounds \citep[e.g.][]{Planck2015_X}. In the language of moment expansion, these first-guess pivots from the literature typically result from fitting a zeroth-order model to the effective SED of the foreground emission:
\begin{align}
\label{eq:zeroth}
 I(\nu)  &= A_1\,f_{\rm sync}(\nu,\beta_s) + A_2\, f_{\rm dust}(\nu,\beta_d, T_d)\,,
\end{align} 
where the pivots, i.e. $\beta_s,\beta_d \textrm{ and } T_d$, are fit in each pixel and then averaged.
As such, the zeroth-order model Eq.~\eqref{eq:zeroth} might not provide the best fit at the largest angular scales, where averaging effects and higher-order moment corrections become relevant to the SED. 
As pointed out above, in Fig.~\ref{Fig:power_spectra} for \pico\ {\tt d1s1} one can see that the performance of \texttt{cMILC} in reducing foreground contamination is spectacular on a large range of multipoles but non-optimal at the lowest multipoles $\ell \lesssim 15$, which are most relevant for constraining $r$. This suggests that first-guess pivots derived from traditional zeroth-order moment fits (Eq.~\ref{eq:zeroth}) might not be optimal for cleaning foregrounds at the largest angular scales, where averaging effects, and thus higher-order moments, become relevant. Another complication is that the optimal pivots will generally depend on the type of method that is employed, given that analysis choices vary but directly introduce averaging effects.

To investigate this point further, we
thus modified the pivot values in the first needlet band to evaluate the impact on the performance of \texttt{cMILC} at low $\ell$. In particular, we found that lowering the pivot value for the dust temperature down to $\overline{T}_d=16$\, K actually improves the performance of \texttt{cMILC} in cleaning foregrounds at low $\ell$, and shows a more consistent picture of monotonic reduction of the bias on $r$ when deprojecting more and more moments with \texttt{cMILC}. 
The results for \pico\ {\tt d1s1} simulation with the updated pivot temperature are shown in Fig.~\ref{Fig:optimal_pivot}. A progressive reduction of the bias on $r$, without much noise penalty, is clear from cMILC01 to cMILC08, while it actually breaks for higher-order constraints (cMILC09, cMILC10, cMILC11), suggesting that a proper optimisation of {\it all} the pivots would probably be required instead of this simple ad-hoc variation of the pivot temperature. 
Still, these results highlight how the performance of \texttt{cMILC} is sensitive to the choice of pivots at low $\ell$, and the importance of choosing the most appropriate pivot parameters for the moment expansion. While a full optimisation of the pivots using sophisticated parametric pivot fitting algorithms deserves an in-depth work that is beyond the scope of this paper, hereafter we highlight a few important aspects. 

We attempted optimizing the pivots by extracting mean polarisation SEDs from our simulation. However, for polarisation the definition of an average SED is not as straightforward, given that no polarisation 'monopole' exists. This makes it hard to define a concrete metric for the pivot optimization. This metric is also expected to directly depend on other analysis choices such as real-space versus harmonic-space analysis.

Besides, further optimisation of \texttt{cMILC} would probably be possible by using local pivots across different regions of the sky. While here we have been using uniform pivot parameters across the sky throughout our analysis, the all-sky average temperature and spectral index of the dust and synchrotron differ from the local mean index and temperature in a certain region of the sky, e.g., in the BICEP2 region. Therefore, a possible improvement of the \texttt{cMILC} method would be to adopt different pivot parameters in different sky regions and perform a local moment expansion. One open question is then how to optimally combine the results from the various regions to constraint on $r$.

Finally, further optimisation of \texttt{cMILC} could potentially be achieved for the localisation of filtering in both pixel space and harmonic space, through the investigation of different needlet windows in terms of number, shape, and width in order to determine whether the performance of the foreground cleaning would benefit from more localisation of the filter in pixel space or more localisation in harmonic space. All this outlines a high-dimensional optimisation problem that will be carried out in the future.

\subsection{Hybrid moments across multipole ranges} 
\label{subsec:hybrid}
The relative contribution of the different moments of dust and synchrotron, e.g., ${A_{\nu_s}(\beta_s - \overline{\beta}_s)\partial_{\beta_s} f_{\rm sync}}$ versus ${A_{\nu_d}(\beta_d - \overline{\beta}_d)\partial_{\beta_d} f_{\rm dust}}$, to the overall foreground power spectrum is likely to vary across multipoles. In particular, we experienced that the first-order synchrotron index moment had more constraining power at low multipoles $\ell \leq 50$, while the first-order dust index moment was helping more at $\ell > 50$. 
Given that \texttt{cMILC} is performed on a needlet frame (see Fig.~\ref{Fig:needlets}), it allows us to set different combinations of moment constraints for different multipole windows in order to optimise the trade-off between residual foreground biases and noise degradation across multipoles. 

To explore this idea further, we thus implemented a hybrid version cMILC12 (see Table~\ref{tab:nom}), which deprojects the moments ${f_{\rm sync}, f_{\rm dust}, \partial_\beta\,f_{\rm sync}, \partial_T\,f_{\rm dust}}$ in the first two needlet bands and the moments ${f_{\rm sync}, f_{\rm dust}, \partial_\beta\,f_{\rm dust}, \partial_T\,f_{\rm dust}}$ at higher multipoles $\ell > 50$. As shown in Fig.~\ref{Fig:r_stat} and Fig.~\ref{Fig:optimal_pivot}, the hybrid version cMILC12 provides the best trade-off and result of the analysis, with both the lowest bias on $r$ due to residual foregrounds and the smallest statistical uncertainty due to low noise degradation: $r = (0.4\pm 0.5)\times 10^{-4}$ consistent with an unbiased recovery of $r=0$, and a cumulative systematic and statistical errors of ${\delta r = \sqrt{r^2+\sigma(r)^2} < 6\times 10^{-5}}$.
\texttt{cMILC} is thus quite flexible by allowing for different combinations of moments across multipoles. 

With future foreground observations, it will be possible to use \texttt{cMILC} for diagnostics of the most relevant moments of the foreground emission for each ranges of multipoles. 
Furthermore, given that the various moments of the non-Gaussian foreground contamination are spatially correlated with each other, it would in principle be possible to compress the number of moments into a small set of independent components ranked by their relevance to optimise foreground cleaning with \texttt{cMILC}. These ideas also have to be worked out more cleanly in the future.

\subsection{Ground-based surveys}
\label{subsec:ground}
For future ground-based CMB surveys, like SO \citep{SO2019} and CMB-S4 \citep{CMB-S4_2016}, it is more difficult to directly probe the reionization peak of the primordial CMB $B$-mode at the lowest multipoles $\ell \lesssim 15$ because only a fraction of the sky from the ground is accessible. Hence, future ground-based CMB experiments will have to rely mainly on information from the recombination peak (i.e. $30 \leq \ell \leq 200$) to constrain $r$. In addition, the number of available frequency bands for ground-based experiments is limited by atmospheric windows, which will limit the number of foreground moments that can be deprojected by \texttt{cMILC}.

However, as evident from the left panels of Fig.~\ref{Fig:power_spectra}, the deprojection of only two or three foreground moments with \texttt{cMILC} is already very successful around the recombination peak, with the reduction of the residual foreground contamination of \texttt{cMILC} in comparison to \texttt{NILC} being most significant at intermediate multipoles  $30 \leq \ell \leq 200$. Therefore, \texttt{cMILC} would be particularly helpful also for future ground-based CMB surveys in removing residual foreground biases on $r$. With upcoming ground-based experiments like CCAT-prime \citep{Aravena:2019tye}, which will provide wider frequency coverage owing to improved atmospheric transparency, we may anticipate significant benefits of using \texttt{cMILC}.
However, dedicated forecasts are required taking differences in the frequency coverage and angular resolution into account.

\subsection{Other observables and figures of merit} 
\label{subsec:other}
While in this work we focused on component separation for $B$-modes, with $r$ defining the figure of merit, \texttt{cMILC} is also directly applicable to CMB temperature and $E$-mode analyses relevant to other observables. As shown in Fig.~\ref{Fig:power_spectra}, the potential of \texttt{cMILC} in removing foregrounds through moments significantly improves towards small angular scales.
The details may change with further optimizations of \texttt{cMILC}, but generally we expect that \texttt{cMILC} could also help improve constraints on cosmological parameters such as $N_{\rm eff}$ and $\sum m_\nu$, which depend on damping tail physics \citep[e.g.,][]{Abazajian2015}. Additional benefits may become apparent for studies of the Sunyaev-Zeldovich (SZ) power spectrum \citep[e.g.,][]{Komatsu2002, Hill2013, Bolliet2019}, which again rely on information gleaned from small scale fluctuations.

Our method could also provide gains in future searches for primordial non-Gaussinity and CMB lensing/delensing analyses. As shown for the one-point statistics of foreground residuals in Fig.~\ref{Fig:sky_stat}, \texttt{cMILC} allows significantly reducing \textit{non-Gaussian} foreground contamination in the CMB map by deprojecting moments. Since CMB lensing \citep[e.g.][]{Hu2002} and bispectrum estimators  \citep[e.g.][]{Bucher2010} rely on the extraction of non-Gaussian features (no matter if primordial or generated by lensing), they are prone to non-Gaussian foreground residuals in the CMB map \citep[e.g.][]{Hill2018,vanEngelen2014}, and thus to be potentially biased depending on the sensitivity of future experiments. This aspect deserves further investigation but is beyond the present scope.

\section{Conclusions}
\label{sec:conc}

For robust measurements of the tensor-to-scalar ratio $r$ as a signature of primordial gravitational waves from inflation, it is essential to minimise both statistical and systematic errors due to residual foregrounds in component separation analyses. 
Statistical uncertainties $\sigma(r)$ arise from the cosmic/sample variance of the residual lensing, foreground, and noise contaminations after component separation and delensing, while systematic errors, leading to biases on $r$, arise from the residual foreground contamination to the CMB $B$-mode power spectrum after component separation. Hence, reliable foreground cleaning is achieved by minimizing cumulative systematic and statistical errors ${\delta r = \sqrt{\left(r-r^{\,\rm true}\right)^2+\sigma^2(r)}}$.

In the present work, we developed the new semi-blind component separation method \texttt{cMILC}, which deprojects the main statistical moments of the foregrounds without altering the CMB $B$-mode signal, thus allowing to reduce systematic errors (or biases) on $r$ due to residual foreground contamination, while still mitigating statistical uncertainties (see Figs.~\ref{Fig:r_stat}, \ref{Fig:r_stat_mhd}, \ref{Fig:optimal_pivot}).
By applying \texttt{cMILC} to sky simulations with varying foreground complexity for experimental settings similar to those of \litebird and \pico, we have identified specific combinations of foreground moments that optimise the trade-off between residual foreground biases and noise degradation for constraints on $r$. In particular, the \texttt{cMILC} method allows us to reach the sensitivity goals on $r$ of both \litebird and \pico mission concepts, by removing biases on $r$ without excessive noise penalties, thus overcoming limitations of \texttt{NILC}.

We show that for LiteBIRD a deprojection of the first-order moment of the dust spectral index (Table~\ref{tab:nom}: cMILC06) is needed for unbiased detection of $r=0$ (see Fig.~\ref{Fig:r_stat}). In this case, the cumulative systematic and statistical errors (i.e. bias + variance) is ${\delta r = \sqrt{r^2+\sigma^2(r)} \lesssim 10^{-3}}$, where the blind \texttt{NILC} method would exceed $\delta r \gtrsim 3\times 10^{-3}$ due to more than $3\sigma$ residual foreground bias on $r=0$. 
For \pico, deprojecting the first-order moments of the dust spectral index and dust temperature (Table~\ref{tab:nom}: cMILC08) provides the best unbiased results on $r$ without much noise penalty (see Figs.~\ref{Fig:r_stat}, \ref{Fig:optimal_pivot}). This yields $r=\left(0.7\pm0.9\right)\times 10^{-4}$, and hence a cumulative systematic and statistical error as low as ${\delta r \simeq 10^{-4}}$. 

While our analysis shows that for constraints on $r$ in principle only first moments are needed, \pico is further sensitive to the second moment of the dust temperature. This conclusion is expected to be a strong function of the complexity of the foreground simulation, and thus may not hold for more complicated skies.
In the presence of multiple dust modified blackbodies along the line-of-sight (Fig.~\ref{Fig:r_stat_mhd} in Sect.~\ref{subsubsec:mhd}), \texttt{cMILC} still performs well and helps reducing residual foreground contaminations and the bias on $r$ without paying much noise penalty.
Our analysis further shows that the optimization of pivots plays an important role (Sect.~\ref{subsec:optimal_pivot}), requiring more investigation.
These findings highlight the potential of \texttt{cMILC} as a robust foreground cleaning method but also diagnostic tool, enabling to test the sky complexity in future analysis.

The ability of \texttt{cMILC} to peel off foregrounds through the deprojection of their spectral moments is not limited to the search for primordial $B$-modes and constraints on $r$, but could as well be exploited for CMB temperature and $E$-mode analyses to improve constraints on other cosmological parameters like $\sum m_\nu$ and $N_{\rm eff}$. 
By getting rid of non-Gaussian residuals (Fig.~\ref{Fig:sky_stat} and Sect.~\ref{subsubsec:stats}), \texttt{cMILC} is also of great interest to searches for primordial non-Gaussianity and in the CMB lensing reconstruction. \texttt{cMILC} would also be helpful for the extraction of other faint cosmological signals such as SZ effects and anisotropic primordial spectral distortions.
Forecasts for ground-based experiments should consider these additional observables in the analysis.

We also stress that moments of the foreground emission result not only from integrating multiple contributions along the line-of-sight, but also arise from several averaging processes across the sky like beam convolutions, spherical harmonic transforms, $Q,U$ to $E,B$ transforms, and other filtering processes (e.g. needlets). 
In fact, future CMB surveys will deliver sky maps of varying native beam resolutions at different frequencies. At low frequencies, the beams are typically larger than at high frequencies, and hence averaging processes are intrinsically larger at the map level. If not properly taken into account when combining maps, this will lead to inevitable decorrelation across frequencies.
In addition, while most of the averaging effects can be propagated analytically, line-of-sight averaging is unavoidable, thus requiring a more general SED treatment even at the pixel level.
All the above aspects pose a challenge for foreground modelling and the search for primordial $B$-modes that cannot be ignored.
Robustly deprojecting the moments of dust and synchrotron emissions that arise from averaging/decorrelation effects is thus crucial to avoid systematic errors from residual foreground contamination that will bias the faint primordial $B$-mode signal, hence the tensor-to-scalar ratio. 
The \texttt{cMILC} method provides an avenue forward this direction. 

While here we specifically focused on a comparison to \texttt{NILC}, not only will it be important to extend this comparison to other methods but one should also think about augmenting other methods using moments. For instance, we anticipate that moment expansion approaches can help parametric component separation methods in the search for primordial $B$-modes, providing a systematic way of extending the list of expected foreground parameters.
The optimal method will also strongly depend on the observable under consideration, which further motivates more extensive comparisons of various cleaning methods in the future, ultimately preparing us for the analysis challenges in the years to come.
 
\small

\section*{Data Availability}
The simulated data used in this article were accessed from the repository /project/projectdirs/pico/ of the National Energy Research Scientific Computing Center (NERSC), a U.S. Department of Energy Office of Science User Facility operated under Contract No. DE-AC02-05CH11231.
%

\section*{Acknowledgements}
This project has received funding from the European Research Council (ERC) under the European Union's Horizon 2020 research and innovation programme (grant agreement No 725456, CMBSPEC). JC was also supported by the Royal Society as a Royal Society University Research Fellow at the University of Manchester. We thank Eiichiro Komatsu and Colin Hill for valuable comments on the manuscript, and the anonymous referee for their comments and suggestions. Some of the results in this paper have been derived using the \healpix\ package \citep{Gorski2005} and the \psm\ package \citep{Delabrouille2013}.

\bibliographystyle{mn2e}
\bibliography{bmode_moments_cilc}

\appendix

\section{Flexibility of cMILC concerning pivot parameter values}
\label{sec:flexi}

\begin{figure}
  \begin{center}
    \includegraphics[width=\columnwidth]{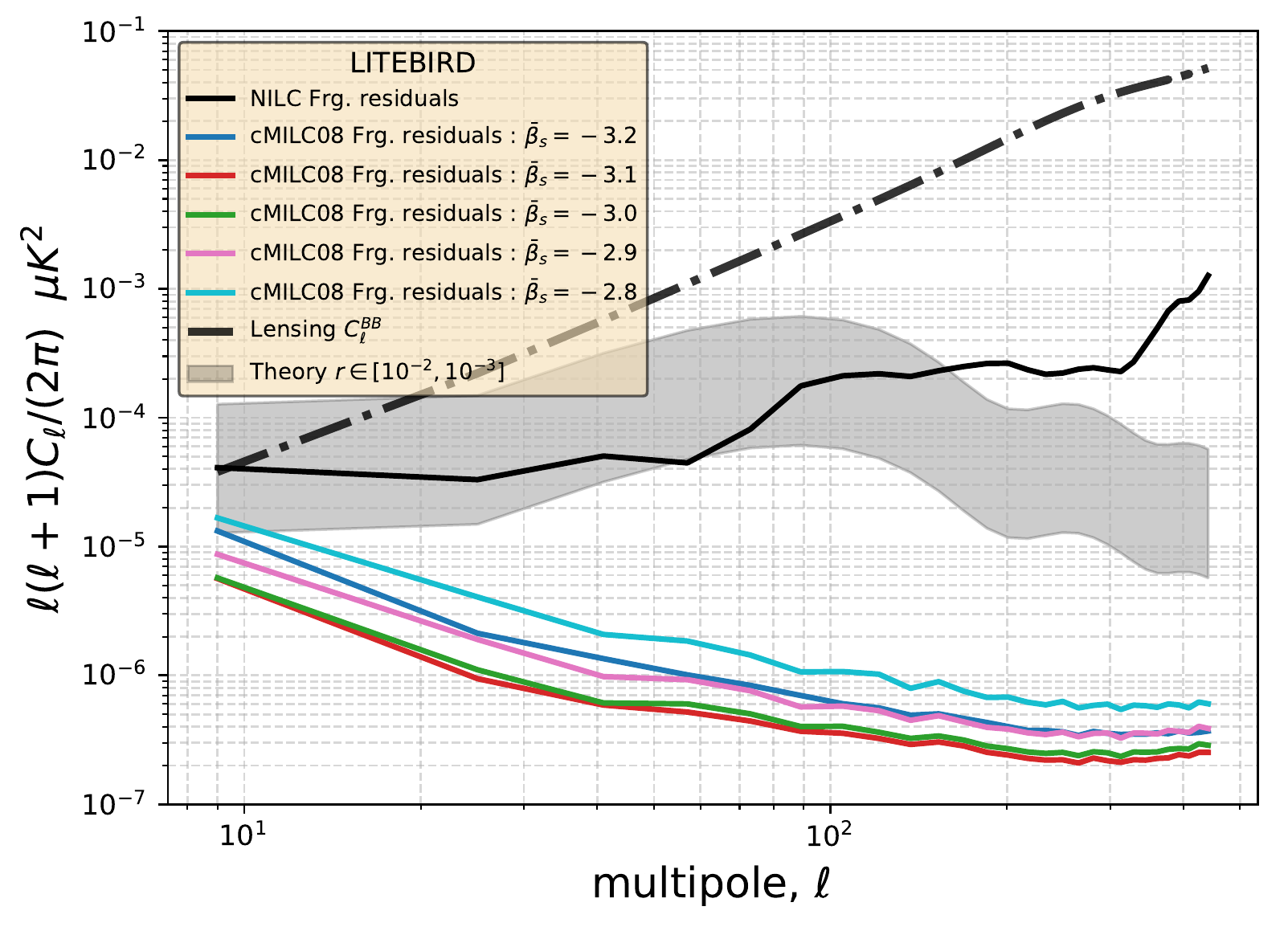}
    \\
    \includegraphics[width=\columnwidth]{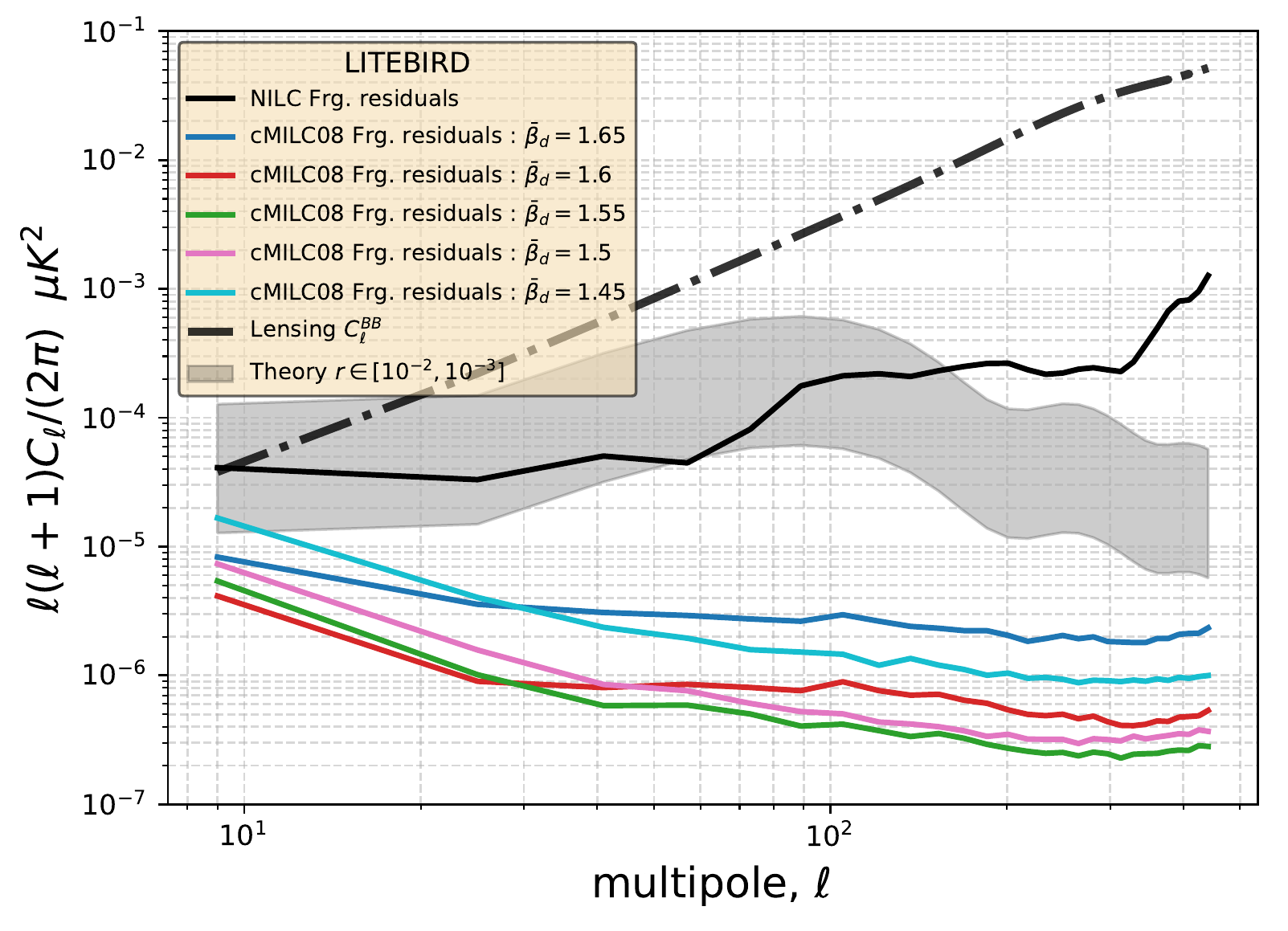}
    \\
    \includegraphics[width=\columnwidth]{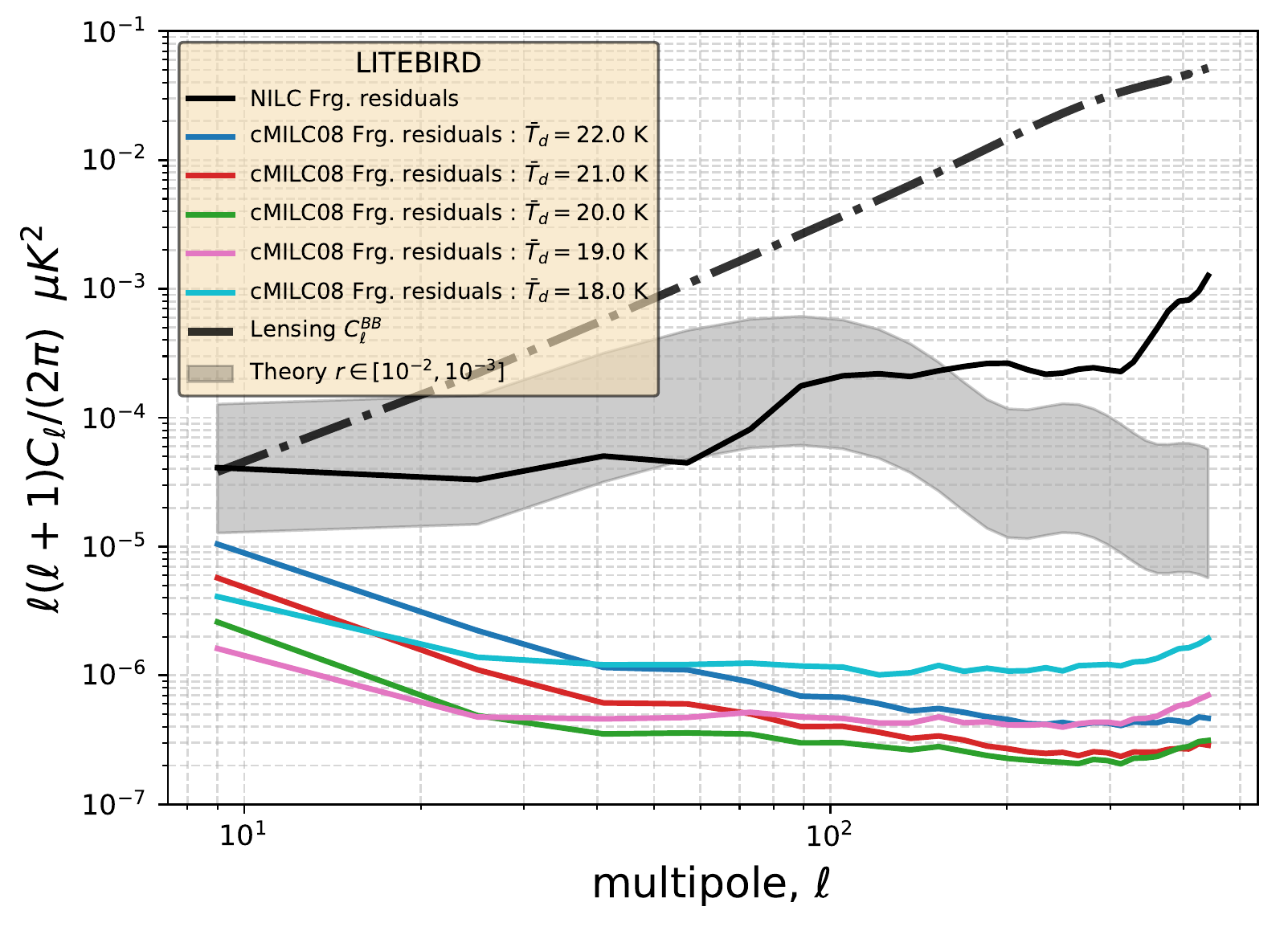}
   \end{center}
\caption{Performance of \texttt{cMILC} versus \texttt{NILC} in terms of residual foreground contamination when assuming different pivot values for $\overline{\beta}_s$, $\overline{\beta}_d$, and $\overline{T}_d$.}
\label{Fig:varpivot}
\end{figure}

By construction, \texttt{cMILC} allows for some flexibility in the assumed pivot values of the foreground parameters. For a set of assumed pivot values, the nulling constraints of \texttt{cMILC}  (Eq.~\ref{eq:cmilc}) enable to null any part of the foreground emission that projects onto the moments centred on these pivots, without altering the CMB signal due to the conservation constraint. Should there be any small departures in the data from the assumed pivot values then a part of the foreground contamination cannot be fully deprojected, but such unconstrained foreground contamination is anyway handled by blind variance minimization, like \texttt{NILC} would proceed. To illustrate the margin on the choice of pivot values, in Fig.~\ref{Fig:varpivot} we show the performance of \texttt{cMILC} in terms of residual foreground contamination for different pivot values, as compared to \texttt{NILC}.

\section{Experimental configurations}
\label{sec:experimental}
Summary of experimental parameters for \litebird (Table~\ref{tab:ltb}) and \pico (Table~\ref{tab:pico}).

\begin{table}
\caption{Instrumental specifications of \litebird \protect\citep{Litebird2019} for the updated design \protect\citep{Sugai2020} with sensitivities taken from \protect\cite{Ghigna2020}. }
  \label{tab:ltb}
 \centering
  \begin{tabular}{ccc}
\toprule
Frequency   & Beam FWHM & Sensitivity \\
$[\rm GHz]$   & $[\rm arcmin]$  & $[\rm \mu K.arcmin]$ \\
\midrule 
      40 & 69& 39.76 \\ 
      50 & 56& 25.76 \\ 
      60 & 48& 20.69\\ 
      68 &  43& 12.72 \\ 
      78 & 39& 10.39 \\ 
      89 & 35& 8.95 \\ 
      100 & 29& 6.43 \\ 
      119 & 25& 4.30 \\ 
      140 &  23& 4.43 \\ 
     166 &  21& 4.86\\ 
     195 &  20&  5.44 \\ 
     235 &  19& 9.72 \\ 
     280 &  24& 12.91 \\ 
     337 &  20& 19.07 \\ 
     402 &  17& 43.53 \\ 
\bottomrule
   \end{tabular}
\end{table}

\begin{table}
\caption{Instrumental specifications of  \pico~\protect\citep{PICO2019} for the baseline design.}
  \label{tab:pico}
 \centering
  \begin{tabular}{ccc}
\toprule
Frequency   & Beam FWHM & Sensitivity \\
$[\rm GHz]$   & $[\rm arcmin]$  & $[\rm \mu K.arcmin]$ \\
\midrule
      21 & 38.4 &   23.9\\ 
      25 & 32.0 &   18.4\\ 
      30 & 28.3 &   12.4\\ 
      36 & 23.6 &   7.9\\ 
      43 & 22.2 &    7.9\\ 
      52 & 18.4 &    5.7\\ 
      62 & 12.8 &    5.4\\ 
      75 & 10.7 &    4.2\\ 
      90 &  9.5 &    2.8\\ 
     108 &  7.9 &    2.3\\ 
     129 &  7.4 &    2.1\\ 
     155 &  6.2 &    1.8\\ 
     186 &  4.3 &    4.0\\ 
     223 &  3.6 &    4.5\\ 
     268 &  3.2 &    3.1\\ 
     321 &  2.6 &   4.2\\ 
     385 &  2.5 &   4.5\\ 
     462 &  2.1 &  9.1\\ 
     555 &  1.5 &  45.8\\ 
     666 &  1.3 &  177\\ 
     799 &  1.1 & 1050\\ 
\bottomrule
   \end{tabular}
\end{table}

\bsp	
\label{lastpage}
\end{document}